\documentclass{article}
\usepackage[utf8]{inputenc}

\pdfoutput=1
\usepackage{amssymb}
\usepackage{amsmath}
\usepackage[dvips]{graphicx}
\usepackage{setspace}
\usepackage{slashed}
\usepackage{mathtools}
\usepackage{amsfonts}
\usepackage{fancyhdr}
\usepackage{xcolor}
\usepackage{graphicx}
\usepackage{rotating}
\usepackage{comment}
\usepackage{color}
\usepackage{subcaption}
\usepackage[percent]{overpic}
\usepackage{cite}
\usepackage{braket}
\usepackage{moresize}
\usepackage{dsfont}

\definecolor{darkgreen}{rgb}{0,0.5,0}
\definecolor{darkblue}{rgb}{0,0,0.6}
\definecolor{purple}{rgb}{0.4,.2,0.7}
\newcommand{\p}{\partial}

\newcommand{\be}{\begin{equation}}
\newcommand{\ee}{\end{equation}}

\usepackage[colorlinks=true,citecolor=darkgreen,linkcolor=black,urlcolor=purple]{hyperref}

\usepackage{pdfsync}

\makeatletter
\newcommand*{\defeq}{\mathrel{\rlap{%
                     \raisebox{0.3ex}{$\m@th\cdot$}}%
                     \raisebox{-0.3ex}{$\m@th\cdot$}}%
                     =} 
\makeatother

\DeclareMathOperator{\Tr}{Tr}
\def\be{\begin{eqnarray}}
\def\ee{\end{eqnarray}}

\newcommand{\bea}{\begin{eqnarray}}
\newcommand{\eea}{\end{eqnarray}}
\def\ben{\begin{equation}}
\def\een{\end{equation}}

    \let\p=\phi \let\r=v

\def\be{\begin{equation}}
\def\ee{\end{equation}}
\def\ba{\begin{eqnarray}}
\def\ea{\end{eqnarray}}

\def\bal#1\eal{\begin{align}#1\end{align}}
\def\bs#1\es{\begin{split}#1\end{split}}

\renewcommand{\p}{\partial}

\allowdisplaybreaks  

\numberwithin{equation}{section}

\def\p{{\phi}}

\newcommand{\bz}{\bar{z}}

\def\be{\begin{equation}}
\def\ee{\end{equation}}
\def\ba{\begin{eqnarray}}
\def\ea{\end{eqnarray}}
\def\bal#1\eal{\begin{align}#1\end{align}}

\def\r{\rightarrow}

\def\r{\right}

\usepackage{tikz}
\usetikzlibrary{positioning,arrows}
\usetikzlibrary{decorations.pathmorphing}
\usetikzlibrary{decorations.markings}
\tikzset{
particle/.style={postaction={decorate}},
graviton/.style={decorate, decoration={snake, amplitude=0.8 mm, segment length=1.5 mm, pre length=0.8 mm, post length=0.8 mm}},
photon/.style={
        decoration={complete sines, amplitude=0.15cm, segment length=0.2cm},
        decorate    
    },
gluon/.style={
        decoration={coil, aspect=0.75, mirror, segment length=1.5mm},
        decorate
    }
}
 
\def \be {\begin{equation}}
\def \ee {\end{equation}}

\renewcommand{\p}{\partial}

\newcommand{\bh}{\bar{h}}

\renewcommand{\O}{\mathcal{O}}
\newcommand{\bL}{\bar{L}}
\newcommand{\id}{\mathbf{1}}
\newcommand{\CL}{D}

\usepackage{framed}

\begin{document}
\onehalfspacing

\begin{center}

~
\vskip5mm

{\LARGE  {
Statistics of three-dimensional black holes \\
\vspace{0.2cm}
from Liouville line defects
}}

\vskip7mm

Jeevan Chandra$^{D_\Sigma}$, Thomas Hartman$^{D_\Sigma}$, and Viraj Meruliya$^{L_\Sigma}$ 

\vskip5mm

{\it $^{D_\Sigma}$Department of Physics, Cornell University, Ithaca, New York, USA \\
\it $^{L_\Sigma}$Department of Physics, McGill University, Montreal, Quebec, Canada
}

\vskip5mm

\end{center}

\vspace{2mm}

\begin{abstract}
\noindent

Black holes and wormholes in the gravitational path integral can be used to calculate the statistics of heavy operators.  An explicit example in higher dimensions is provided by thin shells of matter. We study these solutions in 3D gravity, and reproduce the behavior of black holes and wormholes from the dual CFT using the large-$c$ conformal bootstrap. The CFT operator that creates a thin shell black hole is a line defect, so we begin by using the bootstrap to study the statistics of line defects, both at finite $c$ and in the holographic large-$c$ limit.  The crossing equation leads to a universal formula for the average high-energy matrix elements of the line defect in any compact, unitary 2d CFT with $c>1$. The asymptotics are controlled by a line defect in Liouville CFT at the same value of the central charge. At large $c$, three distinct quantities are related: The statistics of line defects in holographic CFTs, the individual matrix elements of a line defect in Liouville CFT, and the on-shell action of black holes and wormholes in 3D gravity.  The three calculations match for black holes, and if the statistics of the line defects are assumed to be approximately Gaussian, then a class of wormholes is also reproduced by the dual CFT.

 \end{abstract}

\pagebreak
\pagestyle{plain}

\setcounter{tocdepth}{2}
{}
\vfill
\tableofcontents


\date{}

\newcommand{\btau}{\bar{\tau}}

\section{Introduction}

Gravity, as a low-energy effective field theory, captures certain statistical information about the quantum theory in the UV. The most famous example is the black hole entropy: The area of the horizon tells us the density of states above the Planck scale, coarse-grained over a large number of nearby microstates. In some cases in string theory the microstates can be enumerated \cite{Strominger:1996sh}, but generally they are unknown, and presumably they cannot be found without specifying the UV completion. 

More recently, it has been discovered that higher topology contributions to the gravitational path integral, calculated in the low-energy effective field theory of gravity coupled to matter, capture higher statistics of the underlying quantum theory. The double cone calculates spectral statistics \cite{Cotler:2016fpe,Saad:2018bqo,Cotler:2020ugk}; the higher topologies of JT gravity match the statistics of a dual random matrix ensemble \cite{Saad:2019lba}; replica wormholes calculate the von Neumann entropy of Hawking radiation \cite{Penington:2019kki,Almheiri:2019qdq}; and on-shell wormholes supported by pointlike matter calculate the statistics of heavy local operators in two dimensions \cite{Penington:2019kki, Saad:2019pqd, Stanford:2020wkf} and three dimensions \cite{Chandra:2022bqq,Chandra:2023dgq,Chandra:2023rhx}. In higher dimensions, from the point of view of low energy gravity the physics is very similar, but explicit calculations are more difficult. A tractable example, in any dimension, is provided by black holes supported by spherically symmetric thin shells of matter, which exhibit the same phenomena, and the corresponding wormholes have been shown to have Gaussian statistics at leading order \cite{Chandra:2022fwi,Sasieta:2022ksu, Balasubramanian:2022gmo, Bah:2022uyz}. 

In this paper, we study in detail one of the simplest examples of ensemble averaging in more than two bulk dimensions: Thin shell black holes in 3D gravity. In this example, the results on the gravity side can be matched precisely to CFT calculations, and with spherical symmetry, there are simple analytic formulas. The black holes themselves have been studied for a long time (e.g. \cite{Balasubramanian:2010ce,Keranen:2015fqa}), and in the Vaidya limit of null gravitational collapse, they have been reproduced in the dual CFT \cite{Anous:2016kss} from an $n$-point correlation function in the $n \to \infty$ limit. Here we develop a more general approach to study thin shells, both on the gravity side and in CFT, and apply it to both black holes and wormholes. 

Thin shell black holes in the bulk are created by line operators in the dual CFT. We will therefore begin with a general bootstrap analysis of these line operators. This analysis applies to all unitary, compact 2d CFTs with Virasoro chiral algebra, at any central charge $c>1$, assuming only the existence of a certain type of line defect, $\CL_\Sigma$. (From now on these will be called simply `compact CFTs' with the other qualifiers assumed.) We study the crossing equation for the defect 2-point function $\langle \CL^\dagger_\Sigma \CL_\Sigma\rangle$, and use the short-distance limit to derive a universal formula for the average matrix elements of the line operator at high energies:
\begin{align}\label{introC0}
\overline{ | \langle i | \CL_\Sigma | j\rangle|^2} \approx C_0^{\CL}(h_i, h_j) C_0^{\CL}(\bh_i, \bh_j)
\end{align}
where $i,j$ label primary states, $(h,\bh)$ are conformal weights, and the average is over nearby states, as in the microcanonical ensemble. The function $C_0^{\CL}$ is thus a Cardy-like formula \cite{Cardy:1986ie, Collier:2019weq} or Tauberian theorem \cite{Pappadopulo:2012jk} for line defects, controlling the asymptotic spectral density. It plays a similar role for line defects as the universal OPE function (known as $C_0$) does for local operators in 2d CFT \cite{Cardy:2017qhl,Kraus:2016nwo,Collier:2019weq}. An analogous Cardy-like formula governing the universal asymptotics of the OPE coefficients between global conformal primaries in CFT$_d$ has recently been derived in \cite{Benjamin:2023qsc} using the framework of thermal effective field theory \cite{Bhattacharyya:2007vs, Banerjee:2012iz}. It would be interesting to understand if this framework could be used to determine the statistics of thin shell black holes in AdS$_{d+1}$ from CFT$_{d}$.

In principle, the function $C_0^{\CL}$ is defined as a fusion kernel for the line defect. In practice, it is easier to formulate an auxiliary problem in Liouville CFT that calculates this fusion kernel for us. The Liouville CFT is non-compact, and the asymptotic formula \eqref{introC0} does not apply to Liouville itself, but one can nonetheless use Liouville to calculate it, because of its close connection to the representation theory of the Virasoro algebra.  As reviewed in section \ref{secbootstrap} below, there is a similar derivation of the Cardy density of states and the $C_0$ OPE function from Liouville CFT.  Like the Cardy formula, the behavior \eqref{introC0} is universal to all compact CFTs at very high energies, but becomes especially powerful in holographic CFTs where it (conjecturally) holds at lower energies above the black hole threshold \cite{Hartman:2014oaa}. 

\begin{figure}[t]
\centering
\begin{tikzpicture}
\draw  (0,10) rectangle ++(4.5,1.7);
\draw  (7.5,10) rectangle ++(4.5,1.7);
\draw  (3,7) rectangle ++(5.9,1.7);
\draw [thick,<->] (3.5,9.7) --  (4.8,8.9);
\draw [thick,<->] (8.4,9.7) --  (7.1,8.9);
\draw [thick,<->] (5,11) -- (7,11);
\node [align=left] at (2.3, 11) { Thin shell black holes\\and wormholes in $AdS_3$};
\node [align=left] at (9.7, 11) { Statistics of line defects  \\  $\CL_\Sigma$ in the dual CFT$_2$ };
\node [align=left] at (6.0, 7.8) {Line defects $L_\Sigma$ in Liouville CFT};
%
\end{tikzpicture}
\caption{\small 
There is a general relationship between universal behavior of black holes in 3d gravity, universal bootstrap data in 2d CFT, and calculations in an auxiliary Liouville theory. We study this relationship for thin shell black holes, which are created by line operators, $\CL_\Sigma$. Ordinary black hole thermodynamics probes the variance of the matrix elements $\langle i|\CL_\Sigma|j\rangle$, and wormholes probe their higher moments. To show that these agree, we reduce both calculations to an auxiliary problem involving a line defect in Liouville CFT at the same central charge. \label{fig:threetopics}
}
\end{figure}
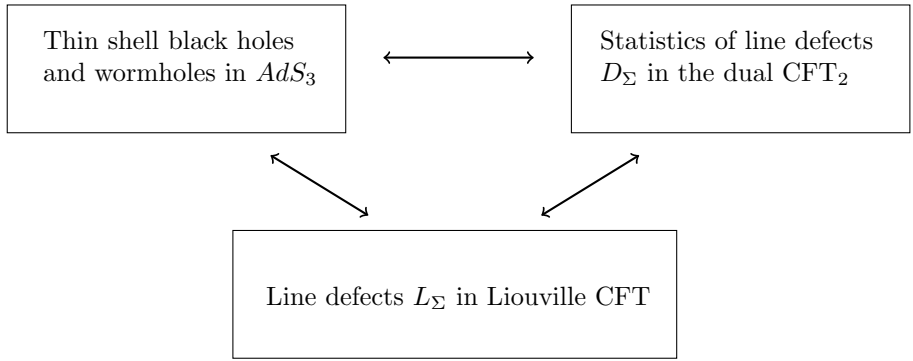

We will therefore study three related problems in this paper: Line defects in Liouville CFT, line defects in compact CFTs, and thin shell solutions in 3D gravity.  To reproduce the behavior of thin shell black holes and wormholes from the dual CFT, we will show that both the gravity calculations and the dual CFT calculations reduce to the same Liouville quantity, which can be found explicitly in the semiclassical limit. The threads connecting these three topics are summarized in figure \ref{fig:threetopics}.
To reproduce 3D wormholes from an ensemble average over CFT data, we reinterpret \eqref{introC0} as the variance of the ensemble, with the average taken over the ensemble of matrix elements rather than over states in a fixed theory. This variance, together with the eigenstate thermalization hypothesis (ETH) and the assumption that the leading statistics are Gaussian, defines an ensemble of CFT data that can be used to calculate averaged observables. These averages match the spherically symmetric wormholes of 3D gravity.

This is analogous to previous work on local operators.
In \cite{Chandra:2022bqq}, it was shown that wormholes in 3D gravity supported by point-like matter can be interpreted as averaged correlation functions of local operators in large-$c$ CFT, with the average taken over an ensemble of OPE coefficients (see also \cite{Belin:2020hea,Cotler:2020ugk,Collier:2023fwi,Abajian:2023bqv,Collier:2024mgv}). We will generalize this result from local operators to line defects. The advantage of working with line defects is that corresponding black holes and wormholes are spherically symmetric, so the gravity solutions are much simpler, and many of the complications from Virasoro descendants drop out of the CFT calculation. This leads to simple analytic formulas for the wormhole amplitudes and CFT statistics. It also makes a direct connection to the spherically symmetric wormhole solutions in any spacetime dimension considered in \cite{Chandra:2022fwi, Sasieta:2022ksu, Balasubramanian:2022gmo, Bah:2022uyz}.

Our results on bootstrapping line defects may be of independent interest, outside of holography --- the line defect we study in Liouville CFT is a 2d version of the defect studied recently in the $O(N)$ model \cite{Allais:2014fqa,Cuomo:2021kfm}. It is interesting to note that the $g$-function of the Liouville line defect sets the magnitude of global charge violation in the dual gravitational theory.

\subsection{Summary}
In the rest of this introduction we summarize the results in each of the three areas --- Liouville CFT, compact CFTs, and gravity ---  and then tie them together with a discussion of the wormholes that calculate the statistics of line defects in holographic CFTs.

\subsubsection{Line defects in Liouville CFT}
In Liouville CFT, we study a line operator defined by integrating the Liouville field over a curve $\Sigma$,
\begin{align}\label{introLsigma}
L_{\Sigma} = \exp\left[{\frac{m}{2\pi b} \int_\Sigma d\ell \phi} \right]. 
\end{align}
The coefficient $m$ has units of mass, so the defect is not conformal, but it can be studied as a continuum limit of a product of Liouville vertex operators. Note that we use $L_\Sigma$ for a line defect in Liouville theory, and $\CL_{\Sigma}$ for a line defect in a compact CFT such as a holographic CFT dual to 3D gravity. All correlation functions involving $L_\Sigma$ are calculations in Liouville CFT, whereas correlators involving $\CL_\Sigma$ are calculated in a compact CFT.

The main result of our Liouville analysis is to find the matrix elements of the defect, $\langle h' | L_\Sigma| h\rangle$, in the semiclassical limit $c \to \infty$. In this limit, the Liouville CFT becomes a theory of hyperbolic surfaces $e^{\Phi} |dz|^2$, with $\Phi$ the classical Liouville field. Inserting the line defect produces an interface where two hyperbolic surfaces are glued together along $\Sigma$, subject to a junction condition. Therefore, in the semiclassical limit, one can study correlators of $L_\Sigma$ by constructing hyperbolic surfaces glued along interfaces. Correlation functions of the line defect are calculated semiclassically by evaluating the classical Liouville action on these joined surfaces.

In section \ref{secLiouville}, we study one-point functions $\langle L_\Sigma\rangle$ and two-point functions $\langle  L_\Sigma^\dagger L_\Sigma\rangle$ on the sphere, the torus, the disk, and the annulus. Each of these correlators is calculated semiclassically by gluing together various hyperbolic surfaces. The results are used to find the matrix elements of $L_\Sigma$ in the large-$c$ limit. 

Each of these Liouville observables plays a role in the applications that follow. The one-point function of the line defect on the disk, $\langle L_\Sigma \rangle_{\rm disk}$, is used to calculate the fusion kernel and infer the universal function $C_0^{\CL}$ introduced above, determining the asymptotic matrix elements of line defects in compact CFTs. This relation is exact, for any $c>1$. At large $c$, the disk 1-point function is also related to the free energy of a thin shell black hole in AdS$_3$. The correlation functions of the Liouville defect on the sphere, $\langle L_{\Sigma}\rangle_{\rm sphere}$ and $\langle L_{\Sigma}^\dagger L_{\Sigma}\rangle_{\rm sphere}$, are related to 2-boundary wormhole amplitudes in 3D gravity.  Liouville correlators on the annulus are related to single-boundary solutions of 3D gravity with thin shells inserted in BTZ, and  Liouville correlators on the torus are related to two-boundary wormholes.

\subsubsection{Universal behavior of line defects in compact CFTs}
In compact CFTs, we study a line defect $\CL_\Sigma$ with the following defining properties: $(i)$ It transforms like a continuous product of local operators, like the Liouville defect $L_\Sigma$ defined in \eqref{introLsigma}, and $(ii)$ two nearby defects $D_\Sigma^\dagger D_\Sigma$ can locally fuse to the vacuum state running along the defect. To be more precise, the vacuum state appears with nonzero spectral weight when the correlator is cut on a rectangle through the defects, as follows:
\begin{align}
\vcenter{\hbox{
 \begin{overpic}[scale=0.4,grid=false, tics=20, trim=100 170 150 150, clip]{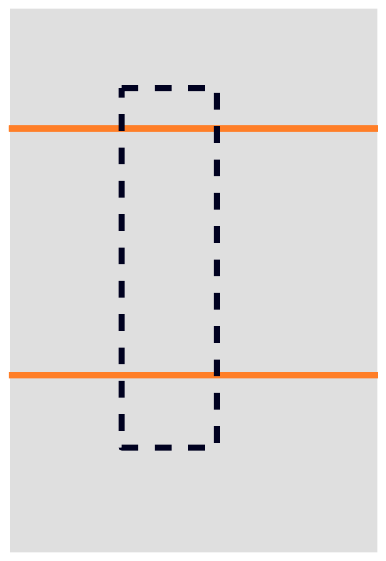}
    \put (29, 35) {$\CL_\Sigma^\dagger$}
    \put (29, 15) {$\CL_\Sigma$}
    \put (58, 25) {$\mathds{1}$}
    \end{overpic}
 }}
\end{align}
The latter property ensures that when two defects approach each other, the dominant contribution to the correlator is a Virasoro identity block, generalized to massive line defects, in which one projects onto the vacuum representation along any cut of this form.
Denote this identity block by ${\cal F}_{\id}^{\CL}$. (We will actually study two different versions of this identity block -- on the sphere and on the torus -- but here we will not make the distinction.) In section \ref{secbootstrap}, by generalizing a method of the Zamolodchikovs from local operators to defects, we derive a formula relating this Virasoro block --- applicable to compact CFTs --- to the Liouville defect. The relation takes the form
\begin{align}
{\cal F}_{\id}^{\CL} = {\cal N} \langle L_{\Sigma} \rangle_{\rm disk}
\end{align}
with a proportionality constant ${\cal N}$ derived below. Then, we expand the Liouville one-point function in the dual (boundary) channel. This provides a Liouville formula for the fusion kernel $C_0^{\CL}$. The formula is exact for $c>1$, though we will only evaluate the Liouville correlator in the large-$c$ limit thereby providing a simple closed expression for $C_0^{\CL}$. The explicit formulas can be found in \eqref{finalCD1} and \eqref{finalCD2} below. 

The large-$c$ limit of the identity conformal block is known to be closely connected to two other topics: The Schottky uniformization of Riemann surfaces, and 3D gravity \cite{Hartman:2013mia,Faulkner:2013yia}. In section \ref{secblocks} we describe the connection between our results on line defects, and the uniformization of Riemann surfaces glued across interfaces. This provides a second, complementary derivation of the defect identity block ${\cal F}_{\id}^{\CL} $ in the large-$c$ limit using uniformization and the monodromy method, generalizing the continuum monodromy method developed in \cite{Anous:2016kss,Anous:2017tza} to beyond the Vaidya limit.

\subsubsection{Thin shell black holes in 3d gravity from line defects}

Next we turn to the problem of thin shell black holes in 3D gravity. Non-rotating thin shells in AdS have been studied previously in \cite{Keranen:2015fqa, Balasubramanian:2011ur, Anous:2016kss,Chandra:2022fwi,Sasieta:2022ksu,Bah:2022uyz}. The wormholes built from these geometries can be interpreted as an ensemble average in the boundary CFT, with Gaussian statistics \cite{Chandra:2022fwi}. This provides a higher-dimensional instance of the West Coast model of black hole interiors, which was first developed in two-dimensional gravity \cite{Penington:2019kki}.
However, in \cite{Chandra:2022fwi}, there was no closed formula for the variance of the ensemble --- it was defined implicitly by a gravitational action. We will fill this gap for 3D gravity.

In section \ref{secblackholes} we study the single-boundary black holes in detail. We calculate the on-shell action of the thin shell black hole and match it to the CFT calculations. The basic relationship for the case with zero angular momentum is:
\begin{align}\label{bhrel}
e^{-I_{BH}} = \langle {\CL}_\Sigma^\dagger \CL_\Sigma\rangle = \left| {\cal F}_{\id}^{\CL} \right|^2 =  e^{-\frac{c}{3}S_L}
\end{align}
In words,
\begin{align}
\mbox{Thin shell black hole} &= \mbox{2-point function in the dual CFT}  \notag \\
&= \left| \mbox{Defect identity block}\right|^2 \notag \\
&= \left|\mbox{Liouville line defect}\right|^2 \ . 
\end{align}
For local operators, there is a known correspondence between 3D geometries with conical defects, correlation functions in large-$c$ CFT, Virasoro identity blocks, and Liouville \cite{Hartman:2013mia,Faulkner:2013yia}; \eqref{bhrel} generalizes the correspondence to extended operators.
On the left of \eqref{bhrel} is the on-shell action of the black hole. The second quantity is the 2-point function of line defects in the dual compact CFT. This is given to leading order by the defect identity block; and the defect identity block is calculated by a certain Liouville action of a line defect on a disk, with ZZ boundary conditions. 

We also study rotating black holes, where the final results are analogous, though the calculations are more involved. The rotating black hole appears in the correlation function $\langle \CL_\Sigma^\dagger e^{i\theta J} \CL_{\Sigma}\rangle$. 
Its on-shell action matches the squared identity block at complex modulus, which is related to an analytically continued Liouville correlator. 

\subsubsection{Statistics of line defects from 3d wormholes}

In Jackiw-Teitelboim gravity in two dimensions, the statistics of black hole pure states can be studied using wormholes supported by end-of-the-world branes  \cite{Penington:2019kki, Gao:2021uro}. This is known as the West Coast model. End-of-the-world branes do not create on-shell Euclidean wormholes in higher dimensions, but the West Coast model was extended to higher dimensions using spherically symmetric dust shells in \cite{Chandra:2022fwi}. This provides a higher dimensional realization of ensemble averaging, in the following sense. In AdS$_{d+1}$ for $d>2$, spherically symmetric black holes supported by matter in the interior can be turned into $k$-boundary wormholes using a cut-and-glue procedure. These wormholes are solutions of the Einstein equations coupled to matter sources. The procedure applies to thin shells and to Oppenheimer-Snyder type solutions, where a spherically symmetric perfect-fluid star is glued to a Schwarzschild black hole spacetime.  The on-shell action of the $k$-boundary wormhole agrees with an ensemble average in the dual CFT, assuming Gaussian statistics at leading order \cite{Chandra:2022fwi}.  Related results have been obtained for finite temperature \cite{Sasieta:2022ksu}, multiple shells \cite{Balasubramanian:2022gmo}, flat asymptotics \cite{Bezrukov:2015ufa}, and light shells with large momentum that collapse to form a black hole after Lorentzian time evolution \cite{Bah:2022uyz}. 

In section \ref{secblackholes} we explain this in detail for AdS$_3$ and extend it in two ways. First, we can be more explicit than \cite{Chandra:2022fwi} because we have derived a closed formula for the matrix elements of the black hole-creating operator, $\CL_\Sigma$. Second, we include the effects of rotation. Another advantage of 3D gravity is that we have actually derived the variance of the ensemble from the dual CFT, whereas in higher dimensions, the variance is an input from the bulk theory \cite{Chandra:2022fwi}.

To state our results, let us first review the West Coast model \cite{Penington:2019kki}, as applied to higher dimensional thin shells in \cite{Chandra:2022fwi}. Consider any theory of gravity+matter that admits spherically symmetric, matter-supported black holes with a single asymptotic boundary. These black holes are pure states in the quantum theory, because they have only a single spacetime boundary, unlike the eternal black hole. The Euclidean path integral of a pure-state black hole is depicted
\begin{align}
 \vcenter{ \hbox{\includegraphics[width=0.9in]{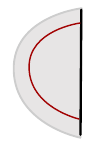} }}
\end{align}
The figure shows the $(\tau,r)$ directions in AdS$_{d+1}$, with spherical symmetry so there is an $S^{d-1}$ at each point; the solid black line on the right is the AdS boundary $S^d$, and the matter source is the red line. (It is drawn here as a thin shell, but the construction also allows for thick shells or extended fluid sources.) In the quantum theory, the state created by this matter insertion is
\begin{align}
|\Psi\rangle = \sum_n \psi_n|n\rangle
\end{align}
where the sum is over energy eigenstates. The contribution of the black hole to the gravity path integral is
\begin{align}\label{zbhi}
Z_{\rm grav} \approx \sum_n |\psi_n|^2 \ .
\end{align}
This relation would be exact in a UV-complete bulk theory, but of course the low energy gravitational theory does not know the exact coefficients $\psi_n$. According to \eqref{zbhi} it knows the average of $|\psi_n|^2$, smeared over a large number of nearby microstates,
\begin{align}
\overline{|\psi_n|^2} \approx \psi^2(E)
\end{align}
The smooth function $e^{S(E)} \psi^2(E)$ is extracted from the on-shell action of the black hole by inserting the matter at Euclidean time $\tau_0$ and doing a Legendre transform to fix the energy.

So far, this is just standard black hole thermodynamics. The gravitational path integral calculates the smoothed density of states $S(E)$ and the average squared matrix element, $\psi^2(E)$. The new results pertain to higher statistics and higher topologies. The result of \cite{Chandra:2022fwi} (building on the 2d results in \cite{Penington:2019kki, Stanford:2020wkf, Saad:2019pqd}) is that the higher moments of the $\psi_n$ are calculated by thin-shell wormholes. The $k$-boundary wormhole calculates the the $2k^{\rm th}$ moment:
\begin{align}\label{intro6boundary}
\vcenter{\hbox{
\begin{overpic}[width=1.2in,grid=false]{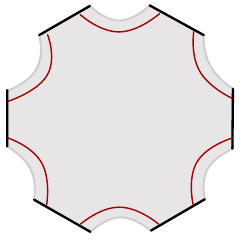}
\put (47,17) {$i_1$}
\put (71,29) {$i_2$}
\put (72,62) {$i_3$}
\put (47,76) {$i_4$}
\put (23,62) {$i_5$}
\put (23, 29) {$i_6$}
\end{overpic}
}} \quad \Rightarrow \quad \overline{ \psi_{n_1}^{i_1*} \psi_{n_1}^{i_2} \psi_{n_2}^{i_2*} \psi_{n_2}^{i_3}
\cdots \psi_{n_{k}}^{i_k*} \psi_{n_{k}}^{i_1} }
\end{align}
(shown for $k=6$). The arrow means that one can extract the moment, as a smooth function of energy, by an appropriate Legendre transform. The $i_\ell$ are `flavor' indices that may stand for an actual flavor of the bulk particles under an approximate global symmetry, or for the different ways a similar black hole state can be created kinematically.

The gravity calculation of the wormhole action agrees with the ensemble prediction if we take the statistics of the expansion coefficients to be Gaussian at leading order \cite{Chandra:2022fwi}:
\begin{align}
\overline{\psi_{m}^* \psi_n}  &\approx \psi^2(E_m) \delta_{mn} \\
\overline{\psi_{m}^* \psi_n \psi_p^* \psi_q} &\approx
\psi^2(E_m) \psi^2(E_p) \delta_{mn}\delta_{pq} + 
 \psi^2(E_m) \psi^2(E_n) \delta_{mq}\delta_{np}
\end{align}
and similarly for higher moments.
This ansatz, in the spirit of the eigenstate thermalization hypothesis \cite{Deutsch:1991msp,Srednicki:1994mfb}, was discussed previously in the context of wormholes in e.g. \cite{Penington:2019kki, Saad:2019pqd, Pollack:2020gfa, Belin:2020hea, Chandra:2022bqq}. 
Since the ensemble is Gaussian, it is determined entirely by the variance, $\psi^2(E)$. Matching the wormhole amplitudes to ensemble averages does not actually require knowing an explicit formula for the variance --- the single-boundary black hole action defines the variance implicitly, and then it is a general property of spherically symmetric wormholes that they match the Gaussian ensemble \cite{Chandra:2022fwi}. This is the leading known contribution; non-Gaussianities are expected at higher orders \cite{Belin:2021ryy}. These results are quite general, applying to any spherically symmetric pure-state black hole. (The standard $k$-boundary wormholes are only on-shell for $k \leq k_{max}$ where $k_{max}$ depends on the matter, but for $k>k_{max}$ there are on-shell solutions with Euclidean time-folds \cite{Bah:2022uyz}.) 

With this background it is now simple to state the results of section \ref{secwormholes}: We determine the ensemble for 3D black holes created by the line defect $\CL_\Sigma$, including an explicit formula for the variance that is derived independently on the gravity side and the CFT side, and extend the results to rotating black holes and wormholes. In particular, the variance of the ensemble is calculated by a wormhole solution with one thin shell and two conical defects going through the wormhole:
\begin{align}
 |C_0^D(h_i,h_j)|^2 & \approx Z_{\text{grav}}\left [\quad
\vcenter{\hbox{
\begin{overpic}[width=2in,grid=false]{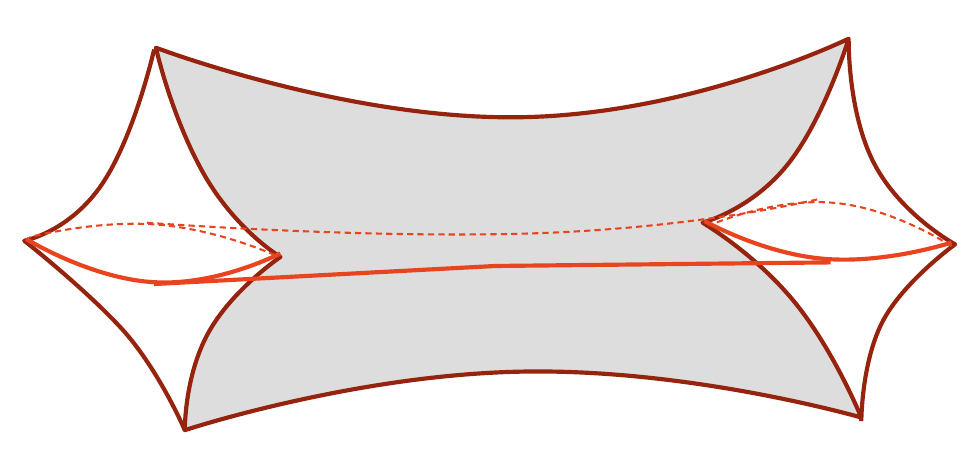}
\put (50,3) {$j$}
\put (50,38) {$i$}
\put (-8,20) {$D_\Sigma$}
\put (100,20) {$D_\Sigma$}
\end{overpic}
}}\quad \right ]
\end{align}
This is the defect version of a wormhole discussed in \cite{Chandra:2022bqq}, with three conical defects going through the wormhole, which calculates $|C_0|^2$ for local operators.
We also consider the ensemble average of line defect correlation functions such as
\begin{align}
\overline{
\langle \CL_\Sigma^{\dagger} e^{-\tau_0^{(1)} L_0 - \btau_0^{(1)} \bL_0} \CL_\Sigma \rangle
\cdots
\langle \CL_\Sigma^{\dagger} e^{-\tau_0^{(k)} L_0 - \btau_0^{(k)} \bL_0} \CL_\Sigma \rangle
}
\end{align}
This average in the boundary CFT matches the on-shell action of the $k$-boundary wormhole in 3D gravity, assuming Gaussian statistics for the line defect matrix elements. The wormhole for this observable is similar to the one shown in \eqref{intro6boundary} above, generalized to have nonzero angular momentum and different moduli on each boundary. These averages can be reduced to calculations in Liouville CFT using the bootstrap results described above, and this is how we do the explicit comparison between gravity and CFT.

As another application, consider the simplest thin shell wormhole: A two-boundary wormhole with a single thin shell going through. This configuration has a saddlepoint that contributes to the ensemble-averaged square of the defect one-point function, so the basic relation between the boundary CFT, Liouville CFT, and gravity in this case is
\begin{align}\label{introGG}
\overline{|\langle D_\Sigma\rangle|^2} = {\cal N}^2 |\langle L_\Sigma\rangle|^2  = Z_{\rm grav} \ . 
\end{align}
The defect has an approximate global symmetry in the bulk, so $\overline{\langle D_\Sigma\rangle} = 0$. Therefore \eqref{introGG} is a non-perturbative violation of the approximate global symmetry on the gravity side, as in \cite{Chen:2020ojn,Hsin:2020mfa,Chandra:2022fwi,Bah:2022uyz}. We learn that the Liouville defect one-point function $\langle L_\Sigma\rangle$, commonly referred to as the $g$-function and much studied in other theories and in the context of RG flows \cite{Affleck:1991tk}, determines the magnitude of global charge violation in the bulk.

In this paper we use the language of ensemble averaging, but all of the results can also be interpreted in terms of coarse-grained states in a single CFT, as in \cite{Chandra:2022fwi}. 
From the low energy point of view, both interpretations are equally valid, and one does not need to address the thornier question of factorization vs.~ensemble averaging in the full quantum theory. Those are questions for the UV completion.

\subsection{Outline and notation}
In section \ref{secLiouville} we define and study the line defect $L_\Sigma$ in Liouville CFT. The matrix elements $\langle h'|L_\Sigma|h\rangle$ are calculated semiclassically using classical solutions of Liouville+defects on the sphere, disk and torus. In section \ref{secbootstrap} we use these results to study the fusion kernel of a line defect $\CL_\Sigma$ in compact CFT. The universal formula $C_0^{\CL}$ for the asymptotic, averaged matrix elements of such a defect is derived from an auxiliary Liouville problem. In section \ref{secblocks} we provide another perspective on this calculation based on the monodromy method for calculating Virasoro identity blocks and the uniformization of Riemann surfaces. In section \ref{secblackholes} we study thin shell black holes in 3D gravity and calculate their action, with and without rotation. In section \ref{secwormholes} we study multiboundary wormholes, calculate their action, and match the results to a Gaussian ensemble of line-defect matrix elements. Notation:
\begin{align*}
\mbox{Central charge:} \qquad & c = 1 + 6(b + b^{-1})^2\\
\mbox{Conformal weights:} \qquad & h = \frac{c-1}{24} + P^2 = \frac{c}{24}(1 + \gamma^2) \\
\mbox{Line defect in Liouville:} \qquad &L_\Sigma = \exp\left[\frac{m}{2\pi b}\int_\Sigma \phi \right]\\
\mbox{Semiclassical matrix elements in Liouville:} \qquad & |\langle h|L_\Sigma|0\rangle| \approx  \exp\left[\frac{c}{6} c_L(\gamma)\right] \\
&|\langle h'|L_\Sigma |h\rangle| \approx \exp\left[\frac{c}{6}c_L(\gamma', \gamma)\right] \\
\mbox{Line defect in compact CFT:} \qquad &\CL_\Sigma \\
\mbox{Asymptotic matrix elements in compact CFT:} \qquad & C_0^{\CL} \\
\mbox{Defect identity block:} \qquad & {\cal F}_{\id}^{\CL} \\
\mbox{Semiclassical block:} \qquad & {\cal F}_{\id}^{\CL} \approx \exp\left[-\frac{c}{6} f_{0}^{\CL}\right]\\
\mbox{Semiclassical limits in compact CFT:} \qquad & C_0^{\CL}(h',h) \approx \exp\left[\frac{c}{6}c_D(\gamma',\gamma)\right] \\
& C_0^{\CL}(h,0) \approx \exp\left[\frac{c}{6}c_D(\gamma)\right]
\end{align*}

\section{Liouville CFT with line defects} \label{secLiouville}

The action of Liouville theory, up to boundary terms, is 
\begin{align}
S = \frac{1}{\pi} \int d^2 z [ \p \phi \bar{\p}\phi + \pi \mu e^{2b\phi} ]\ . 
\end{align}
We set $\mu = \frac{1}{4\pi b^2}$, which corresponds to setting $\ell_{\rm AdS} = 1$ in 3D gravity. The central charge is parameterized as
\begin{align}
c = 1+6Q^2 , \quad Q = b + b^{-1} \ . 
\end{align}
The local primary operators are the vertex operators 
\begin{align}
V_\alpha(x) = e^{2\alpha \phi(x)}
\end{align}
with conformal weights $h_\alpha =  \bh_\alpha = \alpha(Q-\alpha)$. 
In this section we will study the line operator
\begin{align}\label{quantumL}
L_\Sigma = \exp\left[{\frac{m}{2\pi b} \int_\Sigma d\ell \phi } \right] \ .
\end{align}
The parameter $m$ has units of mass. We will consider cases where the defect is an infinite line or a circle.
By discretizing the integral, the line operator $L_\Sigma$ can be obtained as a limit of an infinite product of nearby vertex operators, each with infinitessimal conformal weight,
\begin{align}\label{discreteLsigma}
L_\Sigma = \lim_{n \to \infty} \prod_{j=1}^n V_{\frac{m \ell}{4\pi n b}}(z_i) \ , 
\end{align}
with the points $z_i$ discretizing $\Sigma$ and $\ell = \int_{\Sigma} d\ell$.

These line defects are not topological or conformal. For a discussion of topological line defects in Liouville CFT, see \cite{Sarkissian:2009aa, Petkova:2009pe} (and \cite{Chang:2018iay} for related work). As noted in the introduction, in higher dimensions, line defects of the form $e^{\int \phi}$ play a role in condensed matter and statistical physics, particularly to model impurities in systems like antiferromagnets and superconductors \cite{Affleck:1991tk,sachdev2000quantum,Allais:2014fqa,Cuomo:2021kfm,Cuomo:2022xgw,Rodriguez-Gomez:2022gif,Beccaria:2022bcr,Aharony:2023amq,Zhou:2023fqu}.

Although we only discuss the dynamics of the line defect \eqref{quantumL} in this paper, our analysis can be readily generalized to other rotationally invariant line defects, for example to the family of line defects obtained by intergrating a higher power of the Liouville field,
\begin{equation}
  L^{(n)}_{\Sigma}= 
\exp\left[{\frac{m_n}{2\pi b^{2-n}} \int_\Sigma d\ell \phi^n } \right]  \ .
\end{equation}
The power of the Liouville coupling $b$ is chosen so that in the large-$c$ limit, the defect adds a source to the classical Liouville equation. 
As in \eqref{quantumL}, these defects are not conformal since the parameter $m_n$ is dimensionful. It would be interesting to understand the RG flows triggered by these line defects and the nature of the infared.

In the large-$c$ limit, defined by taking $b \to 0 $ with $\Phi = 2 b \phi$ held fixed, Liouville becomes a classical theory of hyperbolic surfaces. The action becomes $S \approx \frac{c}{6}S_L$, with the bulk classical Liouville action
\begin{align}
S_L = \frac{1}{4\pi} \int d^2 z \left( \p \Phi \bar{\p}\Phi + e^{\Phi} \right)  \ .
\end{align}
The equation of motion, with sources corresponding to vertex operators $e^{\frac{c}{6} \eta_i \Phi(z_i)}$, is the Liouville equation
\begin{align}
\p {\bar \p} \Phi = \frac{1}{2}e^{\Phi} - 2\pi \sum_i \eta_i \delta^{(2)}(z-z_i) \ .
\end{align}
A solution to this equation describes a hyperbolic metric $e^{\Phi} |dz|^2$. If the sources have $0< \eta_i < \frac{1}{2}$, then they produce conical defects with deficit angle $4 \pi \eta_i$.

\begin{figure}[t]
  \centering
  \begin{overpic}[scale=0.4,grid=false, tics=20, trim=250 180 350 150, clip]{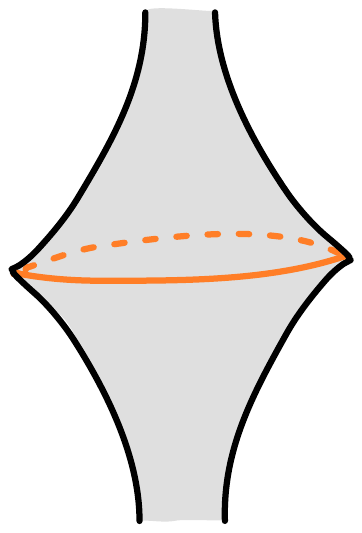}
  \end{overpic}
  \caption{A line defect in Liouville theory. The defect creates a fold in the surface.}
  \label{fig:line_defect_fold}
\end{figure}

In the semiclassical limit, the line operator $L_\Sigma$  produces an interface that can be visualized as a `fold' in the surface. An example is shown in figure \ref{fig:line_defect_fold}. Inserting the line operator $L_{\Sigma} $ adds a source term
\begin{align}\label{SLdef}
S_L  =  \frac{1}{4\pi} \int d^2 z \left( \p \Phi \bar{\p}\Phi + e^{\Phi} \right)   - \frac{m}{4\pi} \int_\Sigma d\ell \Phi \ . 
\end{align}
Solutions to this equation can be described locally near the defect as two hyperbolic surfaces glued along $\Sigma$. Geometrically, the equations of motion derived from \eqref{SLdef} are the constant-curvature condition $R[e^{\Phi} \delta_{\mu\nu} ] = -2$, 
i.e. the Liouville equation
\begin{align}\label{Leqn}
\p \bar{\p}\Phi = \frac{1}{2}e^{\Phi}  \qquad (z \notin \Sigma) \ , 
\end{align}
and a junction condition that specifies that the extrinsic curvature of $\Sigma$ in the metric $e^{\Phi}|dz|^2$ has discontinuity 
\begin{align}
K|_+ - K|_- &= - m e^{-\Phi/2}
\end{align}
at the interface.
Placing the operator along the line $y=y_0$ with $z=x+iy$, the latter condition is
\begin{align}\label{ydisc}
\p_y \Phi |_{+} - \p_y \Phi|_{-} = - 2m
\end{align}
where $+$ is the limit from above and $-$ is the limit from below.  The Liouville field itself is continuous, $\Phi|_+ = \Phi|_-$.

In what follows, we will determine the structure constants of translation invariant (or rotationally invariant) line defects $L_\Sigma$ in the semiclassical limit, by solving the Liouville equation and the junction conditions, and interpreting the results in terms of the matrix elements of $L_\Sigma$. This is done by gluing together two basic solutions of the Liouville equation. The first is the hyperbolic disk. Rather than working on the usual Poincare disk it is more convenient to use the cylinder frame, 
\begin{align}
z = x + i y , \qquad x \sim x + 2\pi \ .
\end{align}
The hyperbolic disk is diffeomorphic to the semi-infinite cylinder $y < 0$ or $y >0$, with 
\begin{align}\label{Ldisk}
ds^2_{\rm disk} = \frac{|dz|^2}{\sinh^2 y} \ . 
\end{align}
The second building block is the hyperbolic cylinder, 
\begin{align}\label{Lcyl}
ds^2_{\rm cyl} = \frac{r_H^2 |dz|^2}{\cos^2(r_H y)} \ .
\end{align}
This is a hyperboloid with a minimal geodesic of length $2\pi r_H$, with $r_H$ a free parameter which will later be identified as the Schwarzschild radius of the dual black hole.

\subsection{One defect on the sphere} \label{defsphone}
Let us start with the vacuum 1-point function $\langle L_\Sigma\rangle$. 
Provided $m > 2$, there is a classical saddlepoint. We work on the cylinder, $z \sim z+2\pi$, and place the defect on the circle Im $z = 0$. By an exponential mapping, this is equivalent to a defect on the complex plane inserted on the unit circle.

\begin{figure}[t]
  \centering
  \begin{overpic}[scale=0.3,grid=false, tics=20, trim=150 200 150 170, clip]{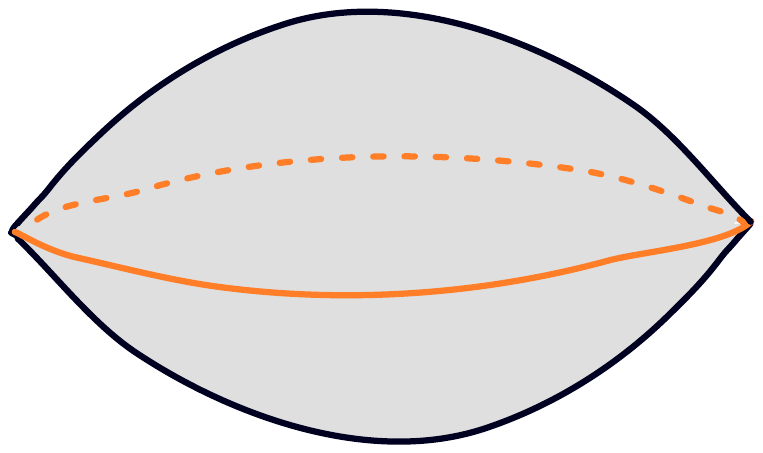}
  \end{overpic}
  \caption{One point function of a line defect on the sphere. The Liouville solution is constructed by gluing two hyperbolic disks.}
  \label{fig:sphere_1_pt}
\end{figure}

The classical solution $e^{\Phi}|dz|^2$ is constructed from two hyperbolic disks glued on a circular interface. See figure \ref{fig:sphere_1_pt}. Each disk is a portion of the geometry in \eqref{Ldisk}.

Thus the Liouville solution used to calculate $\langle L_\Sigma\rangle$ is:
\begin{align}\label{phi1pt}
\Phi = \begin{cases}
-2\log \sinh\left(y+ A\right) & y > 0\\
-2\log \sinh\left( A- y \right) & y< 0
\end{cases}
\end{align}
with $z = x + i y$, and $A>0$. The junction condition \eqref{ydisc} sets 
\begin{align}
A = \frac{1}{2}\log \frac{m+2}{m-2} \ .
\end{align}
The stress tensor associated with the Liouville solution \eqref{phi1pt} is 
\begin{equation}
  T^{\Phi}(z)=\frac{1}{2}\partial^2\Phi-\frac{1}{4}(\partial \Phi)^2=\frac{1}{4}+\frac{m}{2}\delta(y)
\end{equation}
so the presence of the line defect shows up as a $\delta$-function in the stress tensor, with the mass parameter $m$ determining the energy density on the defect.

The Liouville action on the cylinder requires an infrared regulator (see e.g.~\cite{Zamolodchikov:1995aa}). The regulated action, including a single defect $L_\Sigma$ is
\begin{align}\label{cyl1paction}
S_L = \frac{1}{4\pi}\int_{\Gamma} d^2z\left( \p \Phi \bar{\p}\Phi +e^{\Phi}\right) - \frac{m}{4\pi} \int_\Sigma dz \Phi+ \frac{1}{4\pi}\int_{\Gamma_+\sqcup \Gamma_-} \!\! dz \Phi + T+2(1-\log2)
\end{align}
where $\Gamma$ is the region Im $z \in [-T,T]$ and $\Gamma_{\pm}$ are the IR boundaries at Im $z = \pm T$ (both oriented toward increasing Re $z$), with $T \to \infty$ at the end. A constant factor $(1-\log 2)$ is added for each IR boundary. This factor is a choice of normalization for the vacuum state. Although the vacuum state itself is not normalizable in Liouville, the action on the disk is finite, and the choice of constant in \eqref{cyl1paction} corresponds to $\langle 1 \rangle_{disk} = 1$ (classically), as we shall see below in section \ref{diskonepoint}.\footnote{The convention in \cite{Zamolodchikov:2001ah} is to add this factor on the disk, but not on the sphere. Therefore our convention for the vacuum normalization agrees with ZZ on the disk but differs by this factor on the sphere.} To understand the boundary term in \eqref{cyl1paction}, note that the on-shell variation of this action is
\begin{align}
\delta S_L|_{eom} = \frac{1}{8\pi} \int_{\Gamma_+\sqcup \Gamma_-} \!\! dz(\p_n \Phi + 2)\delta\Phi \ .
\end{align}
The asymptotic behavior $\p_y \Phi \to \mp 2$ as $y \to \pm \infty$ ensures that this vanishes for arbitrary variations.

Evaluating the action \eqref{cyl1paction} on the solution \eqref{phi1pt} we find 
\begin{align}
S_L &= m(\log 2+1) - \frac{1}{2}(m+2)\log(m+2) - \frac{1}{2}(m-2)\log(m-2) \ .
\end{align}
Thus the one-point function of the line operator \eqref{quantumL} on the cylinder has leading semiclassical behavior (for $m>2$):
\begin{align}\label{defect1}
\langle L_\Sigma\rangle &\approx \exp\left(-\frac{c}{6}S_L\right)\\
&= \left((2e)^{-m}(m-2)^{m/2-1}(m+2)^{m/2+1}\right)^{c/6} \ . \notag
\end{align}
This quantity is known as the $g$-function of the line defect  (up to normalization) \cite{Affleck:1991tk, Cuomo:2021rkm}. It would be interesting to reproduce it from a calculation in the quantum Liouville theory and to study its behavior under the renormalization group flow to the infrared. The $g$-function is positive, and as a function of the mass, it is bounded from below with the minimum value occurring at $m=2\sqrt{2}$.

\subsection{Two defects on the sphere} \label{twodefsph}

\begin{figure}[t]
  \centering
  \begin{overpic}[scale=0.35,grid=false, tics=20, trim=250 140 350 150, clip]{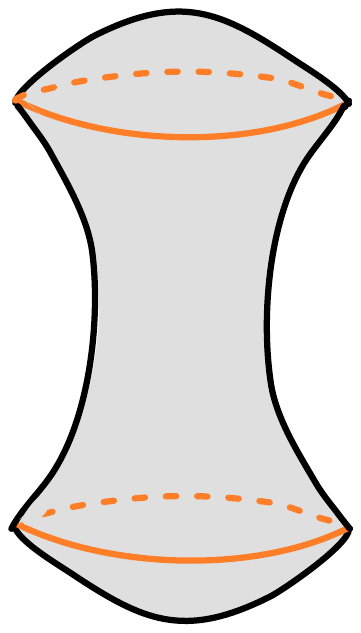}
  \end{overpic}
  \caption{Two point function of line defects on the sphere. The Liouville solution is constructed from a hyperbolic cylinder glued to two hyperbolic disks.}
  \label{fig:sphere_2_pt}
\end{figure}

Now consider the 2-point function, $\langle L_{\Sigma} L_{\Sigma'}\rangle$. The two defects can have different $m$'s but for simplicity we set them equal. 
The corresponding classical solution is a hyperbolic `soup can': a hyperbolic cylinder glued to two endcaps, each of which is a hyperbolic disk. See figure \ref{fig:sphere_2_pt}. 

We will again work on the cylinder with $z =x+iy$, $x \sim x + 2\pi$. We insert the two defects on the circles $y = \pm \tau_0$, with $\tau_0 > 0$.  
The solution to the Liouville equation in the three regions, constrained to have the obvious symmetries, is
\begin{align}\label{phi2cyl}
\Phi = \begin{cases}
-2\log \sinh(y-\tau_0 + A) & y>\tau_0 \\
-2\log \left(\frac{1}{r_H}\cos(r_H y) \right)& |y| < \tau_0 \\
-2\log \sinh(A-y-\tau_0 ) & y < -\tau_0
\end{cases}
\end{align}
The solution is non-singular provided $A>0$ and $0<r_H<\frac{\pi}{2\tau_0}$.
Continuity of $\Phi$ across the interfaces requires
\begin{align}
\sinh(A) = \frac{1}{r_H}\cos(r_H \tau_0)
\end{align}
and the junction condition \eqref{ydisc} imposes
\begin{align}\label{junc2cyl}
m = r_H \tan(r_H \tau_0) +\coth A \
=
r_H \tan(r_H \tau_0) + \sqrt{1 + \frac{r_H^2}{\cos^2 (r_H \tau_0)}} \ .
\end{align}
This last equation implicitly determines $r_H$ as a function of $\tau_0$ and $m$. It has a real solution whenever $m>1$. 
In anticipation of the comparison to 3D gravity, let us define the parameter
\begin{align}
 r_0 = \frac{1}{\sinh A} = \frac{r_H}{\cos(\tau_0 r_H)} \ .
\end{align}
The junction condition \eqref{junc2cyl} can be re-expressed as 
\begin{align}
m = \sqrt{1+r_0^2} + \sqrt{r_0^2 - r_H^2} \ . 
\end{align}
The Liouville action on the cylinder with two defects is
\begin{align}
S_L = \frac{1}{4\pi} \int_{\Gamma}d^2 z (\p \Phi \bar{\p} \Phi + e^{\Phi}) - \frac{m}{4\pi}\int_{\Sigma} dz \Phi - \frac{m}{4\pi} \int_{\Sigma'} dz \Phi + 
\frac{1}{4\pi}\int_{\Gamma_+\sqcup \Gamma_-} \!\! dz \Phi + T+2(1-\log 2)
\end{align}
with the IR regulator as in \eqref{cyl1paction}. Evaluating this on the solution \eqref{phi2cyl} gives
\begin{align}
S_L = 2\left( -  \frac{\tau_0 r_H^2}{2}
 + \frac{\tau_0}{2} - \sinh^{-1}\frac{1}{r_0}+ m - m \log r_0 \right) \ .
\end{align}
The two-point function, writing the defect along $y = \tau_0$ as $L(\tau_0)$, is
\begin{align}\label{defect2}
\langle L(\tau_0)L(-\tau_0)\rangle = e^{-\frac{c}{6}S_L} \ .
\end{align}
When the two defects approach each other they fuse to a single defect with twice the mass, and indeed taking this limit in \eqref{defect2} gives \eqref{defect1} with $m \to 2m$.
%
For any number of parallel defects on the cylinder, the condition to have a classical solution is $\sum_i m_i > 2$. 

We can interpret this result in terms of the matrix elements $\langle h|L_\Sigma|0\rangle$, where $|h\rangle$ is a Liouville primary (which is necessarily scalar) of weight $h$. We have normalized the  action such that the path integral is related to the amplitude by
\begin{align}\label{LLop}
\langle L(\tau_0)L(-\tau_0)\rangle = e^{-cT/6} \langle 0 | e^{-(T-\tau_0)H} L_\Sigma e^{-2\tau_0 H} L_\Sigma e^{-(T-\tau_0)H}|0\rangle
\end{align}
where $H = L_0 + \bL_0 - \frac{c}{12}$ is the cylinder Hamiltonian; the prefactor removes the IR-divergent contribution from the Casimir energy. (When the operator $L_\Sigma$ appears in an operator expression such as \eqref{LLop}, the surface $\Sigma$ is at Im $z=0$, with all of the position dependence accounted for by the action of the Hamiltonian.) Therefore the spectral decomposition is
\begin{align}\label{spectral1pt}
\langle L(\tau_0) L(-\tau_0)\rangle &=  \frac{1}{2} \int_{\mathbb{R}}dP  |\langle h_P| L_\Sigma |0\rangle|^2 \left(e^{-4\tau_0 h_P} + \mbox{descendants}\right)
\end{align}
where the integral runs over the normalizable spectrum of the Liouville CFT, with weights
\begin{align}\label{liouvillehP}
h_P = \frac{c-1}{24} + P^2 \ . 
\end{align}
In the semiclassical limit, we parameterize conformal weights by $\gamma$ with
\begin{align}\label{gammaweight}
h = \frac{c}{24}(1 + \gamma^2) \ .
\end{align}
Descendants are subleading in $1/c$ in rotationally invariant states, so we can simply use the leading exponential (i.e.~the scaling block) in \eqref{spectral1pt}.  
The semiclassical matrix elements $c_L$ are defined by the leading exponential behavior at large $c$ in
\begin{align}
|\langle h|L_\Sigma|0\rangle|^2 &= \exp\left[ \frac{c}{3}c_L(\gamma) + \cdots \right]\\
|\langle h'|L_\Sigma | h\rangle|^2 &= \exp\left[ \frac{c}{3}c_L(\gamma,\gamma') + \cdots \right] 
\end{align}
and calculated from the saddlepoint in \eqref{spectral1pt}. 
The saddlepoint equation for \eqref{spectral1pt} is 
\begin{align} \label{sadgeo}
1+\gamma^2 = \p_{\tau_0} S_L = r_H^2+1
\end{align}
and this leads to\footnote{Here and throughout the paper, the branch cuts in $\cot^{-1} z$ are placed along $z \in \pm(i,i\infty)$, so that on the real axis $\cot^{-1}z$ is continuous and has image $(0,\pi)$. The function $c_L(\gamma)$ is analytic on the domain Re $\gamma > 0$. For $m<1$, the continuation to the imaginary axis defines a continuous function on $\gamma \in (0, i(1-m))$, having Im $c_L(\gamma) = -  \frac{\pi}{2}|\gamma|$.}
\begin{align}\label{CL0}
c_L(\gamma) &= \frac{\pi \gamma}{2}  - m(1+\log(2m)) - \gamma \cot^{-1} \left( \frac{m^2 - \gamma^2-1}{2m\gamma}\right) \\
&\qquad 
+ \frac{m-1}{2}\log\left[ (m-1)^2+\gamma^2\right]
 + \frac{m+1}{2}\log\left[ (m+1)^2+\gamma^2\right]
\end{align}
This result is reliable where the saddle exists, which restricts to $0<\gamma^2<m^2-1$. Since the length of the minimal geodesic around the waist of the cylinder in the Liouville metric is $2\pi r_H$, the equation \eqref{sadgeo} shows that the saddlepoint weight in the two-point Liouville correlator of line defects matches with the geodesic length in the hyperbolic metric.
This generalizes the result of \cite{Teschner:2003em} conjectured in \cite{Verlinde:1989ua} about the relation between the saddlepoint weight in the Liouville correlator of local vertex operators expanded using conformal blocks in some channel to the length of the minimal geodesic wrapping the corresponding waist of the backreacted Riemann surface (see also \cite{Hadasz:2005gk, Harrison:2022frl, Colville:2023nry}), to line defects on the sphere. The relation \eqref{sadgeo} also provides a generalization of Polyakov's conjecture, which states that the Liouville action is the generating functional for Fuchsian uniformisation of Riemann surfaces and was proved by Zograf and Takhtajan in \cite{MR806856,MR882831,MR889594}, to Riemann surfaces with line defects. For the present case, the relation between the Liouville theory on a sphere with two parallel defects and its uniformisation into the upper half plane can be made manifest by deriving the uniformisation map using the Liouville stress tensor 
\begin{equation}
  T^{\Phi}(z)=\frac{1}{2}\partial^2\Phi -\frac{1}{4}(\partial \Phi)^2=-\frac{r_H^2}{4}\Theta(|y|<\tau_0)+\frac{1}{4}\Theta(|y|>\tau_0)+\frac{m}{2}\delta(y-\tau_0)+\frac{m}{2}\delta(y+\tau_0)
\end{equation}
The uniformisation map $w(z)$ can be derived from the above stress tensor by solving the Schwarzian equation $\{w,z\}=2T^{\Phi}(z)$ to give\footnote{With the requirement of inversion symmetry, $w(\overline{z})=\frac{1}{\overline{w(z)}}$, the solution \eqref{uniffuchs} is determined upto a PSL$(2,\mathbb{R})$ redundancy. So, the Liouville field it corresponds to is unique.}
\begin{equation} \label{uniffuchs}
w= \begin{cases}
& \frac{1}{B}\tan(\frac{z-i\tau_0+iA}{2}) \qquad y>\tau_0 \\
& \tanh(z\frac{r_H}{2}+i\frac{\pi}{4}) \qquad |y|<\tau_0 \\
& B\cot(\frac{z+i\tau_0-iA}{2}) \qquad y<-\tau_0
\end{cases}
\end{equation}
As explained in detail in section \ref{secblocks}, we can determine the parameters $A$ and $B$ by continuity of the map and its first derivative across the marked points $z=\pm i\tau_0$. It is easy to check using the relation $e^{\Phi}=-\frac{4|w'(z)|^2}{(w(z)-\overline{w(z)})^2}$ that the above uniformisation map is consistent with the Liouville solution \eqref{phi2cyl}. We have thereby shown that we can view $r_H$ as an accessory parameter for the Fuchsian uniformisation of the two-defect sphere into the upper half plane. In section \ref{secblocks} we discuss Schottky uniformisation, in which the image of the uniformising map is the full complex plane.

\subsection{Structure constants}
To access the general structure constants with two different weights, $\langle h | L_\Sigma | h'\rangle$,
we now consider the two-point function on the torus, $\langle L(\tau_0)L(-\tau_0)\rangle_{\beta}$. The setup is similar, with defects at $\pm \tau_0$, but now
\begin{align}
z \sim z + 2\pi \sim z + i \beta \ . 
\end{align}
The classical solution $e^{\Phi}|dz|^2$ consists of two finite-length hyperbolic cylinders, glued at their ends. Hence the Liouville field is
\begin{align}
\Phi = \begin{cases}
-2 \log \left( \frac{1}{r_H} \cos(r_H y)\right) & |y| < \tau_0 \\
-2\log\left( \frac{1}{r_H'}\cos(r_H' (\frac{\beta}{2}-|y|))\right) &  \tau_0 < |y| < \frac{\beta}{2} \ .
\end{cases}
\end{align}
Continuity and the junction condition determine $(r_H, r_H')$ in terms of $(\tau_0, \beta)$,
\begin{align}
r_0 &= \frac{r_H}{\cos(\tau_0 r_H)} = \frac{r_H'}{\cos((\frac{\beta}{2}-\tau_0)r_H')}\\
m &= r_H \tan(\tau_0 r_H)  + r_H' \tan(r_H' (\frac{\beta}{2}-\tau_0))
= \sqrt{r_0^2-r_H^2} + \sqrt{r_0^2 - r_H'^2} \ .
\end{align}
The classical action is
\begin{align}
S_L = \frac{1}{4\pi} \int_{\Gamma}(\p \Phi \bar{\p}\Phi + e^{\Phi})  - \frac{m}{4\pi } \int_\Sigma dz \Phi - \frac{m}{4\pi} \int_{\Sigma'} dz \Phi \ ,
\end{align}
with $\Gamma$ the fundamental domain of the torus. Evaluating this on the solution above gives
\begin{align}
S_L = 2\left( m - \frac{r_H^2}{2}  \tau_0 -\frac{ r_H'^2}{2} (\frac{\beta}{2}-\tau_0) - m \log r_0 \right) \ .
\end{align}
To calculate the structure constants we start from the spectral decomposition,
\begin{align}\label{spectorus}
\langle L(\tau_0) L(-\tau_0) \rangle_\beta
&= \Tr \left (e^{-(\beta-2\tau_0)H} L_\Sigma e^{-2\tau_0 H} L_\Sigma \right ) \\
&= \frac{1}{4}e^{c \beta/12}  \int_{\mathbb{R}} dP \int_{\mathbb{R}} dP' |\langle h_{P'}|L_\Sigma|h_P\rangle|^2 \left(e^{-4h_P \tau_0 - 4h_{P'} (\frac{\beta}{2}-\tau_0)}  +\mbox{descendants} \right) \ . \notag
\end{align}
Descendants are subleading at large $c$. 
Setting this equal to $e^{-cS_L/6}$ and inverting it to solve for the matrix elements, we obtain the semiclassical matrix elements
\begin{align}\label{CLans}
c_L(\gamma,\gamma') &= 
\frac{\pi}{2}(\gamma + \gamma')
- \gamma \cot^{-1}\left( \frac{m^2+\gamma'^2-\gamma^2}{2m\gamma}\right)
- \gamma' \cot^{-1} \left( \frac{m^2 + \gamma^2 - \gamma'^2}{2m \gamma'}\right)\\
& \quad  + \frac{m}{2}\log\left( \frac{m^2}{4} + \frac{1}{2}(\gamma^2+\gamma'^2) + \frac{1}{4m^2}(\gamma^2-\gamma'^2)^2\right) - m \ .  \notag
\end{align}
This derivation is applicable in the regime $\gamma^2>0, \gamma'^2 > 0, m^2 > | \gamma^2-\gamma'^2|$ where there is a classical solution, but the formula holds for all $\gamma, \gamma'>0$ as we will see below from the calculation on the disk.

\bigskip

\noindent\textit{Derivation of \eqref{CLans}:}
To make the expressions symmetric between $h$ and $h'$ we define $\tau_0' = \frac{\beta}{2} - \tau_0$. The junction condition can be restated as 
\begin{align}
\frac{m}{r_0}  &=   \sin (\tau_0 r_H) + \sin(\tau_0' r_H') \ . 
\end{align}
Varying this implies
\begin{align}
m \, d(\log r_0) &= - r_H d(\tau_0 r_H) - r_H' d(\tau_0' r_H') \ .
\end{align}
Using this relation and \eqref{gammaweight} it is easily shown that the saddlepoint equations for \eqref{spectorus} are 
\begin{align}
\gamma^2 =\p_{\tau_0} S_L(\tau_0, \tau_0') = r_H^2 ,  \qquad
\gamma'^2 =\p_{\tau_0'} S_L(\tau_0, \tau_0') = r_H'^2  \ .
\end{align}
These relations show that the saddlepoint weights in the Liouville correlator match with the lengths of the respective minimal geodesics on the torus.
The semiclassical OPE coefficient is then
\begin{align}
c_L(\gamma, \gamma') =- \frac{1}{2}S_L  + \frac{\gamma^2}{2}\tau_0 + \frac{\gamma'^2}{2}\tau_0'
\end{align}
evaluated at the saddlepoint, which gives \eqref{CLans}.

\subsection{1-point function on the disk} \label{diskonepoint}

\begin{figure}[t]
  \centering
  \begin{overpic}[scale=0.35,grid=false, tics=20, trim=250 180 350 170, clip]{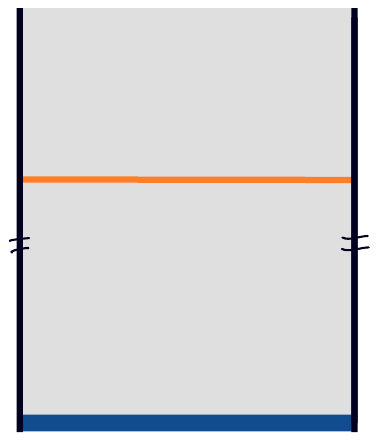}
  \put (100, 62) {$y=\tau_0$}
  \put (100, 13) {$y=0$}
  \put (55, -2) {ZZ}
  \end{overpic}
  \caption{One-point function of line defect inserted on the disk with ZZ boundary condition.}
  \label{fig:disk_1_pt}
\end{figure}

Next we consider the one-point function of the line operator on a disk, with ZZ boundary conditions \cite{Zamolodchikov:2001ah,Fateev:2000ik} on the unit circle. Some relevant properties of Liouville theory on a disk are reviewed in appendix \ref{liouvilleformulas}. In the cylinder frame, we restrict to $y > 0$ and the ZZ boundary condition is
\begin{align}
\Phi \to -2 \log y  \qquad (y \to 0) \ . 
\end{align}
With a defect at $y = \tau_0$ (figure \ref{fig:disk_1_pt}), the classical Liouville field is
\begin{align} \label{Lioudisk}
\Phi =\begin{cases}
-2\log \sinh(y-\tau_0+A) & y>\tau_0\\
-2\log \left(\frac{1}{r_H} \sin(r_H y) \right)& 0< y < \tau_0
\end{cases}
\end{align}
Define $r_0 = \frac{1}{\sinh A}$ as before. 
The continuity and junction equations are
\begin{align}\label{zz1junc}
r_0 &= \frac{r_H}{\sin(r_H \tau_0)} \\
m &= -r_H \cot(r_H \tau_0) + \sqrt{r_0^2+1}\ . \notag
\end{align}
Physical solutions, with $m>0$ and non-singular Liouville field, come in two types. There are solutions with real $r_H>0$ and $r_H \tau_0 < \pi$, and solutions with imaginary $r_H $. In the latter case, the junction condition can be written as
\begin{align}
m = \sqrt{r_0^2+1} - \sqrt{r_0^2 + |r_H|^2} \ . 
\end{align}
This implies that the imaginary-$r_H$ branch exists only for $m < 1$ and satisfies
\begin{align} \label{imbr}
r_H \in i(0, 1-m) \ . 
\end{align}
The classical action with a ZZ boundary condition and a single defect is
\begin{align} \label{LiouvdiskZZ}
S_L  = \frac{1}{4\pi} \int_{\Gamma} d^2 z \left( \p \Phi \bar{\p} \Phi + e^{\Phi}\right)
 - \frac{m}{4\pi} \int_\Sigma dz \Phi + \frac{1}{4\pi} \int_{\Gamma_+} dz \Phi + \frac{T}{2} - \frac{1}{\epsilon} + 1 - \log 2
\end{align}
where $\Gamma$ is the region $y \in (\epsilon, T)$, $\Gamma_+$ is the boundary at $y=T$, and at the end we take $\epsilon \to 0$, $T \to \infty$.  The last term $(\log 2 - 1)$ subtracts the action of the half-cylinder with no operator insertions, so this corresponds to the normalization condition $\langle 0|ZZ\rangle = 1$ standard in the Liouville literature \cite{Zamolodchikov:2001ah}.   The on-shell action is
\begin{align}\label{SLdisk}
S_L=  m  -\frac{1}{2}(r_H^2-1)\tau_0 - \sinh^{-1}\frac{1}{r_0} - m \log r_0 \ .
\end{align}
Let us compare to the spectral decomposition,
\begin{align}\label{spectral1disk}
\langle L(\tau_0)\rangle_{ZZ} = e^{-cT/12}  \langle 0| e^{-(T-\tau_0)H}L_{\Sigma} e^{-\tau_0 H} | ZZ\rangle \ , 
\end{align}
where $\langle L(\tau_0)\rangle_{ZZ}$ denotes the 1-point function on the disk that we just calculated semiclassically, and $|ZZ\rangle$ is the boundary state corresponding to the ZZ boundary condition. The ZZ state has an expansion in Ishibashi states $|P\rangle\!\rangle$, with a wavefunction that is known exactly:
\begin{align}
|ZZ\rangle &= \int_{\mathbb{R}} dP |P\rangle\!\rangle \langle\!\langle P|ZZ\rangle \\
\langle\!\langle P |ZZ \rangle &= -is_L(P)^{-1/2} \rho_0(P)^{1/2}
\end{align}
where $\rho_0$ is the Cardy density of states and $s_L$ is the Liouville reflection coefficient, which is a pure phase for normalizable states.
Now we use this to evaluate \eqref{spectral1disk}, assuming $m>1$, because otherwise light states (with imaginary $r_H$) appear in the spectral decomposition.
In the semiclassical limit, since $L_{\Sigma}$ is rotationally invariant we can ignore the distinction between Ishibashi states and primary states. Thus
\begin{align}\label{zzplug1}
\langle L(\tau_0) \rangle_{ZZ} &\approx 
\int_{\frac{c}{24}}^\infty dh \langle 0 | L_{\Sigma} | h\rangle \langle h |ZZ\rangle e^{-2h \tau_0} \ .
\end{align}
This expression is valid for $m>1$ since in this case the saddlepoint weight is above the threshold of $\frac{c}{24}$.
For normalizable states $\ket{h}$, the matrix element $\langle 0 |L_{\Sigma} | h \rangle$ was calculated from the cylinder 2-point function in \eqref{CL0} up to a phase. In order to be be consistent with \eqref{SLdisk} the phase needs to cancel with the reflection factor in the ZZ wavefunction, so
\begin{align}
\langle h|L_\Sigma |0\rangle \approx - i s_L(P)^{-1/2} e^{ \frac{c}{6}c_L(\gamma(h)) } \ .
\end{align}
The result holds for $m>1$, and $h > \frac{c}{24}$. For $m<1$, there are subthreshold states with $h < \frac{c}{24}$ in the spectral decomposition, corresponding to the imaginary-$r_H$ branch in \eqref{Lioudisk}, so the contour in \eqref{zzplug1} must be deformed to include them. The result for the semiclassical matrix element can be extended to $m<1$ and $h < \frac{c}{24}$ by rewriting it in the form
\begin{align}\label{finalc1}
\langle h |L_{\Sigma} | 0 \rangle &\approx -i s_L(P)^{-1/2} \rho_0(P)^{-1/2}e^{\frac{c}{6}c_L(\gamma(h))+\frac{1}{2}S(h)} 
\end{align}
where $h = \frac{c-1}{24} + P^2$ and 
\begin{align}\label{leadingCardy}
S(h) = 2\pi  \sqrt{ \frac{c}{6}(h - \frac{c}{24})}  = \frac{\pi  c \gamma}{6} 
\end{align}
 is the Cardy entropy.  The case where $\ket{h}$ is below the threshold of $\frac{c}{24}$ is derived in appendix \ref{seclineconical}. The result \eqref{finalc1} holds to leading order in the semiclassical limit, i.e. in the sense $\lim_{c \to \infty} \frac{\log (LHS)}{\log (RHS)} = 1$. For normalizable states, the factor of $\rho_0(P)^{-1/2}$ cancels against $e^{\frac{1}{2}S(h)}$ in this approximation, but for states below $\frac{c}{24}$ they do not cancel even at leading order.

Putting it all together, the conclusion is that for $m>0$, the spectral decomposition of the disk 1-point function is
\begin{align}\label{LZZspectral}
\langle L(\tau_0) \rangle_{ZZ}  &\approx \int_{h_{\text{min}}}^\infty dh \exp\left[ \frac{c}{6}c_L(\gamma(h)) + \frac{1}{2}S(h) - 2 h \tau_0 \right]
\end{align}
where the lower limit of the integral at large-$c$ is given by
\begin{equation} \label{hmin}
 h_{\text{min}}=\begin{cases}
 \frac{c}{24} & m \geq 1 \\
 \frac{c}{24}m(2-m) & 0<m<1
\end{cases}
\end{equation}
Using the expression for $c_L(\gamma)$ obtained previously in \eqref{CL0}, the saddlepoint equation for \eqref{zzplug1} is equivalent to the junction condition \eqref{zz1junc} with $\gamma = r_H$. Evaluated at the saddle we find
\begin{align}
\langle L(\tau_0) \rangle_{ZZ} \approx e^{-\frac{c}{6}S_L}
\end{align}
with $S_L$ the classical action in \eqref{SLdisk}. When $c_L(\gamma)$ was derived in \eqref{CL0} the argument was only valid for a limited range of $\gamma$, but we have now shown that the same expression holds for the full range of $\gamma$ corresponding to $h > h_{min}$.

\subsection{1-point function on the annulus}

\begin{figure}[t]
  \centering
  \begin{overpic}[scale=0.4,grid=false, tics=20, trim=250 200 250 100, clip]{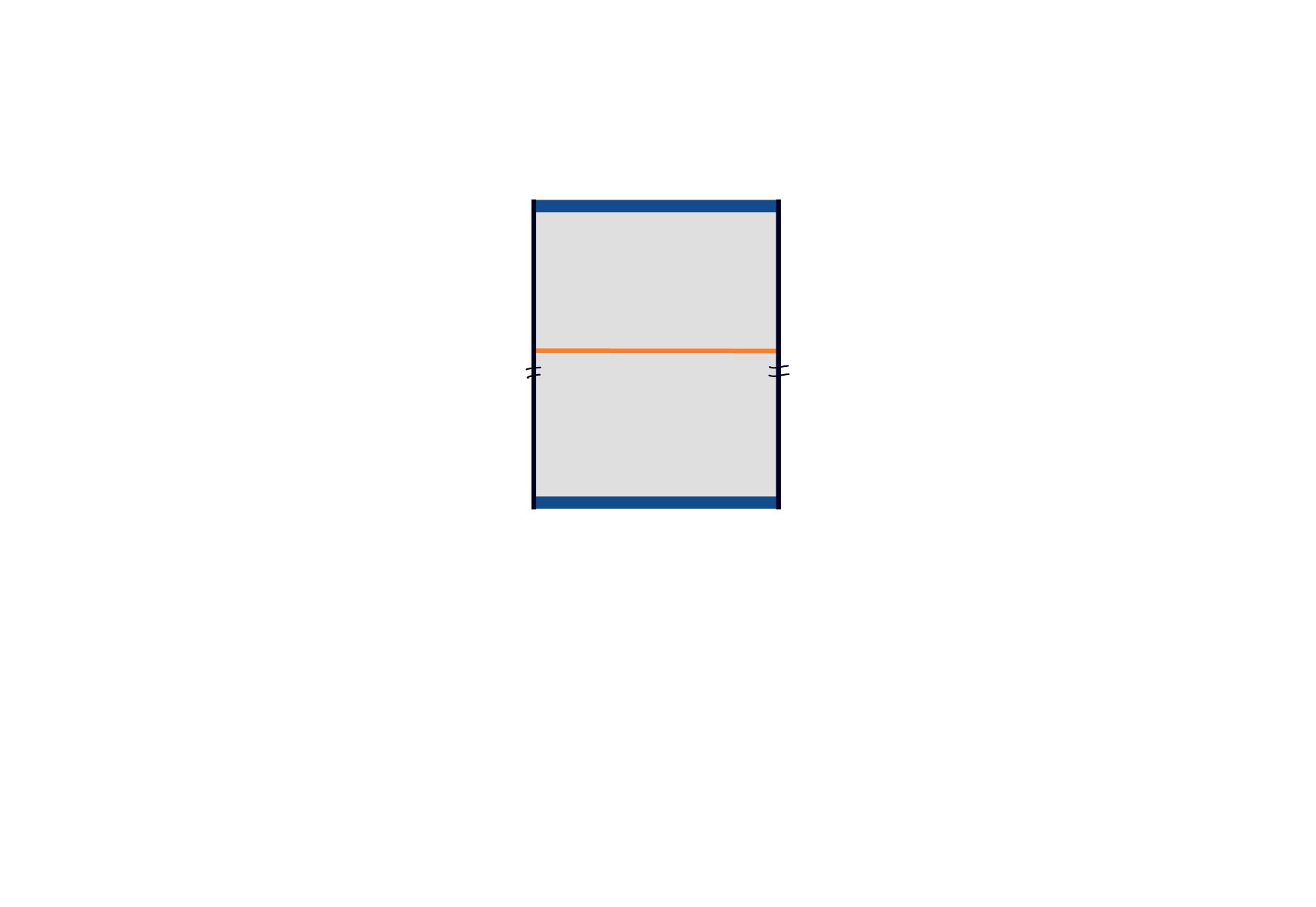}
  \put (46, 75) {ZZ}
  \put (77, 70) {$y=\beta/2$}
  \put (77, 43) {$y=\tau_0$}
  \put (77, 17) {$y=0$}
  \put (46, 9) {ZZ}
  \end{overpic}
  \caption{One point function of line defect inserted on the disk with ZZ boundary condition.}
  \label{fig:annulus_1_pt}
\end{figure}

Finally we compute the 1-point function on the annulus of length $\frac{\beta}{2}$ in order to extract the phases in the matrix element $\langle h | L_{\Sigma} | h'\rangle$. The ZZ boundary condition is now imposed at both the boundaries,
\begin{equation}
   \Phi \to -2\log y \quad (y\to 0,\frac{\beta}{2}) \ .
\end{equation}
Putting the defect at $y=\tau_0$ (figure \ref{fig:annulus_1_pt}), the Liouville solution takes the form
\begin{equation} \label{Liouann}
\Phi=
\begin{cases}
  -2\log \left (\frac{1}{r_H}\sin(r_Hy) \right ) & 0<y<\tau_0\\
   -2\log \left (\frac{1}{r_H'}\sin(r_H'(\frac{\beta}{2}-y)) \right ) & \tau_0<y<\frac{\beta}{2}
\end{cases}
\end{equation}
The classical Liouville action with a single line defect is given by
\begin{align} \label{LiouvannZZ}
S_L  = \frac{1}{4\pi} \int_{\Gamma} d^2 z \left( \p \Phi \bar{\p} \Phi + e^{\Phi}\right)
 - \frac{m}{4\pi} \int_\Sigma dz \Phi - \frac{2}{\epsilon} 
\end{align}
where $\Gamma$ is the region $y \in (\epsilon,\frac{\beta}{2}-\epsilon)$ with $\epsilon \to 0$. The on-shell action evaluates to
\begin{equation} \label{SLann}
    S_L=-\tau_0 \frac{r_H^2}{2}-(\frac{\beta}{2}-\tau_0)\frac{(r_H')^2}{2}+m-m\log(r_0) 
\end{equation}
where $r_0$ is expressed in terms of $r_H$ and $r_H'$ as
\begin{equation}
    r_0^2=\frac{1}{2}(r_H^2+(r_H')^2)+\frac{m^2}{4}+\frac{(r_H^2-(r_H')^2)^2}{4m^2}
\end{equation}
and the parameters $r_H$ and $r_H'$ entering the solution can be expressed in terms of $\tau_0, \beta$ and $m$ by solving the continuity and junction conditions for the Liouville field,
\begin{equation} \label{juncannulus}
    r_0=\frac{r_H}{\sin(r_H \tau_0)}=\frac{r_H'}{\sin(r_H'(\frac{\beta}{2}-\tau_0))} \qquad -r_H'\cot(r_H'(\frac{\beta}{2}-\tau_0))-r_H\cot(r_H \tau_0)=m \ .
\end{equation}
Requiring that $m$ is positive implies that $r_H, r_H' \in \mathbb{R}_+$.
 Now, we compare the Liouville action \eqref{SLann} to the spectral decomposition of the 1-point function on the annulus, 
\begin{equation}
   \langle L(\tau_0) \rangle_{\rm annulus} = \bra{ZZ} e^{-(\frac{\beta}{2}-\tau_0)H}L_{\Sigma} e^{-\tau_0 H}\ket{ZZ}
\end{equation}
to extract the general matrix elements. In the semiclassical limit, the spectral decomposition is
\begin{equation} \label{ann}
       \langle L(\tau_0) \rangle_{\rm annulus} \approx e^{\frac{\beta c}{24}} \int_{\frac{c}{24}}^{\infty} dhdh' \bra{ZZ}h' \rangle\bra{h'}L_{\Sigma}\ket{h}\langle h \ket{ZZ} e^{-2h\tau_0}e^{-2h'(\frac{\beta}{2}-\tau_0)}
\end{equation}
For normalizable states $\ket{h}$ and $\ket{h'}$, the matrix element $\langle h' |L_{\Sigma} | h \rangle$ was calculated from the torus 2-point function in \eqref{CLans} up to a phase. In order to be be consistent with \eqref{SLann} the phase needs to cancel with the reflection factors in the ZZ wavefunction,
\begin{align} \label{zzplug2}
\langle h' |L_{\Sigma} | h \rangle &\approx s_L(P)^{1/2} \rho_0(P)^{-1/2}s_L(P')^{-1/2}\rho_0(P')^{-1/2} e^{\frac{c}{6}c_L(\gamma(h),\gamma(h'))+\frac{1}{2}S(h)+\frac{1}{2}S(h')} 
\end{align}
Thus \eqref{ann} becomes
\begin{align}\label{spectralAnn}
   \langle L(\tau_0) \rangle_{\rm annulus}  &\approx e^{\frac{\beta c}{24}}\int_{\frac{c}{24}}^\infty dh dh' \exp\left[ \frac{c}{6}c_L(\gamma(h),\gamma(h')) + \frac{1}{2}S(h) +\frac{1}{2}S(h') - 2 h \tau_0 -2 h'(\frac{\beta}{2}-\tau_0)\right] \ .
\end{align}
The saddlepoint evaluation of this integral gives $e^{-\frac{c}{6}S_L}$ with the action \eqref{SLann}.

Let us summarize briefly the results of this section. By considering the defect correlators on the disk, sphere, plane, and annulus, we have shown that at large $c$, the matrix elements $\langle h' | L_\Sigma | h\rangle$ and $\langle h | L_\Sigma | 0\rangle$ are given by \eqref{zzplug2} and \eqref{finalc1} respectively. The phases in these expressions cancel against similar phases in the ZZ wavefunction such that the spectral decomposition is purely real; on the disk \eqref{LZZspectral} and annulus \eqref{spectralAnn}.

\section{Bootstrapping line defects in large-$c$ CFT} \label{secbootstrap} 

We now shift our focus to the study of line defects in unitary, compact large-$c$ 2d CFTs with a discrete spectrum and a normalizable vacuum state. This class of theories includes, as a special case, holographic CFTs dual to 3d gravity, which in addition to the aforementioned properties have a sparse spectrum of low-lying operators. Our goal is to bootstrap the line defects in these theories and to derive the universal contribution of heavy operators to their correlation functions.

\subsection{Background on Liouville, identity blocks, and universal OPEs}\label{secbackground}
To put this goal into a more familiar context, let's review the universal behavior of some more common observables: the density of states $\rho(h,\bh)$ and the universal heavy operator contribution to correlation functions. In any unitary, compact 2d CFT, the asymptotic density of primary states as $h,\bh \to \infty$ is given by the Cardy entropy \eqref{leadingCardy} \cite{Cardy:1986ie},
\begin{align}\label{cardyformula}
\rho(h,\bh) \approx \rho_0(h) \rho_0(\bh) \ , \quad
\rho_0(h) \approx e^{S(h)}
\end{align} 
with $\rho_0$ defined in \eqref{rho0}.
The exact density of states is a sum of delta functions, while the right-hand side is a smooth function of $(h,\bh)$, so to be more precise this equation holds after integrating both sides over a microcanonical energy window \cite{Hartman:2014oaa,Mukhametzhanov:2019pzy}. The Cardy formula is derived by solving the modular bootstrap equation at high temperature. In this limit, the partition function is dominated by the vacuum state in one channel, and thus the Casimir energy determines the asymptotics in the dual channel.

Similar logic was used to study the universal contribution of heavy operators to correlation functions of local operators in \cite{Collier:2019weq}. Consider a limit where the 4-point correlation function is dominated by the identity operator in one channel. In the dual channel, the identity must be reproduced by heavy operators. This leads to a universal formula for OPE coefficients, analogous to the Cardy formula \cite{Collier:2019weq}:
\begin{align}\label{C0formula}
|c_{ijk}|^2 \approx C_0(h_i, h_j, h_k) C_0(\bh_i, \bh_j, \bh_k)
\end{align}
where $C_0$ is a known special function, found from the fusion transformation of the identity conformal block. This relation holds in the limit with at least one operator taken to be heavy, upon averaging over a large number of nearby heavy states.

The Cardy entropy $\rho_0$  in \eqref{cardyformula} and the universal OPE coefficient $C_0$ in \eqref{C0formula} are both closely related to Liouville CFT. To be clear, these formulas do \textit{not} apply to Liouville itself, which does not have a normalizable vacuum state, and does not satisfy the usual Cardy entropy formula \cite{Seiberg:1990eb}. The relation to Liouville is indirect, and it arises because of the role that Liouville plays in the representation theory of the Virasoro algebra and hence the structure of conformal blocks \cite{Ponsot:1999uf}. An elegant way to understand this is using the ZZ boundary state \cite{Zamolodchikov:2001ah}. Define the chiral Virasoro characters
\begin{align}
\chi_0(\tau) &= \frac{1}{\eta(\tau)} q^{- (c-1)/24}(1-q)  \\
\chi_h(\tau) &= \frac{1}{\eta(\tau)} q^{h - (c-1)/24}  \qquad(h>0) \ ,
\end{align}
with $q = e^{2\pi i \tau}$. 
The Cardy density of states $\rho_0$ is by definition the modular transformation kernel appearing in
\begin{align}\label{chifusion}
\chi_0(-\frac{1}{\tau})  = \int_{(c-1)/24}^\infty dh \rho_0(h) \chi_h(\tau) \ . 
\end{align}
See \eqref{rho0} for the exact formula for $\rho_0$. 
Now, the ZZ boundary condition in Liouville is such that the only boundary operator is the identity. This means that if we cut open a path integral  on a slice that that ends on a ZZ boundary, then only the vacuum state can appear in the spectral decomposition. By cutting open the Liouville partition function on the annulus, with ZZ boundary conditions at both boundaries, this implies
\begin{align}\label{fusionRhoL}
\chi_0(-\frac{1}{\tau}) &= \langle ZZ| \exp\left[i \pi  \tau(L_0 + \bL_0 - \frac{c}{12})\right] | ZZ \rangle =
\int_{\mathbb{R}} dP | \langle\!\langle P | ZZ\rangle|^2 \chi_{h_P}(\tau)  \ . 
\end{align}
In the last equation we have expanded the ZZ boundary condition in Ishibashi states, which satisfy $\langle\!\langle P| e^{i \pi \tau H}|P'\rangle\!\rangle = \chi_{h_P}(\tau) \delta(P-P')$, and used the Liouville parameterization \eqref{liouvillehP}.
Comparing to the fusion transformation \eqref{chifusion} gives the Cardy density of states as
\begin{align}\label{rho0zz}
\rho_0(h) =  2 \frac{dP}{dh} \rho_0(P) , \quad \rho_0(P) = | \langle\!\langle P | ZZ\rangle|^2 \ . 
\end{align}
The ZZ wavefunction is calculated in Liouville CFT, but the result is a general formula for $\rho_0$  that can be applied to other CFTs.

Similarly, the Virasoro identity block for a 4-point correlation function has the fusion transformation
\begin{align}\label{C0fusion}
 \vcenter{\hbox{
\begin{tikzpicture}[scale=0.5]
\draw (0,1) -- (1, 0);
\draw (0, -1) -- (1,0);
\draw (1,0) -- (3,0);
\draw (3,0) -- (4,1);
\draw (3,0) -- (4,-1);
\node at (-0.25,1) {$1$};
\node at (-0.25,-1) {$1$};
\node at (4.25,1) {$2$};
\node at (4.25,-1) {$2$};
\node at (2,0.4) {$\id$};
\end{tikzpicture}
}}
\qquad = \int_{\frac{c-1}{24}}^\infty dh \rho_0(h) C_0(h_1, h_2, h) \vcenter{\hbox{
\begin{tikzpicture}[scale=0.5]
\draw (0,0) -- (1,-1);
\draw (1,-1) -- (2,0);
\draw (1,-1) -- (1,-3);
\draw (1,-3) -- (0, -4);
\draw (1,-3) -- (2,-4);
\node at (-0.25,0) {$1$};
\node at (2.25,0) {$2$};
\node at (-0.25, -4) {$1$};
\node at (2.3, -3.9) {$2$};
\node at (1.4,-2) {$h$};
\end{tikzpicture}
}}
\end{align}
This relation defines $C_0$. To derive a Liouville formula for it, consider a correlator (in a compact CFT) of four local operators on the plane, $z \in \mathbb{C}$:
\begin{align}\label{gcora}
G = \langle \O_1(z_1,\bz_1) \O_2(z_2,\bz_2) \O_1(\bz_1, z_1) \O_2(\bz_2, z_2)\rangle
\end{align}
We have inserted the operators in a configuration that is reflection-symmetric about the real axis, with $z_1, z_2 \in LHP$. Denote the full (chiral$\times$anti-chiral) contribution of the identity block in the channel $\O_1 \O_1 \to \id \to \O_2 \O_2$, with points inserted at the positions in \eqref{gcora}, by
\begin{align}
G|_{\id} &= | {\cal F}_{\id} |^2 \ .
\end{align}
Our convention is to include all of the position dependence in the block (not just the cross ratio). The same logic used for the torus character above implies that the chiral block is given by a Liouville correlator on the lower half plane, with ZZ boundary conditions:
\begin{align}\label{F0ZZ}
{\cal F}_{\id} &=  \langle ZZ |\hat{V}_{\alpha_1}(z_1)\hat{V}_{\alpha_2}(z_2) | 0 \rangle \ .
\end{align}
The hatted vertex operators are normalized as
\begin{align} \label{verhat}
\hat{V}_\alpha = \frac{1}{U(P)} V_\alpha , 
\end{align}
where $U(P)$ is the one-point function in the upper half plane (or disk) defined by
\begin{align}
\langle V_\alpha(z)\rangle_{ZZ} = \frac{U(P)}{|z-\bz|^{2h_{\alpha}}} \ . 
\end{align}
The factors of $U(P)$ in \eqref{F0ZZ} ensure that the vacuum state appears with unit coefficient as $\hat{V}_\alpha$ approaches the boundary, so that the identity block has the correct singularity. The one-point function is known exactly, 
\begin{align}\label{Uform}
U(P) &= -ia_b \left( s_L(P) \rho_0(P) \right)^{1/2}  \ , 
\end{align}
where $s_L(P)$ is the Liouville reflection amplitude and $a_b$ is a constant. See appendix \ref{liouvilleformulas} for more details. Now expanding the ZZ state in Ishibashi states gives 
\begin{align}\label{fourtran}
{\cal F}_{\id}
&=  \int_{-\infty}^\infty dP  \langle\!\langle P| \hat{V}_{\alpha_1}(z_1) \hat{V}_{\alpha_2}(z_2)|0\rangle \langle ZZ| P\rangle\!\rangle \\
&= 2i \int_{0}^\infty dP \sqrt{\rho_0(P)} s_L(P)^{1/2} \langle\!\langle  P| \hat{V}_{\alpha_1}(z_1) \hat{V}_{\alpha_2}(z_2)|0\rangle \notag\\
&= \frac{a_b^2}{2\pi} \int_0^{\infty} dP \, \rho_0(P)  \frac{C_L(P_1, P_2, P)}{U(P_1) U(P_2) U(P)} \widetilde{{\cal F}}_P\notag
\end{align}
where $C_L$ is the DOZZ 3-point function \cite{Dorn:1994xn,Zamolodchikov:1995aa}. We have used some formulas from appendix \ref{liouvilleformulas} to simplify the integrand. Comparing to the fusion transformation \eqref{C0fusion}  gives an explicit formula for $C_0$ in terms of Liouville data \cite{Zamolodchikov:2001ah, Collier:2019weq}. Since the correlator in any compact CFT is dominated by the identity in the OPE limit, the conclusion is that Liouville determines the universal asymptotics of the OPE coefficients in compact CFTs \cite{Collier:2019weq}. 

This derivation holds when the external operators have $h > \frac{c-1}{32}$, such that only normalizable states with $P \in \mathbb{R}$ appear in the spectral decomposition of the 4-point function. The fusion kernel for light external operators is obtained from \eqref{fourtran} by analytic continuation in the weights, keeping the extra contributions from light states that arise as poles in the integrand cross the real axis \cite{Collier:2018exn}.

\subsection{Exact universal data of line defects}
We will now apply the same strategy to line defects. Consider a line operator $\CL_\Sigma$, in a compact unitary CFT, with the same conformal transformation properties as the Liouville line operator $L_\Sigma$ defined in \eqref{quantumL}. 

This does not require large $c$, but the main example we have in mind is a line defect in a holographic CFT that creates a dust shell in the bulk. These operators were studied in \cite{Anous:2016kss}. They are defined by a product of $n$ local operators arranged uniformly along some curve, and taking $n \to \infty$ to create a line defect. The $n\to \infty$ limit is taken together with the large-$c$ limit in order to create a line defect with energy $E = \O(c)$. Consider a defect on the cylinder, $z \sim z + 2\pi$, along the circle Im $z = \tau_0$. The line defect is defined in the large-$c$ limit by
\begin{align}\label{defCL}
\CL(\tau_0) = \prod_{j = 1}^n \O_j(i\tau_0 + \frac{2\pi j}{n})
\end{align}
with the number of insertions and the dimensions of each operator $\O_j$ scaling as
\begin{align}\label{dustscalings}
n \sim \sqrt{c} , \qquad h \sim \sqrt{c} \ . 
\end{align}
The local scalar operators $\O_j$ are all distinct operators (i.e. different flavors) but with the same conformal weight. They are holographically dual to the dust particles that make up the thin shell on the gravity side. The scalings in \eqref{dustscalings} ensure that an individual dust particle behaves in the bulk like a probe particle, but together they backreact to create a state of energy $E \sim c$.

The basepoint $z_0=i\tau_0$ in \eqref{defCL} is some particular point along the defect. Because the operators are distinct, the line defect is not translation invariant, so the basepoint is meaningful. We will see that $\langle \CL^\dagger(\tau_0) \CL(-\tau_0) \rangle$ corresponds to a Schwarzschild black hole while shifting the base point as in $\langle \CL^\dagger(\tau_0) e^{i\theta J} \CL(-\tau_0) \rangle$ creates a rotating black hole.

This is one way to produce a line defect with the desired properties, but for now we will work at finite $c$ and simply assume the existence of a line defect $\CL_\Sigma$ in a compact CFT that has the same transformation properties as the Liouville defect $L_\Sigma$. That is, $\CL_\Sigma$ transforms like the continuous infinite product of local operators in \eqref{discreteLsigma}.

To study this line defect in the compact CFT, we will use properties of the line defect $L_\Sigma$ in Liouville CFT, tuned to have the same central charge. Liouville is being used as an auxiliary system to extract the universal OPE data in the compact CFT --- one must be careful not to conflate the compact CFT of interest with the auxiliary Liouville theory. When we write correlators involving $L_\Sigma$ they are calculated in Liouville theory, and when we write correlators involving $\CL_\Sigma$, they are calculated in a compact theory with a normalizable vacuum.

\begin{figure}[t]
    \centering
    \begin{overpic}[scale=0.4,grid=false, tics=20, trim=100 180 100 100, clip]{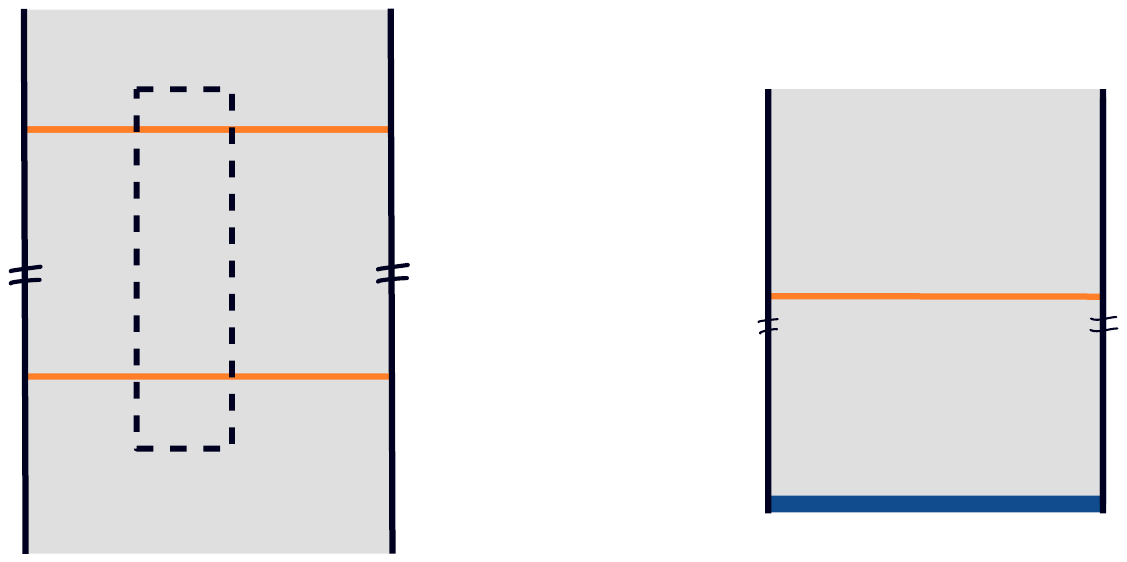}
    \put (-8, 31) {$y=\tau_0$}
    \put (-8, 13) {$y=-\tau_0$}
    \put (25, 25) {$\mathds{1}$}
    \put (43, 20) {$\implies$}
    \put (73, 0) {ZZ}
    \end{overpic}
    \caption{\small The identity block for line defects on the cylinder (or sphere). On the left is the configuration of line defects in the compact CFT for the correlation function $\langle \CL^\dagger(\tau_0) \CL(-\tau_0)\rangle$. The identity block is defined by projecting onto the identity on the dashed rectangle and all similar rectangles cutting the line defects. On the right is the Liouville line defect that calculates the chiral identity block.}
    \label{fig:identity-cylinder}
\end{figure}

Consider the defect two-point function on the cylinder.
We define the identity block for this 2-point function by projecting onto the identity on every rectangle, of any width, centered on the real axis. See figure \ref{fig:identity-cylinder}. Denote the identity contribution in this channel (which we assume to be nonzero) by
\begin{align}
 \langle \CL^\dagger(\tau_0) \CL(-\tau_0) \rangle|_{\id} &= |{\cal F}_{\id}^\CL|^2 \ .
\end{align}
By the same logic applied above to the torus character and the Virasoro block of local operators in section \ref{secbackground}, the chiral defect identity block ${\cal F}_{\id}^\CL$ can be calculated in Liouville CFT. In Liouville, the projection onto the identity is implemented by the ZZ boundary condition. Thus we have
\begin{align}\label{F0cl}
{\cal F}_{\id}^{\CL} &= \langle 0 | \hat{L}(\tau_0) | ZZ \rangle \ . 
\end{align}
This is analogous to \eqref{F0ZZ}, with the local operators replaced by an infinite product creating a line defect.
The hatted line operator is normalized by dividing by the product of disk one-point functions, like the hatted vertex operators in \eqref{verhat}. That is, using the discretized representation \eqref{discreteLsigma},
\begin{align}\label{defLhat}
\hat{L}_\Sigma &= \lim_{n \to \infty} \prod_{j=1}^n \hat{V}_{ \frac{m}{2nb} }(z_i) = {\cal N} L_{\Sigma}  \ ,
\end{align}
where the normalization factor is 
\begin{align}
{\cal N}=  \lim_{n\to \infty}  U(P_n)^{-n}
\end{align}
with $P_n=\frac{i}{2}( b + b^{-1} - \frac{m}{nb})$  the Liouville momentum corresponding to $V_{\frac{m}{2nb}}$.
This can be evaluated using the explicit formula for $U(P)$ from appendix \ref{liouvilleformulas}, giving
\begin{align}
{\cal N} = \left( \frac{\Gamma(b^2)}{4b^2 \Gamma(1-b^2)}\right)^{ \frac{m}{2b^2}}e^{-\frac{m}{b^2}\psi(2 + b^{-2}) - m \psi(1+b^2)} , \quad \psi(x) = \frac{\Gamma'(x)}{\Gamma(x)} \ . 
\end{align}
In the semiclassical limit $b \to 0$, the leading behavior of the normalization factor is
\begin{align}\label{Ncl}
\log {\cal N} \sim -\frac{c}{6} m \log 2 \ . 
\end{align}
Returning to \eqref{F0cl} we have the following exact formula for the defect identity block:
\begin{align}\label{defectIdZZ}
{\cal F}_{\id}^{\CL} = {\cal N}  \langle 0 |  L_\Sigma e^{-2\tau_0 L_0} | ZZ \rangle \ .
\end{align}
We used the conformal boundary condition $L_0 |ZZ\rangle = \bL_0 |ZZ\rangle$ to replace $L_0 + \bL_0 \to 2L_0$. 
This formula will provide the link between correlation functions in large-$c$ CFT, the Einstein action of gravitating thin shells, and Liouville amplitudes. It is exact, though we will only evaluate it explicitly in the semiclassical limit.

The universal asymptotics of the line defect matrix elements are now determined by expanding \eqref{defectIdZZ} in the dual channel. We temporarily restrict to $m>1$ so that only heavy states appear in the spectral decomposition. The expansion in Ishibashi states is 
\begin{align}
{\cal F}_{\id}^{\CL} &= 2i{\cal N} \int_0^\infty dP \rho_0(P)^{1/2} s_L(P)^{1/2} \langle\!\langle P| e^{-2\tau_0 L_0}L_{\Sigma} |0\rangle  \ .
\end{align}
The Ishibashi states in the Liouville expression produce conformal blocks in the dual channel, as in the four-point function discussed above. Therefore we can write the fusion transformation as
 \begin{align}
{\cal F}_{\id}^{\CL}  &= 2\int_0^\infty dP \rho_0(P) C_0^{\CL}(h_P) \widetilde{{\cal F}}_P^{\CL}  \ , 
\end{align}
where $\widetilde{{\cal F}}_P^{\CL}$ is the dual channel block and  
\begin{align}\label{exactCLP1}
C_0^{\CL}(h_P) &= i {\cal N} s_L(P)^{1/2} \rho_0(P)^{-1/2}  \langle h_P | L_{\Sigma} |0\rangle \ .
\end{align}
This formula for the fusion kernel is exact. Since the line defects in $\langle \CL^\dagger(\tau_0) \CL(-\tau_0)\rangle$ fuse to the vacuum in the limit $\tau_0 \to 0$, the vacuum fusion kernel determines the asymptotic structure constants of the line defect at high energy, in a compact CFT:
\begin{align}
\langle h,\bh| {\CL}_{\Sigma} | 0 \rangle \approx C_0^{\CL}(h) C_0^{\CL}(\bh) \ .
\end{align}
This is analogous to the Cardy formula for the density of states and the universal OPE coefficient $C_0$. As in those cases, it holds after averaging over a microcanonical energy window. And like the Cardy formula, in a general CFT the formula holds as $h,\bh \to \infty$ at fixed $c$, but in a holographic theory it holds more broadly for states above the black hole threshold \cite{Hartman:2014oaa}.

\subsubsection{Semiclassical limit}

In the semiclassical limit $c \to \infty$, both the identity block itself, and the universal structure constants, can be calculated from semiclassical Liouville theory. The semiclassical block is defined as the leading term in the exponential in
\begin{align}
{\cal F}^{\CL}_{\id} \approx e^{-\frac{c}{6}f^{\CL}_{0}} \ . 
\end{align}
Using \eqref{defectIdZZ}, we find
\begin{align}
f^{\CL}_{0} = S_L +m \log 2 , 
\end{align}
where $S_L$ is the semiclassical Liouville action of the line defect on a disk, given in \eqref{SLdisk}. Thus\footnote{
We have normalized the identity block for the line defect such that each operator in the infinite product in \eqref{discreteLsigma} is unit normalized. This corresponds to $
\langle \CL^\dagger(\tau_0) \CL(-\tau_0)\rangle \sim (2\tau_0)^{-cm/3}$. 
The solution to the junction conditions \eqref{zz1junc} in the limit $\tau_0 \to 0$ is $r_0 \sim \frac{1}{\tau_0} + \frac{m}{3}$, $r_H \sim \sqrt{ \frac{2m}{\tau_0}}$. Plugging this solution into \eqref{finalfCL} indeed gives the expected limit. \label{opelimsph}
}
\begin{align}\label{finalfCL}
f^{\CL}_0
&=  m(\log2+1)  -\frac{1}{2}(r_H^2-1)\tau_0 - \sinh^{-1}\frac{1}{r_0} - m \log r_0
\end{align}
where $r_0$ and $r_H$ are functions of $m$ and $\tau_0$ determined by the junction conditions \eqref{zz1junc}.

Now consider the matrix elements.
For real $P$, the reflection phase in \eqref{exactCLP1} cancels the phase in the Liouville matrix element $\langle P | L_{\Sigma}| 0\rangle$. Therefore up to a real normalization constant ${\cal N}'$ that is independent of $P$, we may write it as
\begin{align}\label{cphases}
C_0^{\CL}(h_P) = {\cal N}' \rho_0(P)^{-1/2} | \langle h_P | L_{\Sigma}| 0\rangle| \ . 
\end{align}
In the semiclassical limit we parameterize the weights by $P = \sqrt{ \frac{c}{24}}\gamma$ as in section \ref{secLiouville}, and write the semiclassical structure constants as
\begin{align}
\log C_0^{\CL}(h_P) &\sim \frac{c}{6} c_D(\gamma) \\
\log | \langle h_P | L_{\Sigma}| 0\rangle|  &\sim \frac{c}{6}c_L(\gamma) \ . 
\end{align}
The semiclassical Liouville structure function $c_L(\gamma)$ was calculated in \eqref{CL0}. Comparing to \eqref{cphases}, the universal structure constants of a compact CFT at large $c$ and high energy are given by
\begin{align}
c_{\CL}(\gamma) = c_L(\gamma)  -  \frac{\pi \gamma}{2} - m \log 2\ ,
\end{align}
Using \eqref{CL0} we find 
\begin{align}\label{finalCD1}
c_{\CL}(\gamma) &=- m(1+\log(4m)) - \gamma \cot^{-1} \left( \frac{m^2 - \gamma^2-1}{2m\gamma}\right)\notag
+ \frac{m-1}{2}\log\left[ (m-1)^2+\gamma^2\right]\\
&\qquad + \frac{m+1}{2}\log\left[ (m+1)^2+\gamma^2\right]  
\end{align}

In a holographic CFT dual to pure gravity plus dust shells dual to a line defect with $m>1$, this formula holds down to the black hole threshold, $\gamma > 0$ (neglecting potential instabilities for small black holes). For $m<1$ it also applies below the black hole threshold using the formula \eqref{LZZspectral}. Thus the fusion transformation at large $c$, for any $m>0$, is 
\begin{align}\label{fusion1bb}
{\cal F}_{\id}^{\CL} \approx e^{-\frac{c}{6}f_0^{\CL}} \approx \int_{h_{min}}^{\infty} dh\,  \exp\left[S(h) +\frac{c}{6}  c_D(\gamma(h))  - 2 \tau_0 h\right]\ . 
\end{align}
Note that the combination $e^{S(h) +\frac{c}{6} c_D(\gamma(h))} = e^{\frac{1}{2}S(h) +\frac{c}{6} c_L(\gamma(h))}$ remains real when analytically continued to the subthreshold regime $h_{min} < h < \frac{c}{24}$, with $h_{min}$ defined in \eqref{hmin}. Apart from the normalization constant, the fusion transformation \eqref{fusion1bb} is identical to the spectral decomposition of the Liouville defect on the disk \eqref{LZZspectral}.

\subsection{General matrix elements}

\begin{figure}[t]
    \centering
    \begin{overpic}[scale=0.4,grid=false, tics=20, trim=100 200 100 100, clip]{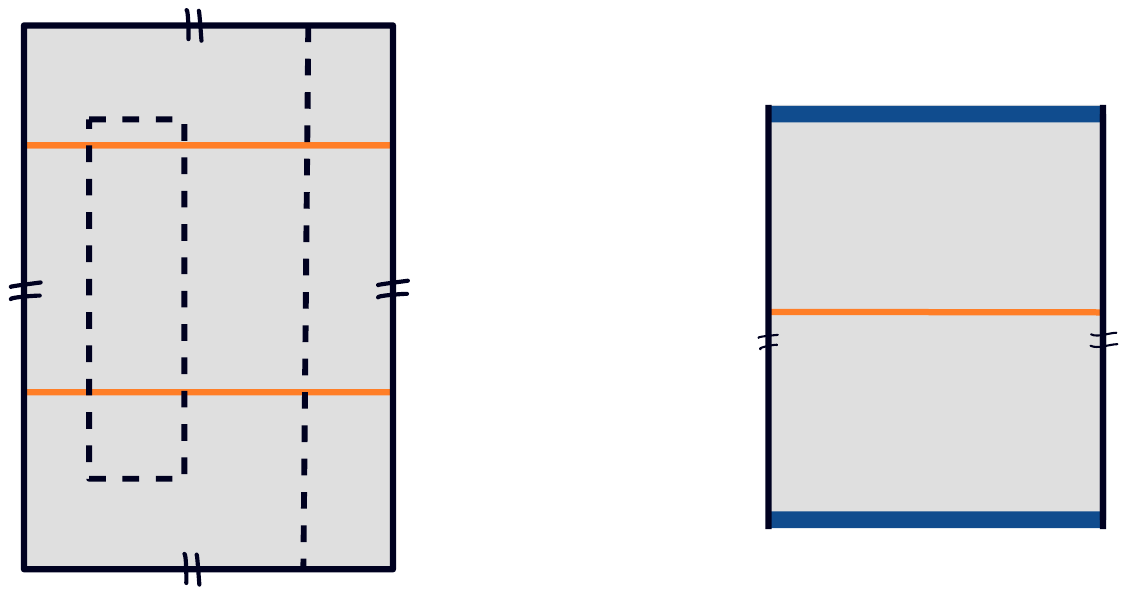}
    \put (-3, 32) {$y=\tau_0$}
    \put (-3, 14) {$y=-\tau_0$}
    \put (26, 25) {$\mathds{1}$}
    \put (35, 25) {$\mathds{1}$}
    \put (50, 20) {$\implies$}
    \put (79, 1) {ZZ}
    \put (79, 38) {ZZ}
    \end{overpic}
    \caption{\small The identity block for line defects on the torus. The left figure defines the identity block in a compact CFT and the right figure defines the Liouville correlator that calculates the chiral block.}
    \label{fig:identity-torus}
  \end{figure}

A similar analysis can be repeated on the torus to derive the structure constants with two independent weights. We shall be very brief. Denote the identity block in the channel described by figure \ref{fig:identity-torus} as
\begin{equation}
  \langle \CL^\dagger(\tau_0)\CL(-\tau_0) \rangle _\beta |_\id=|\mathcal{F}^{\CL}_\id|^2 \ .
\end{equation}
This block can be computed by the following Liouville correlator with ZZ boundaries,
\begin{equation}
  \mathcal{F}^{\CL}_\id=\bra{ZZ}e^{-\tau_0H} \Hat{L}_{\Sigma} e^{-(\frac{\beta}{2}-\tau_0)H} \ket{ZZ} \ .
\end{equation}
Expanding the ZZ states using Ishibashi states, we have
\begin{align}
{\cal F}_{\id}^{\CL} &=4{\cal N} \int_0^\infty dPdP' \rho_0(P)^{1/2} s_L(P)^{1/2}\rho_0(P')^{1/2} s_L(P')^{-1/2} \langle\!\langle P| e^{-\tau_0 H}L_{\Sigma} e^{-(\frac{\beta}{2}-\tau_0)H} |P'\rangle\!\rangle \ .
\end{align}
Therefore we can write the fusion transformation as 
 \begin{align}
{\cal F}_{\id}^{\CL}  &= 4\int_0^\infty dP dP'\rho_0(P) \rho_0(P') C_0^{\CL}(h_P,h_P')\widetilde{{\cal F}}_{P,P'}^{\CL}  \ , 
\end{align}
where $\widetilde{{\cal F}}_{P,P'}^{\CL}$ is the dual channel block and
\begin{align}\label{exactCLP2}
C_0^{\CL}(h_P,h_{P'}) &=  {\cal N} s_L(P)^{1/2} \rho_0(P)^{-1/2}  \rho_0(P')^{-1/2}s_L(P')^{-1/2}  \langle h_P | L_{\Sigma} |h_{P'}\rangle \ .
\end{align}
So far, the analysis is exact. Now, we take the semiclassical limit where the block exponentiates ${\cal F}^{\CL}_{\id} \approx e^{-\frac{c}{6}f^{\CL}_0} $. The semiclassical block can be expressed in terms of the Liouville action of the line defect on the annulus in \eqref{SLann} by\footnote{The normalisation of the block is fixed by noting that as $\tau_0\to 0$ at fixed $m,\beta$, the correlator $\langle \CL^\dagger(\tau_0)\CL(-\tau_0) \rangle_\beta \sim \langle \id \rangle_\beta (2\tau_0)^{-\frac{cm}{3}}$. In this limit, the continuity and junction conditions corresponding to the Liouville solution in \eqref{Liouann} can be solved to give $r_H\sim \sqrt{\frac{2m}{\tau_0}}$, $r_H' \sim \frac{2\pi}{\beta}$ and $r_0 \sim \frac{1}{\tau_0}$. \label{opelimtorus}}
\begin{equation} \label{blocktorus}
  f^{\CL}_0=S_L+m\log2 = -\tau_0 \frac{r_H^2}{2}-(\frac{\beta}{2}-\tau_0)\frac{(r_H')^2}{2}+m-m\log(\frac{r_0}{2}) 
\end{equation}
with the parameters $r_0,r_H,r_H'$ related to $\tau_0,\beta,m$ by the continuity and junction conditions \eqref{juncannulus}.  The universal structure constant in the semiclassical limit is
\begin{align}\label{finalCD2}
c_D(\gamma,\gamma') &= c_L(\gamma,\gamma') - \frac{\pi}{2}(\gamma+\gamma') -m\log 2  \notag \\
&=  
- \gamma \cot^{-1}\left( \frac{m^2+\gamma'^2-\gamma^2}{2m\gamma}\right)
- \gamma' \cot^{-1} \left( \frac{m^2 + \gamma^2 - \gamma'^2}{2m \gamma'}\right)\\
& \quad  + \frac{m}{2}\log\left( \frac{m^2}{4} + \frac{1}{2}(\gamma^2+\gamma'^2) + \frac{1}{4m^2}(\gamma^2-\gamma'^2)^2\right) - m(1+\log2) \notag
\end{align}

\section{Continuum Virasoro blocks and uniformization of surfaces with defects} \label{secblocks}

In the previous section, we computed the semiclassical identity blocks by taking the $c\to \infty$ limit of an exact (in $c$) expression for the identity block written in terms of an appropriate Liouville correlator with a ZZ boundary condition. In this section, we derive the semiclassical blocks directly by solving the monodromy problem for the continuum Virasoro blocks described in \cite{Anous:2016kss} by generalizing Zamolodchikov's monodromy method for the finite-point Virasoro blocks \cite{ZamoRecursion}. We will recover the same expressions for the semiclassical identity blocks derived in the previous section using Liouville technology. This section therefore provides another perspective on those results but it can be skipped on a first reading, as the blocks were already obtained above. In principle, the monodromy problem could be solved to obtain more general continuum conformal blocks (i.e non-identity blocks) unlike the Liouville method used in the previous section which can only be used to compute the continuum identity block. However, in this paper, we will not be solving the general monodromy problem. 

The situation studied in \cite{Anous:2016kss} was a Vaidya spacetime in the bulk, describing a null collapsing shell. In terms of the line defects in the boundary CFT, this corresponds to the limit where the two defects are very close together. In this limit, we will reproduce the results of \cite{Anous:2016kss} for the identity block. The method used here is simpler and more general. Rather than solving the monodromy problem by patching together solutions of the Fuchsian equation, as in \cite{Anous:2016kss}, we will solve the Schwarzian equation directly. This gives both the conformal block and the Schottky uniformization of the Riemann surface. For additional recent work on the monodromy solution with shells, see \cite{Chen:2024hqu}.

The procedure we follow to compute the semiclassical identity block can be summarized as follows: We derive the form of the semiclassical stress tensor of the CFT in the presence of line defects using symmetry and regularity which determines the stress tensor upto an accessory parameter. We fix the accessory parameter by imposing trivial monodromy of a second order ordinary differential equation \eqref{Zamodiff} involving the stress tensor around appropriate non-contractible cycles of the surface (plane or sphere or torus) with defects. We solve this monodromy problem by showing that it is equivalent to uniformizing the corresponding surface in a way which generalizes the Schottky uniformization procedure to surfaces with line defects. This equivalence is a continuum version of the equivalence between the monodromy problem for the semiclassical identity block and the Schottky uniformization of Riemann surfaces punctured by point defects observed in \cite{Hartman:2013mia,Faulkner:2013yia}. (We refer the reader to \cite{Hartman:2013mia,Anous:2016kss} for an introduction to the monodromy method.)  We then derive a Ward identity for the semiclassical block which expresses the partial derivative of the block with respect to the modulus in terms of the accessory parameter. Finally, we integrate this Ward identity to derive the expression for the semiclassical block. 

Trivial monodromy on the complex plane, in a configuration that is reflection-symmetric about the real axis, is equivalent to PSL$(2,\mathbb{R}$) monodromy on the upper half plane. Solutions to the Fuchsian equation with PSL$(2,\mathbb{R}$) monodromy correspond to single-valued, real Liouville fields. Therefore solutions to the trivial monodromy problem on the plane (resp.~torus) produce solutions of the Liouville equation on the disk (resp.~annulus), with defect insertions, and these are the same solutions studied in section \ref{secLiouville}.

As a warmup, we set up the monodromy problem for the continuum identity block in the planar limit and show that it can be solved by uniformizing the plane with defects for the two cases where the modulus is real which corresponds in the bulk to non-rotating planar shells and the case where we complexify the modulus by introducing a twist which corresponds in the bulk to rotating planar shells. We then generalize these ideas to compute the semiclassical continuum identity block on the sphere and on the torus.

\subsection{Uniformization with planar defects}

\begin{figure}[t]
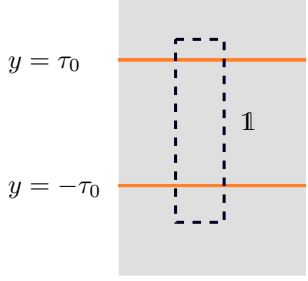

    \centering
    \begin{overpic}[scale=0.4,grid=false, tics=20, trim=100 170 150 150, clip]{figures/id_block_planar.pdf}
    \put (21, 35) {$y=\tau_0$}
    \put (21, 15) {$y=-\tau_0$}
    \put (58, 25) {$\mathds{1}$}
    \end{overpic}
    \caption{\small In the computation of the identity block at a real modulus, we impose trivial monodromy around all the rectangular loops like the one shown in this figure intersecting the operator lines $y=\pm \tau_0$ at the four points $z=x_1\pm i\tau_0, x_2\pm i\tau_0$.}
    \label{fig:planar_monodromy}
\end{figure}

We derive the form of the semiclassical stress tensor of the CFT on the plane in the presence of two infinite defect lines by taking the continuum limit of a large number of local operator insertions uniformly spaced along the lines $\text{Im}z\equiv y=\pm \tau_0$.
We work with the rescaled stress tensor $T(z)\equiv \frac{6}{c}T_{zz}(z)$. Let $T(z)=T^{(1)}(z)+T^{(2)}(z)$ where $T^{(1)}$ and $T^{(2)}$ denote respectively the contribution to the stress tensor from the upper and lower lines of operator insertions. We have
\begin{equation} \label{stresspl}
   T^{(1)}(z)=\sum_{n \in \mathbb{Z}} \left (\frac{6h/c}{(z-z_n)^2}-\frac{a}{z-z_n}\right)
\end{equation}
where we have only written the singular contribution to the stress tensor. The operator insertions are at $z_n=(n+\frac{1}{2})\delta+i\tau_0$ where $\delta$ taken to be a constant is the spacing between successive operator insertions. The operators are all scalar primaries with the same conformal weight $h$ and $a$ is the accessory parameter determined by solving the monodromy problem discussed below. Due to the translational symmetry in the problem, it is also a constant. The sum in \eqref{stresspl} is performed by grouping together the operator insertions at $z=\pm(n+\frac{1}{2})\delta+i\tau_0$ for each $n$ to give
\begin{equation}
    T^{(1)}(z)=\frac{6\pi^2 h}{c\delta^2}\text{sech}^2(\frac{\pi}{\delta}(\tau_0+iz))-i\pi\frac{a}{\delta}\tanh(\frac{\pi}{\delta}(\tau_0+iz)) \ .
\end{equation}
We rescale the conformal weight of the operators and the accessory parameter to define
\begin{equation}
  \frac{\lambda}{2\pi}=\frac{6h}{c\delta} \quad \quad -i\frac{\alpha^2}{2\pi}=\frac{a}{\delta} \ .
\end{equation}
These quantities are defined such that the continuum limit $\delta \to 0$ is taken with the above rescaled quantities held fixed. In this limit, we have
\begin{equation}
   T^{(1)}(z)= \frac{\alpha^2}{2} \text{sgn}(y-\tau_0)+\lambda \delta(y-\tau_0) \ .
\end{equation}
The $\delta$-function can be observed by integrating the stress tensor in a small window around $y=\tau_0$.
We now add to this the contribution from the other line of operator insertions by noting that the accessory parameter switches sign by symmetry.
Therefore, in the continuum limit, the stress tensor with the two lines of operator insertions at $y=\pm \tau_0$ takes the form
\begin{equation} \label{Plstress}
  T(z) \equiv \frac{6}{c}T_{zz}(z)=-\alpha^2 \Theta (-\tau_0<y<\tau_0)+\lambda \delta (y-\tau_0)+\lambda \delta (y+\tau_0) \ .
\end{equation}
Notice that the stress tensor is not a holomorphic function of $z$. In \eqref{Plstress}, we treat $\lambda$ and $\tau_0$ as the input and $\alpha$ is an accessory parameter which can be determined in terms of the input quantities by specifying the monodromy associated with the following second order differential equation
\begin{equation} \label{Zamodiff}
    \psi''(z)+T(z)\psi(z)=0
\end{equation}
around the rectangular cycles intersecting the two lines of operation insertions as illustrated in figure \ref{fig:planar_monodromy}. Following \cite{ZamoRecursion}, the monodromy is related to the internal weight of the Virasoro conformal block in the semiclassical limit. Imposing a given monodromy implicitly determines the accessory parameter as a function of the moduli, and by integrating this relation one obtains the semiclassical block itself.
%

We take $\lambda$ to be a constant, meaning that the local operators are distributed uniformly along the lines $y=\pm \tau_0$. As explained in the next section, this setup is holographically dual to a dust shell in 3d gravity which is homogeneous in the transverse direction.
Since we are currently interested in computing the identity block, we impose trivial monodromy of \eqref{Zamodiff} around each of the rectangular cycles. In the following, we show that the mondoromy problem thus described can be solved by uniformising the upper-half $z$ plane. The strategy is to find a map $w(z)$ that is analytic at the basepoints of the two defects, $z = \pm i \tau_0$, with branch cuts that begin at these points and extend in both directions along both defects. We will see that once we have such a map it automatically provides a solution to the trivial monodromy problem on the rectangle. 
To this end, we define a $w$ coordinate as the ratio of the two linearly independent solutions (with unit Wronskian determinant) to the above differential equation,
\begin{equation}
    w(z)=\frac{\psi_2(z)}{\psi_1(z)}
\end{equation}
in terms of which we can express the stress tensor as a Schwarzian derivative
\begin{equation} \label{Schwarzeqn}
    T(z)=\frac{1}{2}\{w(z),z\}=\frac{1}{2}\left(\frac{w'''}{w'}-\frac{3}{2}\left(\frac{w''}{w'}\right)^2 \right)
\end{equation}
We work with the following solution of the above Schwarzian equation which obeys the reflection symmetry $w(\overline{z})=\overline{w(z)},$
\begin{align} \label{unifpl}
w=\begin{cases}
    z-i\tau_0+iA  \qquad & y>\tau_0 \\
    B\tanh(z\alpha) \qquad & |y|<\tau_0\\
    z+i\tau_0-iA \qquad  & y<-\tau_0
\end{cases}
\end{align}
where $A$ and $B$ are real constants determined by imposing continuity of the map and continuity of its first derivative across either marked points on the two lines of operator insertions $y=\pm \tau_0$ which we choose to be at $z=\pm i\tau_0$,
\begin{equation}
    \begin{split}
        & A= \frac{\sin(2\alpha \tau_0)}{2\alpha}=\frac{\lambda}{\alpha^2+\lambda^2}\\
        & B= \frac{\cos^2(\alpha \tau_0)}{\alpha}=\frac{\alpha}{\alpha^2+\lambda^2}
    \end{split}
\end{equation}
The accessory parameter $\alpha $ can be determined in terms of $\lambda$ and $\tau_0$ by requiring that the jump in the second derivative of the map across either marked points matches with the coefficient of the $\delta$-function in the stress tensor,
\begin{equation} \label{jumppl}
   2i\lambda=\frac{\Delta w''}{w'}\implies  \lambda=\alpha\tan(\alpha \tau_0) \ .
\end{equation}
Interpreting the $w$ coordinate to be uniformising the upper half $z$ plane into the upper half $w$ plane, we can write down an associated Liouville field by\footnote{Imposing reflection symmetry in the solution to the Schwarzian equation breaks the PSL$(2,\mathbb{C})$ redundancy in the solution down to PSL$(2,\mathbb{R}$). Since PSL$(2,\mathbb{R}$) is the isometry group for the Poincare metric on the upper half plane, the Liouville field written down in \eqref{Liouvunif} is uniquely specified by \eqref{Schwarzeqn}.}
\begin{equation}
    e^{\Phi(z,\overline{z})}=-\frac{4|w'(z)|^2}{(w(z)-\overline{w(z)})^2} \ .
\end{equation}
For the present case where the uniformising coordinate is given by \eqref{unifpl}, the Liouville field is given by
\begin{align} \label{Liouvunif}
   \Phi= \begin{cases}
         -2\log(y-\tau_0+A) \qquad y>\tau_0 \\
         -2\log\left(\frac{1}{2\alpha}\sin(2\alpha y)\right) \qquad 0<y<\tau_0
    \end{cases}
\end{align}
To compare to the gravity calculation done for spherical shells below, we need to work in the limit where $r_0,r_H \gg 1$\footnote{More precisely, the planar limit on the gravity side corresponds to $m \to \infty$, $\tau_0 \to 0$ with $m\tau_0=O(1)$. This corresponds to scaling $r_H, r_0 \to \infty$ such that $\frac{r_H}{m},\frac{r_0}{m}=O(1)$ and light and heavy shells correspond respectively to $m<r_H$ and $m>r_H$. In this limit, one can compute the energy of the black hole by solving the transcendental equation,
\begin{equation}
    \sin(r_H \tau_0)=\frac{2m r_H}{m^2+r_H^2} \xleftrightarrow{} \tan(\frac{r_H \tau_0}{2})=\frac{m}{r_H}
\end{equation}}. Requiring that the energies match, we have $2\alpha =r_H$. To verify this, note that the energy density computed by integrating the stress tensor on the real axis is given by
\begin{equation}
    E_{\text{CFT}}=\int \frac{d\theta}{2\pi} T_{yy}=-\frac{1}{2\pi}\int T_{zz}dz-\frac{1}{2\pi}\int T_{\overline{z}\overline{z}}d\overline{z}=\frac{c}{3}\alpha^2 \ .
\end{equation}
On comparing with $E_{\text{grav}}=\frac{c}{12}r_H^2$, we see that $2\alpha=r_H$. We also have $2\lambda=m$ which can be derived by matching the jump condition \eqref{jumppl} with the Israel junction condition,
\begin{equation}
    r_0-r_H \cot(r_H\tau_0)=m \ .
\end{equation}
In the above analysis, we have written down a solution to the Schwarzian equation with respect to marked points on the operator lines chosen on the imaginary axis. But in order to show that the monodromy of the differential equation \eqref{Zamodiff} under \eqref{unifpl} around any rectangular loop is trivial, it is convenient to write down a solution to the Schwarzian equation with respect to a general pair of marked points $x_0\pm i\tau_0$ on the operator lines,
\begin{align}
w=\begin{cases}
    z-x_0+i\tau_0+iA  \qquad & y>\tau_0 \\
    B\tanh(\alpha(z-x_0)) \qquad & |y|<\tau_0\\
    z-x_0+i\tau_0-iA \qquad  & y<-\tau_0
\end{cases}
\end{align}

\begin{figure}[t]
    \centering
    \begin{overpic}[scale=0.35,grid=false, tics=20, trim=70 110 70 110, clip]{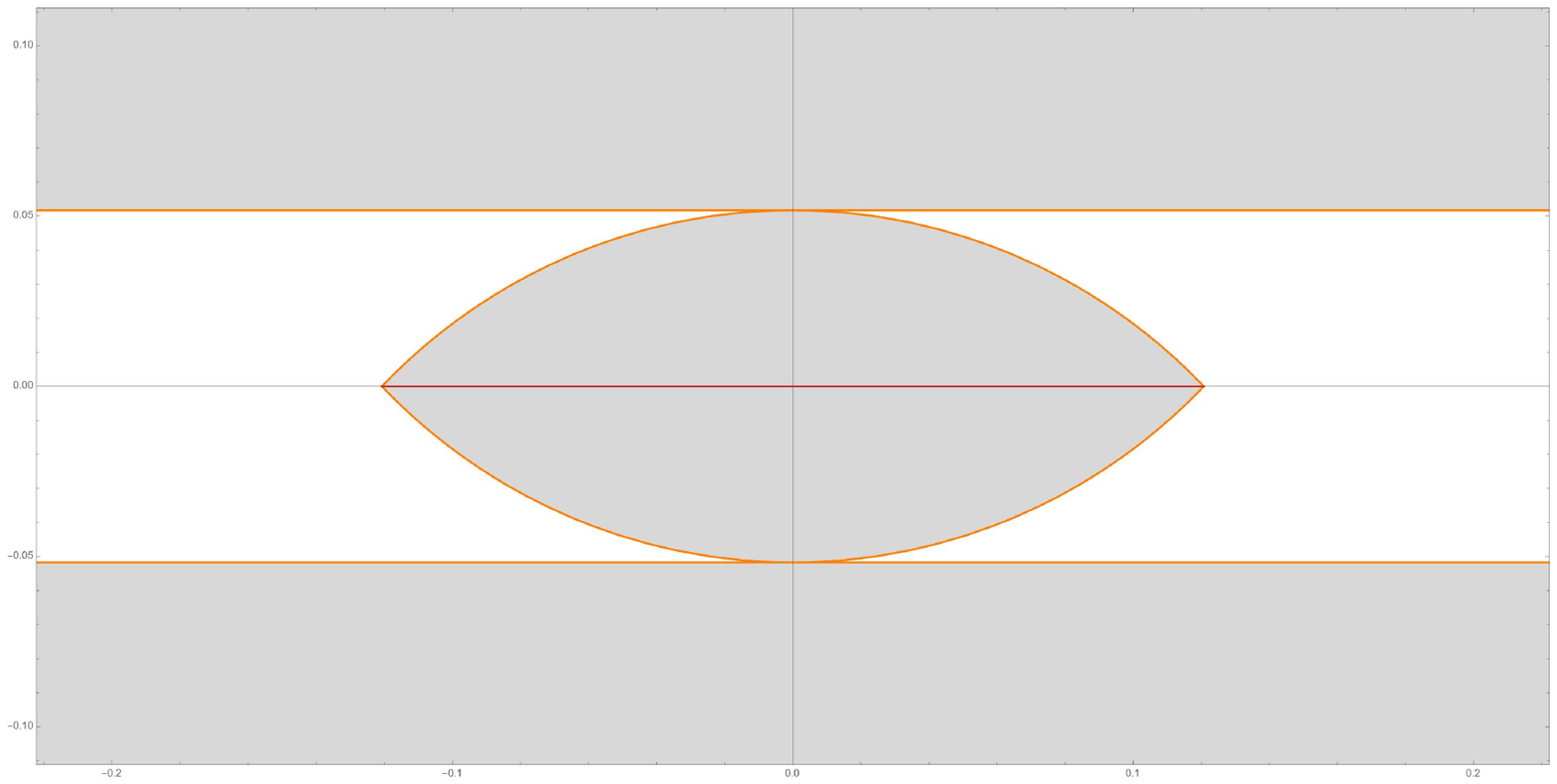}
    \put (95,50) {$w$}
    \put (94,53) {\line(0,-1){5}}
    \put (94,48) {\line(1,0){5}}
    \put (12,24) {$-B$}
    \put (82,24) {$B$}
    \put (52,44) {$iA$}
    \put (49,10) {$-iA$}
    \put (71,35) {\rotatebox{45}{$\longleftrightarrow$}}
    \put (71,19) {\rotatebox{-45}{$\longleftrightarrow$}}
    \put (25,40) {\rotatebox{-45}{$\longleftrightarrow$}}
    \put (25,14) {\rotatebox{45}{$\longleftrightarrow$}}
    \end{overpic}
    \caption{\small 
    The shaded region shows the image of the $z$-plane with the defects at $\text{Im}(z)=\pm \tau_0$ under the map $w$ in \eqref{unifpl}. The image of the real-$z$ axis is the red line segment. The top and bottom of the defects map to the orange lines, which are thus glued together (non-smoothly) as indicated by the arrows.
    When $\tau_0=\frac{\pi}{4\lambda}$ which corresponds to $A=B$, the shaded region in the middle is a circle of radius $B$ centered at the origin. In this case, in the dual geometry, the thin shell passes through the Euclidean horizon. In the figure drawn above, $B>A$ in which case the shell is outside the horizon in the dual geometry.
%
    \label{fig:football}
    }
\end{figure}

The image of the $z$-plane with the defects under this map is shown in figure \ref{fig:football}.
Now consider a rectangular loop described by the four points at which it intersects the two lines of operator insertions, $x_1\pm i\tau_0$ and $x_2\pm i\tau_0$. We can track how the $w$ coordinate changes as we travel around the loop,
\begin{multline}
    z-x_1+i\tau_0+iA\to B\tanh(\alpha(z-x_1))\to z-x_1+i\tau_0-iA \\\to x_2-x_1+B\tanh(\alpha(z-x_2))\to z-x_1+i\tau_0-iA
\end{multline}
Since we ended up with the same $w$ coordinate as we traversed around the rectangular loop, we have thus verified that the monodromy of the differential equation \eqref{Zamodiff} under \eqref{unifpl} around rectangular loops is trivial.

\subsection{Uniformization with shifted planar defects}

\begin{figure}[t]
    \centering
    \begin{overpic}[scale=0.4,grid=false, tics=20, trim=100 190 150 110, clip]{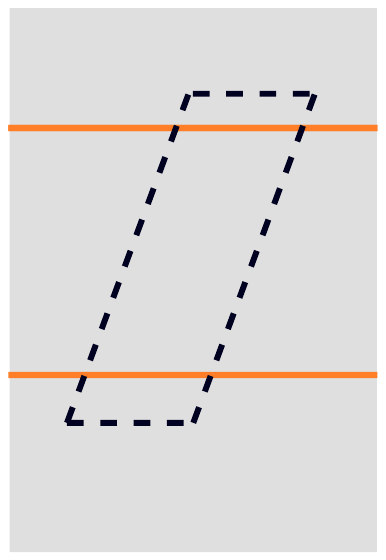}
    \put (20, 34) {$y=\tau_0$}
    \put (20, 15) {$y=-\tau_0$}
    \put (59, 25) {$\mathds{1}$}
    \end{overpic}
    \caption{\small A tilted rectangle intersecting the operator lines $y=\pm \tau_0$ at the four points $z=x_1\pm i\sigma, x_2\pm i\sigma$ (where $\sigma=\tau_0-i\theta$) around which we impose trivial monodromy in the computation of the chiral identity block.}
    \label{fig:planar_twist_monodromy}
\end{figure}

In the previous section, we solved the uniformisation problem associated with the continuum identity block evaluated at a real modulus in the planar limit. Now, we describe and solve the uniformization problem associated with the block evaluated at a complex modulus. This gives a contribution to the correlator $\langle D_{\Sigma}^\dagger e^{2i\theta J}D_\Sigma \rangle$. To this end, we introduce a relative twist of $2\theta$ between the two lines of operator insertions and define a complex modulus $\sigma=\tau_0-i\theta$. The stress tensor takes the same general form,
\begin{equation}
   T(z) =-\alpha^2 \Theta (-\tau_0<y<\tau_0)+\lambda \delta (y-\tau_0)+\lambda \delta (y+\tau_0) \ .
\end{equation}
However, as we shall show, the accessory parameter $\alpha$ will now turn out to be complex. We parametrise the solution to the Schwarzian equation in a way that generalises \eqref{unifpl},
\begin{align} \label{unifplrot}
w=\begin{cases}
    z-i\sigma+iA  \qquad & y>\tau_0 \\
    B\tanh(z\alpha) \qquad & |y|<\tau_0\\
    z+i\sigma-iA \qquad  & y<-\tau_0
\end{cases}
\end{align}
with $A$ and $B$ determined by imposing continuity of the map and of its first derivative across the marked points $z=\pm i\sigma$ on the two operator lines giving
\begin{align}
A= \frac{\sin(2\alpha \sigma)}{2\alpha} \ , \qquad
B= \frac{\cos^2(\alpha \sigma)}{\alpha} \ .
\end{align}
Requiring that the jump in the second derivative of the $w$ map across the marked points gives the coefficient of the delta function in the stress tensor gives
\begin{equation} \label{junrotpl}
   \lambda=\alpha \tan(\alpha \sigma) \ .
\end{equation}
Since $\lambda$ is taken to be a real parameter, $\alpha$ is complex. Using this relation we have
\begin{align}
A=  \frac{\lambda}{\alpha^2+\lambda^2} , \qquad
 B=\frac{\alpha}{\alpha^2+\lambda^2} \ . 
\end{align}
\begin{figure}[t]
    \centering
    \begin{overpic}[scale=0.35,grid=false, tics=20, trim=70 110 70 110, clip]{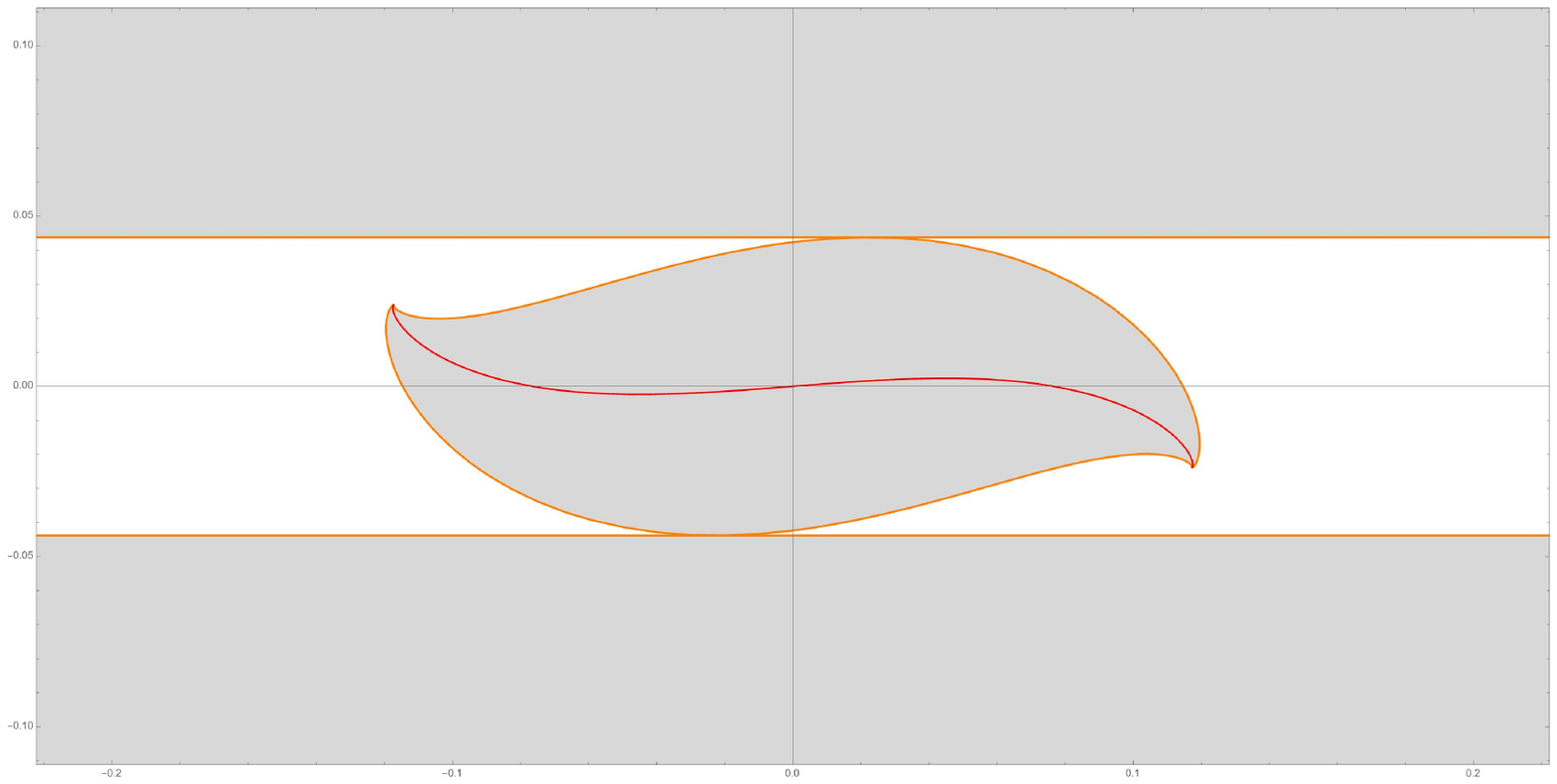}
    \put (95,50) {$w$}
    \put (94,53) {\line(0,-1){5}}
    \put (94,48) {\line(1,0){5}}
        \put (13,33) {$-B$}
        \put (81,18) {$B$}
        \put (58,40) {$iA$}
        \put (38,11) {$-iA$}
    \end{overpic}
    \caption{\small  The image of the $z$-plane with the defects at $\text{Im}(z)=\pm \tau_0$ and a relative shift of $2\theta$, under the map \eqref{unifplrot}. The map $w(z)$ is chosen to be analytic at the basepoints of the defects, $z =\pm( i\tau_0 + \theta)$, which map to $w = \pm i A$. The real-$z$ axis maps to the red curve. Orange lines are the defects and are identified as in figure \ref{fig:football}.
    This can be viewed as the Bers simultaneous uniformization of the upper and lower half planes, viewed as two surfaces with boundaries that have different complex structure moduli.
%
%
    \label{fig:rotated_football}}
\end{figure}
The image of the $z$-plane with the shifted defects under the map in \eqref{unifplrot} is shown in figure \ref{fig:rotated_football}. Now, we can check that the monodromy of the differential equation \eqref{Zamodiff} under \eqref{unifplrot} taken around a tilted rectangular loop as shown in figure \ref{fig:planar_twist_monodromy} described by the four points $x_1\pm i\sigma$ and $x_2\pm i\sigma$ at which it intersects the two lines of operator insertions is trivial,
\begin{multline}
    z-x_1+i\sigma+iA\to B\tanh(\alpha(z-x_1))\to z-x_1+i\sigma-iA \\\to x_2-x_1+B\tanh(\alpha(z-x_2))\to z-x_1+i\sigma-iA
\end{multline}
Thus we have solved the monodromy problem for the chiral identity block evaluated at a complex modulus in the planar limit by uniformising the plane with shifted defects. The Liouville field associated with this uniformisation of the upper half $z$-plane can be defined as 
\begin{equation}
  e^{\Phi} = -\frac{4 w'(z) w'(\bz)}{(w(z)-w(\bz))^2} \ .
\end{equation}
This is complex-valued and is an analytic continuation of the Liouville field in \eqref{Liouvunif}.
Relating to the gravitational description in terms of a black hole sourced by a rotating shell discussed in section \ref{secrotbh}, the parameter $\alpha$ is related to the radii $r_\pm$ by the relation $2\alpha=r_++i|r_-|$ and as usual $\lambda$ is related to the mass of the shell by $2\lambda=m$. Written in terms of $\alpha$, the condition for the shell's trajectory to go behind the horizon of the rotating black hole is Re$(\alpha \sigma)>\frac{\pi}{4}$. To see this, observe that the jump condition \eqref{junrotpl} translates this condition to $m>\sqrt{r_+^2-r_-^2}$. Written in terms of the mass and the spin (imaginary) of the black hole, it gives $m^2>\sqrt{M^2-J^2}$ which matches with \eqref{horcross} in the planar limit where $M\gg 1$.

\subsection{Semiclassical continuum identity block on the sphere}

Away from the planar limit, the analysis is similar. Working in the cylinder frame ($z\sim z+2\pi)$, the major difference from the planar case is that we need to incorporate the Casimir energy density on the circle into the stress tensor,
\begin{equation} \label{stresscyl}
    T(z)=-\alpha^2 \Theta (-\tau_0<y<\tau_0)+\frac{1}{4}\Theta(|y|>\tau_0)+\lambda \delta (y-\tau_0)+\lambda \delta (y+\tau_0)
\end{equation}
with the two circles of operator insertions at $y=\pm \tau_0$.
It is this setup, in the short-distance limit $\tau_0 \to 0$, that was studied in \cite{Anous:2016kss}. The method used there to solve for trivial monodromy (i.e., find $\psi_1$ and $\psi_2$, patch together solutions across the defects, and impose the monodromy condition) can also be applied at finite $\tau_0$, but it is rather cumbersome. It is simpler to solve the Schwarzian equation directly, as we did for the planar defect above.

Up to a PSL$(2,\mathbb{R})$ redundancy, the reflection symmetric $(w(\overline{z})=\overline{w(z)})$ solution to the Schwarzian equation $\{w,z\}=2T(z)$ can be written as
\begin{equation}
    w=\begin{cases}
        \tan(\frac{z-i\tau_0+iA}{2}) \quad \quad & y>\tau_0\\
        B\tanh(z\alpha) \quad \quad & |y|<\tau_0 \\
        \tan(\frac{z+i\tau_0-iA}{2}) \quad \quad & y<-\tau_0
    \end{cases}
\end{equation}
$A$ and $B$ are determined by demanding continuity of the map and continuity of its first derivative across either marked points $z=\pm i\tau_0$ on the two circles of operator insertions,
\begin{equation}
    B\tan(\alpha \tau_0)=-2\alpha \csc(2\alpha \tau_0)\pm \sqrt{1+4\alpha^2 \csc^2(2\alpha \tau_0)}
\end{equation}
To show that the monodromy around rectangular closed loops is trivial, it suffices to check that the monodromy around the half of this loop in the upper-half cylinder (closed in the upper half cylinder to form a loop itself) is in PSL$(2,\mathbb{R})$ or equivalently the corresponding Liouville field should be single-valued. As shown below, the Liouville field constructed from the uniformising map on the upper-half cylinder is single-valued,
\begin{equation}
   \Phi = \begin{cases}
        -2\log \sinh(y-\tau_0+A) \quad \quad & y>\tau_0 \\
        -2\log(\frac{1}{2\alpha}\sin(2\alpha y)) \quad \quad & 0<y<\tau_0
    \end{cases}
\end{equation}
Matching this solution with the corresponding gravitational solution in section \ref{secBTZtoAdS} or equivalently to the Liouville solution in \eqref{Lioudisk}, we see that the relation between the parameters used in the CFT and gravity or Liouville calculations is $2\alpha=r_H$ and $2\lambda=m$. The former relation can be verified by matching the energies of the solution.
We can verify the latter relation $2\lambda=m$ by matching the jump condition with the Israel junction condition as follows,
\begin{equation}
    2i\lambda=\frac{\Delta w''}{w'}=\tan(\frac{iA}{2})+2\alpha\tanh(i\alpha\tau_0)=i\tan(\alpha \tau_0)(B+2\alpha)
\end{equation}
which corresponds to
\begin{equation} \label{monosph}
    2\lambda=\pm\sqrt{1+4\alpha^2 \csc^2(2\alpha \tau_0)}-2\alpha \cot(2\alpha \tau_0)
\end{equation}
When written in terms of quantities defined in the gravity calculation namely the turning radius of the shell $r_0$ and the horizon radius $r_H$,
\begin{equation}
    B=r_H\left (\frac{\pm \sqrt{1+r_0^2}-r_0}{r_0-r_H\cot(r_H\tau_0)}\right )
\end{equation}
As we will show below, the requirement that $\lambda>0$ will remove two of the four solutions above,
\begin{equation}
    B=r_H\left (\frac{ \sqrt{1+r_0^2}-r_0}{r_0- r_H\cot(r_H\tau_0)}\right )
\end{equation}
Substituting the expression for $B$ derived above, we see that
\begin{equation}
    2\lambda = \pm \sqrt{1+r_0^2}-r_H\cot(r_H\tau_0)
\end{equation}
If we impose the restriction that $\lambda>0$, then 
\begin{equation}
    2\lambda=\sqrt{1+r_0^2}-r_H\cot(r_H\tau_0)=m
\end{equation}
The match between $\lambda$ and $m$ can more generally be expected because $2i\lambda =\frac{\Delta w''}{w'}$ and using the relation between the Liouville field and the uniformising coordinate, this is same as the jump in the normal derivative of the Liouville field across the interface. Now, we evaluate the semiclassical idenity block $f_0^{\CL}(\tau_0;m)$ defined by $\mathcal{F}^{\CL}_{\id}=e^{-\frac{c}{6}f_0^{\CL}}$ by integrating the Ward identity derived in Appendix \ref{secwardsph} which we rewrite here,\footnote{Using the Ward identity in \eqref{wardsph}, we see that the dual channel conformal block defined by fixing the monodromy around a cycle homologous to the line defect which corresponds to fixing $E$ in \eqref{wardsph}, in the semiclassical limit is given by the scaling block.}
\begin{equation} \label{wardsph}
    \frac{c}{6}\partial_{\tau_0}f_0^{\CL}=E+\frac{c}{12}
\end{equation}
Noting that $E=\frac{c}{12}r_H^2$, we can turn the above Ward identity into a differential equation for $r_H$,
\begin{equation}
    \partial_{r_H}f_0^{\CL}=\frac{r_H^2+1}{2\Dot{r}_H}
\end{equation}
where the $\Dot{r}_H$ denotes derivative of $r_H$ with respect to $\tau_0$. $\Dot{r}_H$ can be computed using the relation,
\begin{equation}
    1+\frac{r_H^2}{\sin^2(r_H\tau_0)}=\left(\frac{m^2+r_H^2+1}{2m}\right)^2
\end{equation}
which is just a rewriting of the trivial monodromy condition \eqref{monosph}, using which we see that

\vspace{0.3cm}
\begin{multline}
    \frac{r_H^2}{\Dot{r}_H}=\frac{2mr_H}{1+r_H^2-m^2}-\cot^{-1}\left(\frac{1+r_H^2-m^2}{2mr_H}\right)\\-\left(\frac{m^2+r_H^2+1}{1+r_H^2-m^2}\right)\frac{4mr_H^3}{((2mr_H)^2+(1+r_H^2-m^2)^2)}
\end{multline}
This expression as written above holds for both heavy and light shells. 
On integrating, we thus have an expression for the semiclassical block,
\begin{equation} \label{blocksphmono}
  f_0^{\CL}(\tau_0;m)=-\frac{\tau_0 r_H^2}{2}+\frac{\tau_0}{2} -\sinh^{-1}(\frac{1}{r_0})+m-m\log(\frac{ r_0}{2})
\end{equation}
where $r_0$ and $r_H$ can be expressed in terms of $\tau_0$ and $m$ by solving the following equations,
\begin{equation}
   \sqrt{1+r_0^2}-r_H\cot(r_H\tau_0)=m \quad  \quad 1+\frac{r_H^2}{\sin^2(r_H\tau_0)}=\left(\frac{m^2+r_H^2+1}{2m}\right)^2
\end{equation}
The integration constant in the above expression for the block was fixed using the behaviour of the block in the OPE limit described in footnote \ref{opelimsph}. The expression for the block \eqref{blocksphmono} thus calculated using the monodromy method agrees with the corresponding expression for the block in \eqref{finalfCL} calculated by taking the semiclassical limit of a Liouville correlator. In \eqref{blocksphmono}, we have an expression for the block evaluated at a real modulus. We can evaluate the block at a general complex modulus by solving the uniformization problem analogous to the one discussed for rotating planar shells and integrating the Ward identity derived in Appendix \ref{secwardsph}. In practice, this just corresponds to analytically continuing the expression in \eqref{blocksphmono} to complex modulus. It is convenient to express the block in terms of the saddlepoint momentum (related to the saddlepoint weight by $h=\frac{c}{24}(1+\gamma^2)$) when expanded using the dual channel (which is complex when evaluated at a complex modulus $\sigma$),
\begin{multline} \label{sphblock}
    f_0^{\CL}(\sigma;m)=-\frac{1}{2}\bigg [(\gamma-\frac{1}{\gamma})\cot^{-1}\left(\frac{1+\gamma^2-m^2}{2m\gamma}\right)-2m-2m\log(4m)\\+(m+1)\log((m+1)^2+\gamma^2)+(m-1)\log((m-1)^2+\gamma^2) \bigg ]
\end{multline}
where $\gamma(\sigma,m)$ is determined by solving the transcendental equation,
\begin{equation} \label{cftjuncrot}
    1+\frac{\gamma^2}{\sin^2(\gamma\sigma)}=\left (\frac{m^2+\gamma^2+1}{2m} \right )^2
\end{equation} 
The block \eqref{sphblock} is related to the Liouville action on the disk \eqref{SLdisk} analytically continued to complex moduli by
\begin{equation}
  f_0^{\CL}=S_L+m\log 2
\end{equation}

\subsection{Semiclassical continuum identity block on the torus}

In this section, we setup the uniformisation problem for the continuum identity block on the torus, in the channel shown in figure \ref{fig:identity-torus}, and compute the semiclassical block by solving the uniformisation problem. The stress tensor on the torus described by the identifications $z\sim z+2\pi \sim z+i\beta$ in the continuum limit is given by
\begin{equation}
    T(z)=-\alpha^2 \Theta(|y|<\tau_0)-(\alpha')^2 \Theta(\tau_0<|y|<\frac{\beta}{2})+\lambda \delta(y-\tau_0)+\lambda \delta(y+\tau_0) \ .
\end{equation}
We work with the fundamental domain: $\text{Re}z\in(0,2\pi)$, $\text{Im}z\in(-\frac{\beta}{2},\frac{\beta}{2})$.
The reflection symmetric $(w(\overline{z})=\overline{w(z)})$ solution to the Schwarzian equation up to a PSL$(2,\mathbb{R}$) redundancy is given below,
\begin{equation}
    w=\begin{cases}
       \tanh(\alpha'(\frac{i\beta}{2}-z)) \quad \quad & \tau_0<y<\frac{\beta}{2}\\
       B\tanh(z\alpha) \quad \quad & |y|<\tau_0 \\
       -\tanh(\alpha'(\frac{i\beta}{2}+z)) \quad \quad & -\frac{\beta}{2}<y<-\tau_0
    \end{cases}
\end{equation}
Without loss of generality, we assume that $\alpha>\alpha'$.
The above solution and its derivative are continuous across the marked points $\pm i\tau_0$, provided
\begin{equation}
    \alpha' B \tan(\alpha \tau_0)=-\alpha\csc(2\alpha \tau_0)\pm \sqrt{\alpha^2\csc^2(2\alpha\tau_0)-(\alpha')^2} \ .
\end{equation}
This corresponds to the following solution to the Liouville equation on the upper-half torus which is topologically a cylinder,
\begin{equation}
   \Phi= \begin{cases}
      -2\log \left(\frac{1}{2\alpha}\sin(2\alpha y)\right) \quad \quad & y \in (0,\tau_0)\\
      -2\log \left(\frac{1}{2\alpha'}\sin(2\alpha'(\frac{\beta}{2}-y)\right)\quad \quad & y \in (\tau_0,\frac{\beta}{2})
    \end{cases}
\end{equation}
which matches with the Liouville solution written down in the gravity calculation in section \ref{secBTZtoBTZ} or equivalently to the Liouville solution in \eqref{Liouann} with $2\alpha = r_H$ and $2\alpha'=r_H'$. The jump condition $2i\lambda=\frac{\Delta w''}{w'}$ reads
\begin{equation}
    2\lambda= 2\alpha' \tan(\alpha'(\frac{\beta}{2}-\tau_0))+2\alpha \tan(\alpha \tau_0)=\tan(\alpha \tau_0)(2\alpha' B+2\alpha) \ .
\end{equation}
It is convenenient to express $B$ in terms of bulk quantities namely the horizon radii $r_H, r_H'$ and the turning radius $r_0$,
\begin{equation}
    B=\frac{r_H}{r_H'}\left [\frac{\pm \sqrt{r_0^2-(r_H')^2}-r_0}{r_0\pm \sqrt{r_0^2-r_H^2}}\right] \ .
\end{equation}
Note the following relation
\begin{equation}
    \tan(\alpha\tau_0)=\csc(2\alpha\tau_0)-\cot(2\alpha\tau_0)=\frac{r_0- r_H\cot(r_H\tau_0)}{r_H}
\end{equation}
using which we can rewrite the jump condition as
\begin{equation}
    2\lambda=\pm \sqrt{r_0^2-(r_H')^2}-r_H\cot(r_H\tau_0) \ .
\end{equation}
Requiring $\lambda>0$ restricts the above equation to
\begin{equation}
    2\lambda=\sqrt{r_0^2-(r_H')^2}- r_H\cot(r_H\tau_0) \ .
\end{equation}
This matches with the gravity calculation provided we identify $2\lambda=m$. In the following, we integrate the Ward identity for the continuum identity block on the torus $\mathcal{F}_{\id}^D=e^{-\frac{c}{6}f_0^D}$ derived in Appendix \ref{wardtorus} which we rewrite below,
\begin{equation} \label{Wardtorus}
    \frac{c}{6}\partial_{\tau_0}f_0^{\CL}=E-E'
\end{equation}
Note that we are taking the derivative holding $\beta$ fixed. The saddlepoint energies in the above Ward identity are related to the parameters $r_H$ and $r_H'$ by $E=\frac{c}{12}r_H^2$ and $E'=\frac{c}{12}(r'_H)^2$.
The Ward identity \eqref{Wardtorus} and the symmetry condition that the block is invariant under $\tau_0 \xleftrightarrow{} \frac{\beta}{2}-\tau_0$ determines the block with the normalisation fixed using the OPE limit as described in footnote \ref{opelimtorus}. The following expression,
\begin{equation}  \label{torusblockmono}
    f_0^{\CL}(\tau_0,\beta;m)=-\tau_0\frac{r_H^2}{2}-(\frac{\beta}{2}-\tau_0)\frac{(r_H')^2}{2}+m-m\log(\frac{ r_0}{2}) 
\end{equation}
subject to the relations 
\begin{equation} \label{Wardreltorus}
    \begin{split}
        & r_H=r_0\sin(r_H\tau_0) \quad \quad r_H'=r_0\sin(r_H'(\frac{\beta}{2}-\tau_0)) \\
        & r_0^2=\frac{1}{2}(r_H^2+(r_H')^2)+\frac{m^2}{4}+\frac{(r_H^2-(r_H')^2)^2}{4m^2}
    \end{split}
\end{equation}
satsifies these conditions. The expression for the block \eqref{torusblockmono} agrees with \eqref{blocktorus} obtained by taking the semiclassical limit of the corresponding Liouville correlator computed in \eqref{SLann}. To evaluate the block at complex moduli, it is convenient to express it in terms of the saddlepoint momenta when expanded using the dual channel ($h=\frac{c}{24}(1+\gamma^2)$, $h'=\frac{c}{24}(1+\gamma'^2)$),
\begin{multline} \label{torblock}
    f_0^{\CL}(\sigma,\sigma';m)=-\frac{1}{2}\bigg[\gamma\cot^{-1}\left(\frac{\gamma^2-\gamma'^2-m^2}{2m\gamma}\right )+\gamma'\cot^{-1}\left(\frac{\gamma'^2-\gamma^2-m^2}{2m\gamma'}\right )\\+m\log \left(\frac{1}{2}(\gamma^2+\gamma'^2)+\frac{m^2}{4}+\frac{(\gamma^2-\gamma'^2)^2}{4m^2} \right)-2m(1+\log2)\bigg]
\end{multline}
where the saddlepoint momenta treated as functions of complex moduli are determined by solving the following set of transcendental equations,
\begin{equation}
    \begin{split}
        & \gamma=r_0\sin(\gamma\sigma) \quad \quad \gamma'=r_0\sin(\gamma'\sigma') \\
        & r_0^2=\frac{1}{2}(\gamma^2+\gamma'^2)+\frac{m^2}{4}+\frac{(\gamma^2-\gamma'^2)^2}{4m^2}
    \end{split}
\end{equation}
Here, the moduli $\sigma$ and $\sigma'$ are determined by complexifying $\tau_0 \to \sigma$ and $\frac{\beta}{2}-\tau_0 \to \sigma'$. The block \eqref{torblock} is related to the Liouville action on the annulus \eqref{SLann} analytically continued to complex moduli by
\begin{equation}
  f_0^{\CL}=S_L+m\log 2 \ .
\end{equation}

\section{Black holes} \label{secblackholes}

In this section, we describe the Euclidean black hole geometries sourced by thin shells of dust particles either non-rotating or rotating by solving the Israel junction conditions for gluing approporiate 3d geometries. We compute the on-shell gravitational action for each case and observe that it matches with the semiclassical identity block for the line defect in the dual CFT calculated in sections \ref{secbootstrap} and \ref{secblocks}, which in turn also matches with the corresponding Liouville action for the Liouville line defect computed in section \ref{secLiouville}. Thus, the statistics of thin-shell black holes is governed by the dynamics of appropriate line defects in large-$c$ holographic CFTs and in Liouville CFT.

\subsection{Non-rotating black hole: Gluing BTZ to empty AdS} \label{secBTZtoAdS}

We compute the on-shell action of a Euclidean black hole geometry sourced by the backreaction of a thin shell of dust particles. This geometry was constructed in \cite{Keranen:2015fqa,Chandra:2022fwi} by solving the Israel junction conditions across the shell which glues a portion of empty AdS to a portion of the BTZ geometry. The gravitational partition function on this geometry is a contribution to the two-point function $\langle \CL^\dagger(\tau_0) \CL(-\tau_0)\rangle$ of the line operator in the dual CFT on a cylinder. In \cite{Chandra:2022fwi,Bah:2022uyz}, the authors used Schwarzschild coordinates to do the calculations. For the purposes of the present paper, it is convenient to work in a hyperbolic slicing of AdS, so that the relation to the Liouville action written down in \eqref{LiouvdiskZZ} can be made manifest. In fact, as we shall show below, the Israel junction conditions for the present setup reduce to the continuity and junction condition for the Liouville field written down in \eqref{ydisc} by working with the hyperbolic slicing. We start with the ansatz for the metric,
\begin{equation}
    ds^2=d\rho^2+\cosh^2(\rho)e^\Phi |dz|^2
\end{equation}
with $\rho \in (-\infty, \infty)$ and the constant $\rho$ slices are semi-infinite cylinders with $z\sim z+2\pi$ and $\text{Im}(z)\equiv y\in(0,\infty)$. In these coordinates, we parametrise the shell to be along a trajectory $y=\tau_0$ and shall show below that such a trajectory solves the Israel junction conditions. Einstein's equations away from the shell give the Liouville equation $\partial \overline{\partial}\Phi=\frac{e^\Phi}{2}$. The induced metric on the shell is given by
\begin{equation}
    ds^2_{\text{shell}}=d\rho^2+\cosh^2(\rho)r_0^2d\theta^2
\end{equation}
where $r_0$ is the turning radius for the shell in Schwarzschild coordinates and is related to the Liouville field $\Phi_0$ at the interface by $r_0=e^{\frac{\Phi_0}{2}}$.
Continuity of the metric across the shell implies continuity of the Liouville field. The $\rho \rho$ component of the Israel junction condition $\Delta K_{ij}-\Delta K h_{ij}=\sigma(\rho)U_i U_j$ where the velocity field is $U=\partial_\rho$ with $\sigma(\rho)=\frac{m}{r_0\cosh(\rho)}$ gives the condition $\sqrt{1+r_0^2}\pm \sqrt{r_0^2-r_H^2}=m$. Thus, the Israel junction condition reduces to the gluing condition for the Liouville field,
\begin{equation} \label{juncAdStoBTZ}
   \sqrt{1+r_0^2}\pm \sqrt{r_0^2-r_H^2}=m  \xleftrightarrow{} \partial_y \Phi |_+-\partial_y \Phi |_-=-2m
\end{equation}
where the $+$ sign holds for shells which go behind the horizon on the time-symmetric slice which we refer to as `heavy' shells and the $-$ sign holds for shells which are outisde the horizon on the time-symmetric slice which we refer to as `light' shells.  (Light shells were studied in \cite{Keranen:2015fqa,Bah:2022uyz} and heavy shells in \cite{Chandra:2022fwi}.) In the above expression, $r_H$ is the radius of the horizon of the BTZ black hole. Thus, the solution to the Liouville equation consistent with the junction condition is given by
\begin{align}
\Phi =\begin{cases}
-2\log \sinh(y-\tau_0+A) & y>\tau_0\\
-2\log \left(\frac{1}{r_H} \sin(r_H y) \right)& 0< y < \tau_0
\end{cases}
\end{align}
where $A = \sinh^{-1}(\frac{1}{r_0})$. This solution is the same as the one derived in \eqref{Lioudisk}.
Having described the solution, we can now compute the on-shell action,
\begin{equation}
    I=-\frac{1}{16\pi G}\int \sqrt{g}(R+2)-\frac{1}{8\pi G}\int \sqrt{h}(K-1)+m\int d\ell
\end{equation}
where the GHY term is evaluated on the boundaries of the region described below. In the hyperbolic slicing, the AdS boundary is the union of the cylinders at $\rho_c=\pm (-\log (2\Lambda)-\frac{\Phi}{2})$ so that the metric induced on the boundary is the flat metric on the cylinder given by $ds^2_{\text{bdry}}=\Lambda^2|dz|^2$ with the cutoff parameter $\Lambda \to \infty$.  Solving Einstein's equations in the vicinity of the shell, we observe that the Einstein-Hilbert term evaluated in the vicinity of the shell cancels against the shell propagator term. So, in the rest of the analysis, we stick to the region $(\Gamma_+ \cup \Gamma_-)\times \mathbb{R}$ where $\Gamma_+$ is the portion of the cylinder above the interface, $y \in (\tau_0,T)$, and $\Gamma_-$ is the portion of the cylinder below the interface, $y \in (\epsilon, \tau_0)$. Here, the cutoff parameters are such that $T\to \infty$ and $\epsilon \to 0$. 
The Einstein-Hilbert term evaluated in this region gives,
\begin{equation}
    -\frac{1}{16\pi G}\int \sqrt{g}(R+2)=\frac{1}{4\pi G}\int_{\Gamma_+ \cup \Gamma_-} d^2z(\Lambda^2+e^\Phi\log(2\Lambda)-\frac{\Phi}{2}e^\Phi)
\end{equation}
The GHY term evaluated on the AdS boundary gives
\begin{equation}
    -\frac{1}{8\pi G}\int \sqrt{h}(K-1)=-\frac{1}{8\pi G}\int_{\Gamma_+\cup \Gamma_-} d^2z(2\Lambda^2+e^\Phi+\partial \Phi \overline{\partial}\Phi)-\frac{2}{8\pi G}\int \sqrt{\gamma}(\pi - \Theta)
\end{equation}
where the last term is the sum of Hayward terms evaluated at the two corners where the shell intersects the AdS boundary. Note that the Hayward term was not included in \cite{Sasieta:2022ksu,Bah:2022uyz}, but this is not an issue because it only affects the normalization of the shell operator. In our case we must include it in order to find a precise match to the identity conformal block and the Liouville line defect --- our normalization has already been fixed in section \ref{secblocks}.

We express the angle between the AdS boundaries in the empty AdS and BTZ regions as $\Theta=\Theta_1+\Theta_2$ where $\Theta_1$ is the angle between the shell and boundary of empty AdS and $\Theta_2$ is the angle between the shell and boundary of BTZ. These angles can be evaluated to give
\begin{equation}
         \Theta_1 = \frac{\pi}{2}+\sin^{-1}\left(\sqrt{\frac{1+r_0^2}{1+\Lambda^2}}\right )  \quad \quad  \Theta_2 = \frac{\pi}{2}\pm \sin^{-1}\left(\sqrt{\frac{r_0^2-r_H^2}{\Lambda^2-r_H^2}}\right )
\end{equation}
where the $+$ sign is for heavy shells and the $-$ sign is for light shells. Thus, the Hayward term evaluates to $I_{\text{Hay}}=\frac{m}{2G}$. In addition, since we are introducing a large cutoff at $y=T$, we need to add a GHY term at the cutoff surface which is topologically a cylinder $\{y=T\}\times \mathbb{R}$. This term evaluates to give $I_{\text{cut}}=-\frac{1}{8\pi G}\int_{y=T}\sqrt{h}K=\frac{1}{2G}\log(2\Lambda)$.
Adding the terms, we get
\begin{equation}
    I=\frac{1}{2\pi G}\int_{\Gamma_+ \cup \Gamma_-} d^2z\left(\frac{1}{4}(\partial\Phi\overline{\partial}\Phi+e^\Phi)-\frac{1}{2}\overline{\partial}(\Phi\partial\Phi)-\partial\overline{\partial}\Phi(1-\log(2\Lambda))\right)+\frac{m}{2G}+\frac{\log(2\Lambda)}{2G}.
\end{equation}
 Thus, the total on-shell gravitational action can be expressed as a sum of Liouville actions evaluated on $\Gamma_+$ and $\Gamma_-$,
\begin{multline}
    I=\frac{c}{3}\bigg [\frac{1}{4\pi}\int_{\Gamma_+}d^2z( \partial \Phi \overline{\partial}\Phi+e^{\Phi})-\frac{i}{4\pi}\int_{\partial \Gamma_+} dz \Phi \partial \Phi+\frac{1}{4\pi}\int_{\Gamma_-}d^2z (\partial \Phi \overline{\partial}\Phi+e^{\Phi})\\-\frac{i}{4\pi}\int_{\partial \Gamma_-} dz \Phi \partial \Phi+1+m\log(2\Lambda) \bigg]
\end{multline}
Substituting the solution to the Liouville equation, we get
\begin{equation}
    I_{\text{ren}}=\frac{c}{3}\left [ m  -\frac{1}{2}(r_H^2-1)\tau_0 - \sinh^{-1}\frac{1}{r_0} - m \log(\frac{ r_0}{2}) \right ]
\end{equation}
where we have defined the renormalised gravitational action,
\begin{equation} \label{rengrav}
   I_{\text{ren}}=I-\frac{c}{3}m\log(\Lambda)-\frac{c T}{6}-\frac{c}{3\epsilon}-\frac{c\log \epsilon}{3\epsilon}-\frac{c}{3}\log(2)
\end{equation}
The normalisation of the action has been chosen so that in the limit $\tau_0 \to 0$ at fixed $m$ which corresponds to $r_0=\frac{1}{\tau_0}\to \infty$, the action should reproduce the OPE limit of the continuum identity block, so $I_{\text{ren}} \to \frac{c}{3}m\log(2\tau_0)$. It is convenient to express the renomalised action in terms of the horizon radus $r_H$ which is related to the energy of the black hole by $E=\frac{r_H^2}{8G}$,
\begin{multline} 
    I_{\text{ren}}=-\frac{c}{6}\bigg [\left(r_H-\frac{1}{r_H}\right )\cot^{-1}\left(\frac{1+r_H^2-m^2}{2m r_H}\right)-2m-2m\log(4m)\\+(m+1)\log((m+1)^2+r_H^2)+(m-1)\log((m-1)^2+r_H^2) \bigg ]
\end{multline}
where $r_H$ can be expressed in terms of the landing parameter $\tau_0$ and mass $m$ using the Israel junction conditions which is equivalent to the following transcendental equation,
\begin{equation}
  1+\frac{r_H^2}{\sin^2(r_H\tau_0)}=\left ( \frac{m^2+r_H^2+1}{2m}\right )^2 \ .
\end{equation}
The renormalised gravitational action matches with two copies of the semiclassical block computed in \eqref{finalfCL} and is related to two copies of the Liouville action $S_L$ computed in \eqref{SLdisk} by
\begin{equation}
   I_{\text{ren}}=\frac{c}{6}(2f_0^{\CL}(\tau_0;m))=\frac{c}{6}(2S_L)+\frac{c}{3}m \log 2 \ .
\end{equation}
Thus we have shown that to leading order in the semiclassical limit, 
\begin{align}\label{fourOnSphere}
e^{-I_{\rm ren}}  \approx \langle D^\dagger(\tau_0) D(-\tau_0)\rangle_{\rm cyl} \approx {\cal N}^2 \left| \langle L(\tau_0) \rangle_{\rm disk} \right|^2 \approx \left| {\cal F}_{\id}^D\right|^2
\end{align}
where ${\cal N}$ the normalization constant derived in \eqref{Ncl}, and ${\cal F}_{\id}^D$ is the identity block for two defects on the sphere in the channel shown in  figure \ref{fig:identity-cylinder}. The first quantity $e^{-I_{\rm ren}}$ is caclulated on the gravity side; the second $\langle D^\dagger D\rangle$ is in the dual CFT; and $\langle L\rangle_{\rm disk}$ is in Liouville.

For heavy local operators, the correspondence between defect solutions in the bulk, correlators in the dual CFT, and solutions of the Liouville equation was first described in \cite{Hartman:2013mia,Faulkner:2013yia}, and the connection to the ZZ boundary condition was made in \cite{Hulik:2016ifr} (see also \cite{Jackson:2014nla,Verlinde:2015qfa,Mertens:2018fds,Raeymaekers:2022sbu}). The equation \eqref{fourOnSphere} is the spherically-symmetric, line-defect version of this correspondence.


\subsection{Shells below the black hole threshold}

We can generalize the above calculation to the case where we glue a portion of empty AdS$_3$ to a portion of the conical AdS$_3$ across a thin shell. Conical AdS$_3$ is the geometry sourced by a massive particle below the black hole threshold, with a conical deficit at the center of the global AdS$_3$:
\begin{align}
ds^2=(r^2+(1-2\eta)^2)d\tau^2+\frac{dr^2}{r^2+(1-2\eta)^2}+r^2d\phi^2 \ . 
\end{align}
The glued geometry is global AdS$_3$ inside the shell, and conical AdS$_3$ outside the shell. This geometry gives a contribution to the correlator $\langle \CL^\dagger(\tau_0) \CL(-\tau_0)\rangle$. Note that the point particle which sources the conical deficit does not appear in the physical portion of the geometry i.e the geometry is smooth everywhere away from the trajectory of the shell. The Liouville solution that solves Einstein's equations and the Israel junction conditions in the hyperbolic slicing for this case is given by
\begin{align} \label{Liouvdef}
\Phi =\begin{cases}
-2\log \sinh(y-\tau_0+A) & y>\tau_0\\
-2\log \left(\frac{1}{(1-2\eta)} \sinh((1-2\eta) y) \right)& 0< y < \tau_0
\end{cases}
\end{align}
where $A = \sinh^{-1}(\frac{1}{r_0})$ and the turning radius $r_0$ of the shell is given by  
\begin{equation}
   \sqrt{1+r_0^2}-\sqrt{(1-2\eta)^2+r_0^2}=m \implies r_0^2=\left (\frac{m^2+1-(1-2\eta)^2}{2m} \right )^2-1
\end{equation}
In the above solution, $\eta \in (0,\frac{1}{2})$ parametrizes the conical deficit angle $4\pi\eta.$ This solution exists for $\eta>\frac{m}{2}$ which is satisfied only if $m<1$ consistent with the observation made in \eqref{imbr}. Substituting the Liouville solution in \eqref{Liouvdef} and renormalising the action according to \eqref{rengrav}, we get the following renormalised on-shell action,
\begin{multline}
  I_{\text{ren}}= \frac{c}{3}\bigg[ m(1+\log(4m))  +\frac{1}{2}((1-2\eta)^2+1)\tau_0 -\frac{1}{2}(m+1)\log\left((m+1)^2-(1-2\eta)^2\right)\\ -\frac{1}{2}(m-1)\log\left((m-1)^2-(1-2\eta)^2\right)\bigg ]
\end{multline}
Clearly, we observe that both the solution and the action can be obtained from the previous calculation by the analytic contiuation $r_H\to i(1-2\eta)$ or equivalently by analytically continuing the energy of the solution to below the black hole threshold. Thus, we have described a solution sourced only by the thin shell of dust whose energy is below the black hole threshold.

\subsection{Non-rotating black hole: Gluing BTZ to BTZ} \label{secBTZtoBTZ}

It is straightforward to generalize the previous calculation to compute the action of the geometry obtained by gluing two BTZ black holes across a thin shell. 
See \cite{Sasieta:2022ksu} for the construction of this geometry by solving the Israel junction conditions in Schwarzschild coordinates. This is a gravitational calculation of the thermal two-point function $\langle \CL^\dagger(\tau_0)\CL(-\tau_0)\rangle_\beta$ of the shell operator. So, the boundary condition is determined by a flat torus $z\sim z+2\pi \sim z+i\beta$ with two defects at $z=\pm i\tau_0$. We work with a fundamental domain of the torus with $\text{Im}z \equiv y \in (-\frac{\beta}{2},\frac{\beta}{2})$. In the hyperbolic slicing, the geometry is an interval times a finite-length hyperbolic cylinder with a single defect. Recall that the Liouville solution on such a cylinder of height $\frac{\beta}{2}$ with the defect at $y=\tau_0$ as derived in \eqref{Liouann} is given by
\begin{equation}
\Phi=
\begin{cases}
  -2\log \left (\frac{1}{r_H}\sin(r_Hy) \right ) & 0<y<\tau_0\\
   -2\log \left (\frac{1}{r_H'}\sin(r_H'(\frac{\beta}{2}-y)) \right ) & \tau_0<y<\frac{\beta}{2}
\end{cases}
\end{equation}
where $r_H$ and $r_H'$ are the horizon radii of the two BTZ black holes. For later convenience, we assume without loss of generality that $r_H>r_H'$.
Following the steps similar to that in the previous section, the gravitational action can be reduced to\footnote{The Einstein-Hilbert term evaluated in $(\Gamma_+ \cup \Gamma_-)\times \mathbb{R}$ and the GHY term evaluated on the AdS boundary including the Hayward term at the two corners at the intersection of the shell trajectory with the AdS boundary gives the action in \eqref{gravBTZtoBTZ}. Unlike the previous calculation, there is no need to introduce an infrared cutoff since we are working with finite cylinders in this calculation.}
\begin{multline} \label{gravBTZtoBTZ}
    I=\frac{c}{3}\bigg [\frac{1}{4\pi}\int_{\Gamma_+}d^2z( \partial \Phi \overline{\partial}\Phi+e^{\Phi})-\frac{i}{4\pi}\int_{\partial \Gamma_+} dz \Phi \partial \Phi+\frac{1}{4\pi}\int_{\Gamma_-}d^2z (\partial \Phi \overline{\partial}\Phi+e^{\Phi})-\frac{i}{4\pi}\int_{\partial \Gamma_-} dz \Phi \partial \Phi \bigg]
\end{multline}
where $\Gamma_+$ is the region of the cylinder below the interface $y \in (\epsilon,\tau_0)$ and $\Gamma_-$ is the region above the interface $y\in (\tau_0,\frac{\beta}{2}-\epsilon)$ with $\epsilon \to 0$ being a cutoff parameter. On renormalising the action to remove the terms divergent in the cutoff, we get
\begin{equation} \label{rentorus}
    I_{\text{ren}}=\frac{c}{3}\left[-\tau_0 \frac{r_H^2}{2}-(\frac{\beta}{2}-\tau_0)\frac{(r_H')^2}{2}+m-m\log(\frac{r_0}{2}) \right] \ .
\end{equation}
Here, $r_0$ is the turning radius of the shell in Schwarzschild coordinates which can be expressed in terms of $r_H$ and $r_H'$ in a way that is symmetric under $r_H \xleftrightarrow{} r_H'$ as
\begin{equation} \label{r0BTZtoBTZ}
    r_0^2=\frac{1}{2}(r_H^2+(r_H')^2)+\frac{m^2}{4}+\frac{(r_H^2-(r_H')^2)^2}{4m^2} \ .
\end{equation}
This relation can be derived by noting that the Israel junction conditions which translate into the continuity of the Liouville field across the interface,
\begin{equation} \label{contBTZtoBTZ}
    r_0=\frac{r_H}{\sin(r_H \tau_0)}=\frac{r_H'}{\sin(r_H'(\frac{\beta}{2}-\tau_0))}
\end{equation}
and the jump in the normal derivative of the Liouville field across the interface,
\begin{equation} \label{junBTZtoBTZ}
    \sqrt{r_0^2-(r_H')^2}\pm \sqrt{r_0^2-r_H^2}=m
\end{equation}
with the $\pm$ signs respectively apply for `heavy' and `light' shells. In the present context, we refer to shells as heavy (resp. light) if the time-symmetric slice of the resulting geometry has two (resp. one) locally minimal surfaces.\footnote{The following relations derived from \eqref{junBTZtoBTZ} are useful,
\begin{equation}
    \begin{split}
       & \sqrt{r_0^2-(r_H')^2}=\frac{m^2+r_H^2-(r_H')^2}{2m}\\
       & \sqrt{r_0^2-r_H^2}=\frac{|m^2-r_H^2+(r_H')^2|}{2m}
    \end{split}
\end{equation}
}
Heavy shells satisfy $m^2>r_H^2-(r_H')^2$ in which case we can express the continuity condition \eqref{contBTZtoBTZ} as
\begin{equation}
        \frac{\beta}{2}-\tau_0 = \frac{\pi-\sin^{-1}(\frac{r_H'}{r_0})}{r_H'}  \qquad  \tau_0=\frac{\pi-\sin^{-1}(\frac{r_H}{r_0})}{r_H} \ .
\end{equation}
Light shells satisfy $m^2<r_H^2-(r_H')^2$ and the continuity condition \eqref{contBTZtoBTZ} translates to
\begin{equation}
        \frac{\beta}{2}-\tau_0 = \frac{\pi-\sin^{-1}(\frac{r_H'}{r_0})}{r_H'} \qquad \tau_0=\frac{\sin^{-1}(\frac{r_H}{r_0})}{r_H} \ .
\end{equation}
Using these relations, we can express the action \eqref{rentorus} in terms of the horizon radii of the two black holes,
\begin{multline} 
    I_{\text{ren}}=-\frac{c}{6}\bigg[r_H\cot^{-1}\left(\frac{r_H^2-r_H'^2-m^2}{2mr_H}\right )+r_H'\cot^{-1}\left(\frac{r_H'^2-r_H^2-m^2}{2mr_H'}\right )\\+m\log \left(\frac{1}{2}(r_H^2+r_H'^2)+\frac{m^2}{4}+\frac{(r_H^2-r_H'^2)^2}{4m^2} \right)-2m\bigg]
\end{multline}
with the horizon radii expressed in terms of the landing parameters by \eqref{r0BTZtoBTZ} and \eqref{contBTZtoBTZ}.
The renormalised gravitational action matches with two copies semiclassical block computed in \eqref{torblock} and is related to two copies of the Liouville action $S_L$ computed in \eqref{SLann} by
\begin{equation}
   I_{\text{ren}}=\frac{c}{6}(2f_0^{\CL}(\tau_0,\beta;m))=\frac{c}{6}(2S_L)+\frac{c}{3}m \log 2 \ .
\end{equation}
The conclusion is that we have confirmed the general relationship between gravity, the dual CFT, Liouville, and the identity block, as in \eqref{fourOnSphere}, on the torus as well.

\subsection{Rotating black hole: Gluing rotating BTZ to empty AdS} \label{secrotbh}

In this section, we describe the computation of the on shell action of a rotating black hole formed by the collapse of a thin shell of dust. This will provide a bulk contribution to the twisted correlator of the line defect in the dual CFT, $\langle \CL^\dagger(\tau_0) e^{2i\theta J} \CL(-\tau_0) \rangle$. Recall that we are assuming the dust particles making up the thin shell are flavored, so that when the defects are shifted by the insertion of $e^{2i\theta J}$, each particle must travel through as a transverse distance $2\theta$ along the shell.

\begin{figure}
\begin{center}
\begin{overpic}[grid=false]{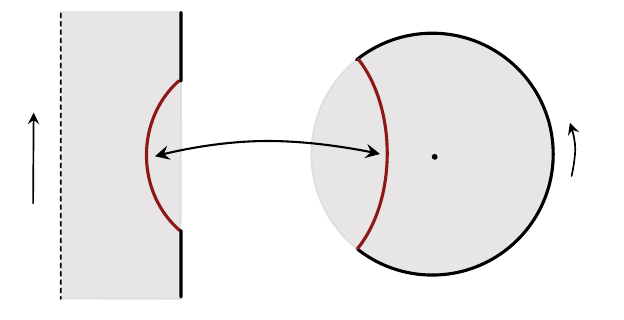}
\put (94, 25) {$\tau$}
\put (2,25) {$\Tilde{\tau}$}
\put (41,25) {\footnotesize glue}
\put (56,45) {$\tau_0$}
\put (53,7) {$-\tau_0$}
\put (77,25) {I}
\put (64,25) {II}
\put (14,25) {III}
\put (18,45) {IV}
\put (18,8) {IV}
\put (9,14)
{
\begin{tikzpicture}[scale=1]
\draw (10,7.6) -- (12,7.6);
\draw (10,5) -- (12,5);
\end{tikzpicture}
}
\put (56,26)
{
\begin{tikzpicture}[scale=1]
\draw (0,-0.2) -- (-1.3,1.5);
\end{tikzpicture}
}
\put (56,11)
{
\begin{tikzpicture}[scale=1]
\draw (0,0.1) -- (-1.3,-1.5);
\end{tikzpicture}
}
\end{overpic}
\end{center}
\caption{Euclidean thin-shell black hole, obtained by gluing a piece of global AdS (left) to a piece of the rotating BTZ black hole (right) across a thin shell (red). Each point on the diagram is $S^1$, except for the dashed line, which is the center of Euclidean global AdS. In this construction, the physical region contains the horizon i.e the fixed point of the $U(1)$ isometry around the time direction of the rotating BTZ black hole. There is a relative twist of $2\theta$ along the transverse $S^1$ between the two landing points of the shell which is not depicted in this figure. The actions of the labelled regions are computed later in the section. \label{fig:thinshell}}
\end{figure}
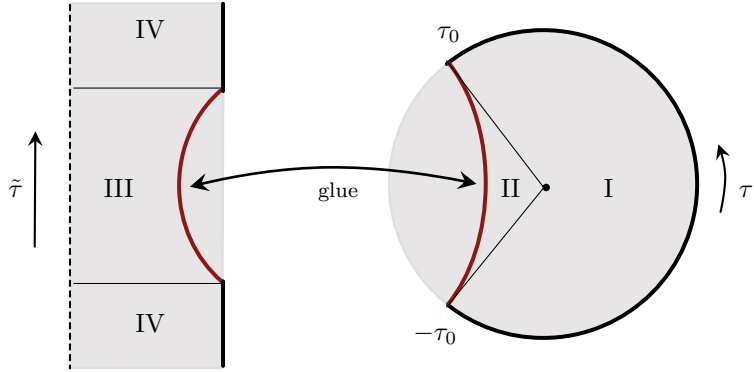

\begin{figure}
\begin{center}
\begin{overpic}[width=1.8in,grid=false]{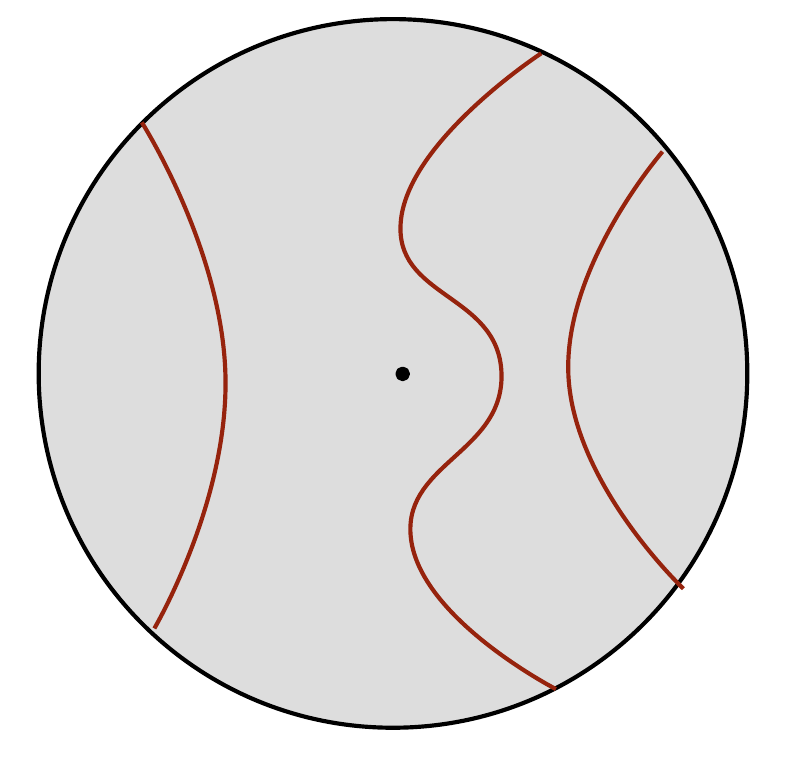}
\put (85,80) {(a)}
\put (68,91) {(b)}
\put (11,84) {(c)}
\end{overpic}
\end{center}
\caption{The three types of trajectories of a thin shell in a rotating BTZ black hole. In the gluing construction to obtain the thin shell black hole, the physical region is to the right of these trajectories. Trajectories of type (a) are outside the horizon and obey $m^2<M+1$. Trajectories of type (b) where the shell turns once are also outside the horizon and obey $M+1<m^2<\sqrt{M^2-J^2}+1$. These trajectories don't exist for a non-rotating black hole. Trajectories of type (c) go behind the horizon and obey $m^2>\sqrt{M^2-J^2}+1$. We refer to  (a) and (b) as `light shells' and (c) as `heavy shells'. \label{fig:shelltrajectories}}
\end{figure}

This computation is similar to the one in \cite{Chandra:2022fwi} adapted for the case of rotating BTZ. We first describe the one-sided rotating black hole solution by solving the Israel junction conditions describing the gluing of empty AdS to rotating BTZ. See figure \ref{fig:thinshell} for a diagrammatic description of the construction. It is convenient to solve the junction conditions by working in a frame which is co-rotating with the black hole along the trajectory of the shell. We will work with the following metric for the Euclidean rotating BTZ black hole,
\begin{equation}
    ds^2 = f(r)d\tau^2 + \frac{dr^2}{f(r)} + r^2\left( d\phi - i\frac{J}{2r^2}d\tau \right)^2
\end{equation}
where\footnote{The mass parameter $M$ and angular momentum parameter $J$ have been rescaled by $8G$ for convenience,
whereas the generator in $e^{2i\theta J}$ has the standard normalization, $J \sim \p_\phi$.
}
\begin{equation}
    f(r) = r^2 - M + \frac{J^2}{4r^2} = \frac{(r^2 - r_+^2)(r^2 - r_-^2)}{r^2}  \ .
\end{equation}
Here $r_+$ is the radius of the horizon.
By solving the junction conditions, we show that $J$ and $r_-$ are purely imaginary so the metric is real. The collapsing shell is a codimension-1 surface which can be parametrized as $\tau(l), r(l), \phi(l,\psi)$ where the coordinates $(l,\psi)$ on the shell are chosen such that
\begin{equation}
    dl^2 = f(r)d\tau^2 + \frac{dr^2}{f(r)} \,, \quad d\psi = d\phi - i\frac{J}{2r^2}d\tau = d\phi - \Omega(l)dl \,, \quad \left(\Omega(l) = i\frac{J}{2r^2}\dot{\tau} \right)
\end{equation}
The induced metric on the shell with the tangent and normal vectors given by
\begin{align}
    &\hspace{2cm} ds^2 = dl^2 + r^2(l) d\psi^2 \\
    \partial_l &= \dot{\tau}\partial_\tau + \dot{r}\partial_r + \Omega\partial_\phi \,, \quad \partial_\psi = \partial_\phi \,,\quad  n = -\dot{r}d\tau + \dot{\tau}dr
\end{align}
using the normal vector we can now compute the extrinsic curvature of the shell.
\begin{equation}
    K_{ab} = \nabla_a n_b = e^{i}_{a}e^{j}_{b}\nabla_i n_j
\end{equation}
\begin{equation}
    K_{l\psi} = \frac{iJ}{2r} \,, \quad K_{\psi\psi} = -r\sqrt{f - \dot{r}^2}
\end{equation}
where we have used $\dot{\tau} = \epsilon \frac{\sqrt{f-\dot{r}^2}}{f} $ where $\epsilon=\pm 1$. The sign is chosen based on how we glue the geometries together. We choose $\epsilon=-1$ when the shell is behind the horizon (heavy shell) and $\epsilon=1$ when the shell is in the region outside the horizon (light shell). This can be seen from the figure \ref{fig:shelltrajectories}. The terminology for the heavy and light shell will be clear as we go through the computation. 
We can perform similar computations for the shell trajectory in the empty AdS region. The metric of AdS in global coordinates is
\begin{equation}
    ds^2 = (1 + r^2)d\tilde{\tau}^2 + \frac{dr^2}{1 + r^2} + r^2 d\tilde{\phi}^2 
\end{equation}
where the $\tilde{\tau},\tilde{\phi}$ are written to distinguish from the BTZ coordinates. There is no $\tilde{r}$ since the continuity of the metric across the shell fixes radial coordinate to be the same on both sides of the shell. Computing the extrinsic curvature in AdS part of the geometry.
\begin{equation}
    K_{l\psi} = 0 \,, \quad K_{\psi\psi} = r\sqrt{r^2 + 1 - \dot{r}^2}
\end{equation}
where we have used $\dot{\tilde{\tau}} = \frac{\sqrt{r^2+1-\dot{r}^2}}{r^2+1}$. In this case, $\tilde{\tau}$ is always increasing along the shell and so the sign is positive. We can now impose the Israel junction condition to glue the two geometries together. These will determine the evolution of the shell trajectories. We start with
\begin{equation}
    \label{jun}
    \Delta K_{ab} - h_{ab}\Delta K = -T_{ab}
\end{equation}
where $\Delta$ denotes the difference between the quantities on the BTZ and AdS region respectively.
For the stress tensor appearing on the right, we take the thin shell to be described by a pressureless fluid which fixes the form of the stress tensor to be
\begin{equation}
    T_{ab} = -\sigma(l) u_a u_b \,,\quad (|u|=1)
\end{equation}
\begin{equation}
    u = u^a \partial_a = \gamma(l) \left( \partial_l + \omega(l)\partial_\psi \right) \,, \quad (\gamma^2 = \frac{1}{1+\omega^2 r^2} )
\end{equation}
Taking the trace of (\ref{jun}) with $h_{ab}$ fixes $\Delta K = T = -\sigma(l)$. The junction conditions for the $(l\psi)$ and $(\psi\psi)$ components are
\begin{align}
    &(l\psi) \implies \sigma \gamma^2 \omega = i\frac{J}{2r^3} \\
    &(\psi\psi) \implies \sigma \gamma^2  = \frac{-\epsilon \sqrt{f-\dot{r}^2} + \sqrt{r^2+1-\dot{r}^2} }{r}
\end{align}
To determine $\sigma$ we use the conservation of the stress tensor
\begin{equation}
    \nabla_a T^{ab} = 0 \implies r(\sigma \gamma^2)'= \sigma\gamma^2(r^2\omega^2 - 1)r' \implies (r\sigma\gamma^2)' = \sigma\gamma^2 \omega^2 r^2 r'
\end{equation}
this equation can be solved by setting $g(r) = \sigma \gamma^2 r$ and using the boundary condition $g(r=\infty) = m$.
\begin{equation}
    \sigma\gamma^2 = \frac{1}{r} \sqrt{m^2 + \frac{J^2}{4r^2}} \,,\quad \omega = \frac{iJ}{r\sqrt{4m^2r^2 + J^2}}
\end{equation}
Plugging this expression in above we get
\begin{equation}
    \frac{1}{r} \sqrt{m^2 + \frac{J^2}{4r^2}} = \frac{-\epsilon\sqrt{f - \dot{r}^2} + \sqrt{r^2+1-\dot{r}^2} }{r}
\end{equation}
using this we can derive the differential equation for $r(l)$.
\begin{equation}
    \dot{r}^2 = r^2 + 1 - \frac{(m^2+M+1)^2}{4\left(m^2 + \frac{J^2}{4r^2}\right)} = \frac{4m^2(r^2-r_{0+}^2)(r^2-r_{0-}^2)}{J^2+4 m^2 r^2}
\end{equation}
the $r$ trajectory doesn't depend on the choice of $\epsilon$. We have also defined the turning points of the shell in the radial direction as $r_{0\pm}$ and are explicitly given as
\begin{equation}
    r_{0\pm}^2 = \frac{\left((m-1)^2+M\right) \left((m+1)^2+M\right)-J^2 \pm \sqrt{\left[\left((m-1)^2+M\right) \left((m+1)^2+M\right)-J^2\right]^2-16 J^2 m^2}}{8 m^2}
\end{equation}
Since $J$ is imaginary, we see that $r_{0-}^2<0$.
Note that $r_{0\pm}$ satisfies the following equation:
\begin{equation} \label{juncrot}
    \sqrt{m^2+\frac{J^2}{4r_{0\pm}^2}} =  -\epsilon \, \sqrt{r_{0\pm}^2-M+\frac{J^2}{4r_{0\pm}^2}} + \sqrt{r_{0\pm}^2+1}
\end{equation}
\begin{equation}
    \dot{r} = \pm 2m\sqrt{\frac{(r^2-r_{0+}^2)(r^2-r_{0-}^2)}{J^2+4 m^2 r^2}}
\end{equation}
where the $\pm$ sign in $\dot r$ comes from whether we are in the $(r_{0+} \rightarrow \infty)$ region $(+)$ or $(\infty \rightarrow r_{0+})$ region $(-)$. The trajectory of the shell is then described by the following equations.

\subsubsection*{Trajectory in the BTZ region}
\begin{equation}
    \dot{\tau} = -\frac{r^3 \left(\frac{J^2}{2 r^2}+m^2-M-1\right)}{(r^2 - r_+^2)(r^2 - r_-^2) \sqrt{J^2+4 m^2 r^2}}
\end{equation}
\begin{equation}
    \label{dt_dr_BTZ}
    \frac{d\tau}{dr} = \mp \frac{r^3 \left(\frac{J^2}{2 r^2}+m^2-M-1\right)}{2m(r^2 - r_+^2)(r^2 - r_-^2) \sqrt{(r^2-r_{0+}^2)(r^2-r_{0-}^2)}}
\end{equation}
We can now integrate the above to find $\tau(r)$ for the shell trajectory. 
\begin{equation}
    \int_{\tau_i}^{\tau} d\tau = \tau - \tau_i = \int_{r_{0+}}^{r} dr' \, \frac{d\tau}{dr'}(r')
\end{equation}
in this integral we choose the $+$ branch for $d\tau/dr$ i.e. the region where the radial coordinate $r$ goes from $r_{0+}$ to $\infty$. Here, the boundary value $\tau_i$ is determined by whether we work with heavy or light shells. From figure \ref{fig:shelltrajectories}, we see that for light shells $\tau_i=0$ and for heavy shells $\tau_i = \beta/2$. We define $\tau_0$ as the point where the shell trajectory intersects the boundary of the Euclidean BTZ. To evaluate $\tau_0$ from above, we send $r\to\infty$ and get
\begin{equation}
    \tau_0 = \tau_i + \int_{r_{0+}}^{\infty} dr' \, \frac{d\tau}{dr'}(r') = \tau_i + [I(r)]_{r_{0+}}^{\infty}
\end{equation}
We can rewrite the condition satisfied by $r_{0+}$ as
\begin{equation}
    \begin{split}
        &\left[\sqrt{m^2+\frac{J^2}{4r_{0+}^2}}  + \epsilon\, \sqrt{r_{0+}^2-M + \frac{J^2}{4r_{0+}^2}} \right]^2 =  r_{0+}^2+1 \\
    \end{split}
\end{equation}
For $\epsilon=-1$, this implies $m^2 - M - 1 > -J^2/2r_{0+}^2 > -J^2/2r^2$ and so the overall sign in (\ref{dt_dr_BTZ}) is negative. Whereas for $\epsilon=1$, we have $m^2 - M - 1 < -J^2/2r_{0+}^2$. This is why we called the $\epsilon=\pm$ case heavy and light shells. Thus the overall sign in (\ref{dt_dr_BTZ}) is again negative.
\begin{equation}
    \frac{d\tau}{dr} = -\frac{r^3 \left(\frac{J^2}{2 r^2}+m^2-M-1\right)}{2m(r^2 - r_+^2)(r^2 - r_-^2) \sqrt{(r^2-r_{0+}^2)(r^2-r_{0-}^2)}}
\end{equation}
It is convenient to express everything in terms of $r_{0\pm}^2, r_{\pm}$ using the relations
\begin{equation}
    M = r_+^2 + r_-^2 \,,\quad J = 2\,r_+ r_- \,,\quad m^2 = \frac{r_+^2r_-^2}{r_{0+}^2r_{0-}^2}
\end{equation}
\begin{equation}
    \frac{d\tau}{dr} = -\frac{r^3 \left(\frac{2\,r_+^2 r_-^2}{r^2} + \frac{r_+^2r_-^2}{r_{0+}^2r_{0-}^2} - (r_+^2 + r_-^2) - 1\right)}{2\frac{r_+ r_-}{r_{0+} r_{0-}}(r^2 - r_+^2)(r^2 - r_-^2) \sqrt{(r^2-r_{0+}^2)(r^2-r_{0-}^2)}}
\end{equation}
The indefinite integral is
\begin{equation}
    I(r) = c_+ \, I_+(r) + c_-\, I_-(r)
\end{equation}
where
\begin{equation}
    I_{+}(r) = \tan^{-1}\left( \sqrt{\frac{r^2-r_{0+}^2}{r^2-r_{0-}^2}} \sqrt{\frac{r_{+}^2-r_{0-}^2}{r_{0+}^2-r_{+}^2}} \right) \,,\quad I_{-}(r) = \tanh^{-1}\left( \sqrt{\frac{r^2-r_{0+}^2}{r^2-r_{0-}^2}} \sqrt{\frac{r_{0-}^2 - r_{-}^2}{r_{0+}^2 - r_{-}^2}} \right)
\end{equation}
\begin{equation}
    c_{\pm} = \frac{(r_+^2r_-^2 - r_{0+}^2r_{0-}^2 \pm r_{0+}^2r_{0-}^2(r_-^2 - r_+^2))}{2(r_+^2 - r_-^2) \sqrt{\pm \frac{r_{\mp}^2}{r_{\pm}^2}r_{0+}^2r_{0-}^2(r_{0+}^2 - r_{\pm}^2)(r_{\pm}^2 - r_{0-}^2)} } 
\end{equation}
from this we can compute $\tau_0$. Note that $I(r=r_{0+}) = 0$, which gives
\begin{equation}
    \tau_0 = \tau_i + c_+ \, \tan^{-1}\left( \sqrt{\frac{r_{+}^2-r_{0-}^2}{r_{0+}^2-r_{+}^2}} \, \right) + c_-\, \tanh^{-1}\left(  \sqrt{\frac{r_{0-}^2-r_{-}^2}{r_{0+}^2-r_{-}^2}} \, \right)
\end{equation}
Similarly, we can compute the twist $\theta$ by integrating $(\Omega + \omega)$ along the shell. We also have different boundary condition for the heavy and light shells. For the heavy shell, $\theta_i = i\frac{\beta r_{-}}{2r_{+}}$ and for the light shell, the boundary condition is $\theta_i = 0$. The integral for $\theta$ is
\begin{equation}
    \begin{split}
        \int_{\theta_i}^{\theta} d\theta &= -\int dl \, (\Omega(l) + \omega(l)) = -\int_{r_{0+}}^{\infty} dr \, \frac{dl}{dr} \, (\Omega(r) + \omega(r)) \\
        \theta &= \theta_i - \frac{iJ}{4m} \int_{r_{0+}}^{\infty} dr \, \frac{r(2r^2 - M- m^2 +1 )}{(r^2 - r_+^2)(r^2 - r_-^2) \sqrt{(r^2-r_{0+}^2)(r^2-r_{0-}^2)}} \\
        &= \theta_i - \left[ \tilde{c}_+ \, I_+(r) + \tilde{c}_-\, I_-(r) \right]_{r_{0+}}^{\infty} \\
    \end{split}
\end{equation}
where $\tilde{c}_+ = -\sqrt{\frac{-r_{-}^2}{r_{+}^2}} \, c_+ \,,\, \tilde{c}_- = \sqrt{\frac{r_{+}^2}{-r_{-}^2}} \, c_- $. Using these relations, we can write $\theta$ as
\begin{equation}
    \theta = \theta_i - \left( \tilde{c}_+ \, \tan ^{-1}\left( \sqrt{\frac{r_{+}^2-r_{0-}^2}{r_{0+}^2 - r_{+}^2}} \, \right) + \tilde{c}_-\, \tanh^{-1}\left(  \sqrt{\frac{r_{0-}^2-r_{-}^2}{r_{0+}^2-r_{-}^2}} \, \right) \right)
\end{equation}

\subsubsection*{Trajectory in the AdS region}
Repeating the same steps as above for the AdS part of the geometry, we get
\begin{equation}
    \dot{\tilde{\tau}} = \frac{r(m^2+M+1)}{(1+r^2)\sqrt{J^2+4 m^2 r^2}}
\end{equation}
\begin{equation}
    \frac{d\tilde{\tau}}{dr} = \frac{r(m^2+M+1)}{2m(1+r^2)\sqrt{(r^2-r_{0+}^2)(r^2-r_{0-}^2)}}
\end{equation}
\begin{equation}
    \begin{split}
        \Tilde{\tau}_0 &= i\tan^{-1}\left(\frac{2 + r_{0+}^2 + r_{0-}^2}{2\sqrt{(1 + r_{0+}^2)(-1 - r_{0-}^2)}}\right) - i\tan^{-1}\left(\sqrt{\frac{1 + r_{0+}^2}{-1 - r_{0-}^2}}\right) = \tanh^{-1}\left(\sqrt{\frac{1 + r_{0-}^2}{1 + r_{0+}^2}}\right)  
    \end{split}
\end{equation}

\subsubsection*{Computing the Action}
We now compute the action of this Euclidean geometry. We divide the contributions in terms of four regions, shown in figure \ref{fig:thinshell}. Region I is the BTZ portion with $|\tau|<\tau_0$, Region II is the BTZ region between $|\tau|>\tau_0$ and the shell for $\epsilon=-1$ and between $|\tau|<\tau_0$ and the shell for $\epsilon=1$, Region III is the AdS region with $|\tilde{\tau}| < \tilde{\tau}_0$, and Region IV is AdS with $|\tilde{\tau}| > \tilde{\tau}_0$. The total action:
\begin{equation}
    I = I_1 - \epsilon I_2 + I_3 + I_4 + I_{\text{Hay}}
\end{equation}
The extra factor of $-\epsilon$ comes in because for heavy shell we need to add the contribution from the region II whereas for the light shells we need to remove them. The last term $I_{\text{Hay}}$ is the Hayward term which is added at the two corners where the shell intersects the AdS boundary as we did the non-rotating case.
\\
\textbf{Region I}: 
BTZ geometry with $|\tau|<\tau_0$.
\begin{equation}
    I_{1} = \frac{2\tau_0}{\beta}I_{BTZ} = -\frac{2\tau_0}{\beta}\frac{\pi^2}{2G}\frac{\beta}{\beta^2 + \chi^2} = - \frac{\tau_0}{4G}(r_+^2 - r_-^2) 
\end{equation}
here $\beta,\chi$ are inverse temperature and the angular potential. They determine the periodicity of the coordinates $(\tau,\phi)\sim(\tau,\phi+2\pi)\sim(\tau+\beta,\phi+\chi)$. We have used the relation $\beta = \frac{2\pi r_{+}}{r_{+}^2-r_{-}^2}$ and $\chi = i\frac{\beta r_-}{r_+}$.
\\
\textbf{Region II}: 
BTZ geometry between $|\tau|>\tau_0\, (\epsilon=-1)$  or $|\tau|<\tau_0\,(\epsilon=1)$ and the shell. 
\begin{equation}
    \begin{split}
        I_{2} &= -\frac{1}{16\pi G}\int d^3x \, \sqrt{g} (R+2) \\
        &= \frac{2}{4\pi G} \int_{\tau_i}^{\tau_0} d\tau \int_{r_+}^{g^{-1}(\tau)} dr \, r \int_{0}^{2\pi} d\phi \,,\quad \\
        &= \frac{1}{2G} \int_{\tau_i}^{\tau_0} d\tau \, [r^2]_{r_+}^{g^{-1}(\tau)}   \\
        &= \frac{\epsilon}{2G} \int_{r_{0+}}^{\infty} dr \, \frac{d\tau}{dr} \, (r^2 - r_+^2) \\
        &= -\frac{\epsilon}{4mG} \int_{r_{0+}}^{\infty} dr \, \frac{r^3 \left(\frac{J^2}{2 r^2}+m^2-M-1\right)}{(r^2 - r_-^2) \sqrt{(r^2-r_{0+}^2)(r^2-r_{0-}^2)}} \\
    \end{split}
\end{equation}
where $\tau = g(r)$ is the shell trajectory. For the case the heavy shell, we integrate from $\tau=\tau_0$ to $\tau=\beta/2$ and for light shells from $\tau=0$ to $\tau=\tau_0$. The integral in the last line is
\begin{equation}
    \begin{split}
        s(r) &= ( c_1 + c_2 )\tanh^{-1}\left(\sqrt{\frac{r^2-r_{0+}^2}{r^2-r_{0-}^2}}\right) + c_3 \tanh^{-1}\left(\frac{r^2-r_-^2 - \sqrt{(r^2-r_{0-}^2)(r^2-r_{0+}^2)} }{\sqrt{(r_{0-}^2 - r_-^2)(r_{0+}^2 - r_-^2)}}\right) - c_1\frac{i\pi}{2} \\
    \end{split}
\end{equation}
we have used identities $\tanh^{-1}(x) = \tanh^{-1}(1/x) - i \pi/2$ for $x>1$. The constants $c_i$ are given by
\begin{equation}
    \begin{split}
        c_1 &= -(1 + r_+^2 + r_-^2) = -(1+M)  \,,\quad c_2 = \frac{r_+^2r_-^2}{r_{0+}^2r_{0-}^2} = m^2 \,,\\
        &\quad c_3 = -\frac{r_{-}^2(-r_+^2r_-^2 + r_{0+}^2r_{0-}^2 + r_{0+}^2r_{0-}^2(r_-^2 - r_+^2))}{r_{0+}^2r_{0-}^2\sqrt{(r_-^2-r_{0-}^2)(r_-^2-r_{0+}^2)}}
    \end{split}
\end{equation}
We are interested in evaluating $s(r=\infty) - s(r=r_{0+})$. The function $s(r)$ diverges as $r\rightarrow\infty$. To regulate this we introduce a cutoff at $r=\Lambda$ for some large $\Lambda$.
\begin{equation}
    \begin{split}
        s(r=\Lambda) - s(r=r_{0+}) &= -(c_1+c_2)\log\left(\frac{\sqrt{r_{0+}^2-r_{0-}^2}}{2\Lambda}\right) + c_3 \tanh^{-1}\left(\sqrt{\frac{r_{0-}^2-r_{-}^2}{r_{0+}^2-r_{-}^2}}\right)  + \mathcal{O}(\Lambda^{-1})
    \end{split}
\end{equation}
\begin{equation}
    I_2 = -\frac{\epsilon}{4mG}\left[ -(c_1+c_2)\log\left(\frac{\sqrt{r_{0+}^2-r_{0-}^2}}{2\Lambda}\right) + c_3 \tanh^{-1}\left(\sqrt{\frac{r_{0-}^2-r_{-}^2}{r_{0+}^2-r_{-}^2}}\right) \right]
\end{equation}
\\
\textbf{Region III}:
AdS geometry between $|\tilde{\tau}| < \tilde{\tau}_0$ and the shell.
\begin{equation}
    \begin{split}
        I_3 &= \frac{2}{4\pi G} \int_{0}^{\tilde{\tau}_0} d\tilde{\tau}\int_{0}^{g_{-}^{-1}(\tau)} dr \, r \int_{0}^{2\pi} d\phi \\
        &= \frac{1}{ 2G} \int_{r_{0+}}^{\infty} dr \, \frac{d\tilde{\tau}}{dr} \, r^2 \\
        &= -\frac{m^2+M+1}{4m G}\log \left(\frac{\sqrt{r_{0+}^2 - r_{0-}^2}}{2\Lambda} \right) - \frac{\Tilde{\tau}_0}{2G} + \mathcal{O}(\Lambda^{-1}) \\
    \end{split}
\end{equation}
\\
\textbf{Region IV}:
AdS geometry between $|\tilde{\tau}| > \tilde{\tau}_0$. The Einstein-Hilbert term in the interior and the GHY term on the boundary of this region cutoff at $|\tau|=T$ evaluate to give
\begin{equation}
    I_4 = -\frac{T-\tau_0}{4G} 
\end{equation}
where $T \to \infty$ is a regulator. The first term in the above expression i.e $-\frac{T}{4G}$ is removed when we define the renormalised action.
\\
\textbf{Hayward term}:
Contribution from the region near the intersection of AdS and BTZ geometry.
\begin{equation}
    I_{\text{Hay}} = -\frac{1}{4\pi G}\int dx \sqrt{\gamma} (\pi - \Theta)
\end{equation}
where $\Theta = \Theta_1 + \Theta_2$. $\Theta_1$ the angle between the shell and AdS boundary. $\Theta_2$ the angle between the shell and BTZ boundary. We compute this angle using the vectors $v = \dot{\tau}\partial_\tau + \dot{r}\partial_r + \Omega\partial_\phi \,, \quad u = \partial_\tau$. We find to leading order in the limit of large cutoff $\Lambda$:
\begin{equation}
        \Theta_1 = \frac{\pi}{2} + \frac{1 + M + m^2}{2m \Lambda} \,, \quad \Theta_2 = \frac{\pi}{2} - \frac{1 + M - m^2}{2m \Lambda}
\end{equation}
The contribution to the action can then be computed as  $(\sqrt{\gamma}=\Lambda)$:
\begin{equation}
    I_{\text{Hay}} = \frac{1}{4\pi G} \int_0^{2\pi} d\phi \, m  = \frac{m}{2 G}
\end{equation}
\ \\
\textbf{Total Action}: $I_{\text{ren}} = I - \frac{m}{2G}\log\Lambda + \frac{T}{4G}$.
\begin{equation} \label{Sren}
    I_{\text{ren}} = -\frac{\tau_0}{4G}(r_+^2 - r_-^2 - 1) - \frac{m}{2G}\log\left(\frac{\sqrt{r_{0+}^2-r_{0-}^2}}{2}\right) - \frac{\tilde{\tau}_0}{2G}  + \frac{c_3}{4mG}  \tanh^{-1}\left(\sqrt{\frac{r_{0-}^2-r_{-}^2}{r_{0+}^2-r_{-}^2}}\right) + \frac{m}{2 G}
\end{equation}
\\
\textbf{Factorization of the Action}:
The above action takes a manifestly holomorphically factorized form when expressed in terms of the complex conformal weights which are related to the mass and (imaginary) angular momentum parameters of the black hole as
\begin{equation}
  M=\frac{1}{2}(\gamma^2+\overline{\gamma}^2) \qquad J=\frac{1}{2}(\gamma^2-\overline{\gamma}^2)
\end{equation}
where the complex saddlepoint momenta $\gamma$ and $\overline{\gamma}$ are related to the conformal weights by $h=\frac{c}{24}(1+\gamma^2)$ and $\overline{h}=\frac{c}{24}(1+\overline{\gamma}^2)$.
In terms of these parameters, we can write down the condition for the trajectory of the shell to cross the Euclidean horizon (denoted $r_+$) of the black hole as
\begin{equation} \label{horcross}
  m^2=\sqrt{M^2-J^2}+1=\gamma\overline{\gamma}+1 \ .
\end{equation}
Shells heavier than the above threshold (at fixed mass and angular momentum of the black hole) go `behind' the Euclidean horizon while those lighter than the above threshold are `outside' the Euclidean horizon.
In either case, we can express the renormalised on-shell action \eqref{Sren} in a manifestly holomorphically factorized form
\begin{equation}
    I_{\text{ren}} = \frac{c}{6}\left( f(\gamma) + f(\overline{\gamma}) \right)
\end{equation}
where the function $f(\gamma)$ is given by
\begin{multline} 
    f(\gamma)=-\frac{1}{2}\bigg [(\gamma-\frac{1}{\gamma})\cot^{-1}\left(\frac{1+\gamma^2-m^2}{2m\gamma}\right)-2m-2m\log(4m)\\+(m+1)\log((m+1)^2+\gamma^2)+(m-1)\log((m-1)^2+\gamma^2) \bigg ]
\end{multline}
This is equal to the chiral identity block on the sphere, given in \eqref{finalfCL} or  \eqref{sphblock}, and can be expressed in terms of the complex modulus $\sigma=\tau_0 - i\theta$ by solving the Israel junction condition \eqref{juncrot} which is equivalent to the junction condition derived using CFT \eqref{cftjuncrot}.\footnote{To show this, it is convenient to work with the complexified horizon radius, $\tilde{r}_{H} = r_+ + i|r_-|$ and complexified turning radius, $\tilde{r}_{0} = r_{0+} + i|r_{0-}|$. Using this, we get $\Tilde{r}_0=\Tilde{r}_H \sin(\Tilde{r}_H \sigma)$ where $\sigma=\tau_0-i\theta$ consistent with the CFT prediction.} Thus, we have shown that semiclassically the gravitational partition function of the Euclidean black hole matches with the product of chiral identity blocks on the sphere, 
\begin{equation}
e^{-I_{\rm ren}}  \approx \langle D^\dagger(\tau_0) e^{2i\theta J} D(-\tau_0)\rangle_{\rm cyl} \approx \left| {\cal F}_{\id}^D\right|^2
\end{equation}

\section{Wormholes} \label{secwormholes}
In this section we will tie together all of the results above on Liouville line defects, universal asymptotics of line defects in compact CFT, and thin shell black holes in 3D gravity in order to describe thin shell wormholes as an ensemble average. As described in the introduction, this is a specific example of the thin-shell wormhole model in \cite{Chandra:2022fwi}, and at the same time it generalizes the CFT ensemble introduced in \cite{Chandra:2022bqq} from point particles to extended objects.

\subsection{Wavefunction for thin shell black holes}
We will follow \cite{Chandra:2022bqq,Chandra:2022fwi} closely, with one difference, which is that we will use microcanonical wavefunctions to simplify the comparison between bulk and boundary. In \cite{Chandra:2022bqq,Chandra:2022fwi} the comparison was done at fixed kinematics, which is analogous to working in the canonical ensemble.

The microcanonical wavefunction of an eternal black hole in $D>2$ bulk dimensions was obtained in \cite{Chua:2023srl} (following earlier results in JT gravity \cite{Harlow:2018tqv}).\footnote{Strictly speaking, the states in \cite{Harlow:2018tqv,Chua:2023srl} fix the transverse metric (or dilaton in two dimensions) rather than the ADM charges, so they are not the exact dual of microcanonical states. But they are equivalent to microcanonical states at stationary saddlepoints and so can be labeled by the corresponding $h,\bh$.} The results carry over immediately to pure state black holes created by matter.  The idea is to define a semiclassical wavefunction in the bulk theory by cutting open the Euclidean path integral. We start with the result of section \ref{secrotbh}, which can be summarized as
\begin{align}\label{summarybh}
\vcenter{
\hbox{
\includegraphics[width=1in]{figures/singlebh.pdf}
}}
\qquad \approx \qquad
\left| \int_{c/24}^\infty dh \rho_0(h) C_0^{\CL}(h) e^{-2\tau_0 h} \right|^2
\end{align}
On the left-hand side of this equation is $e^{-I_{\rm ren}}$ for a rotating BTZ black hole glued across a thin shell to vacuum AdS$_3$. The solid black line on the right is the $S^2$ asymptotic boundary; the geometry caps off on the left.
As usual, `$\approx$' means that we work to leading order in the semiclassical approximation in the exponent. We now define a wavefunction in the bulk theory by imposing a boundary condition labeled by energy and angular momentum, or conformal weights $(h,\bh)$. This boundary condition is imposed on a bulk slice that meets the boundary at $\tau = 0$. It therefore defines a state
$|\CL; h, \bh\rangle$
in the bulk effective theory, which is a state with a thin shell behind the horizon and possessing an extremal surface of energy $h+ \bh$ and angular momentum $h - \bh$. The details of the boundary conditions can be found in \cite{Chua:2023srl} and will not be needed.

The result of \cite{Chua:2023srl}, extended to include matter behind the horizon, is that the wavefunction is calculated by a bulk saddle described as follows. We start with the thin shell black hole with the operator insertion locations tuned so that the saddlepoint lands at $(h,\bh)$. We assume that the kinematics are such that this saddlepoint has a thin shell behind the horizon. Then, this geometry is cut on a bulk slice with a corner at the Euclidean horizon. The result is a wedge geometry:
\begin{align}\label{wedgepic}
\vcenter{\hbox{
\begin{overpic}[grid=false,width=1.4in]{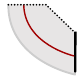}
\put (92,40) {\Huge\}}
\put (105,45) {$\tau_0$}
\put (47,70) {BTZ}
\put (47, 60) {$h,\bh$}
\put (25,30) {vacuum}
\put (70,100) {\vector(-2,-1){12}}
\put (72,98) {horizon}
\end{overpic}
}}
\end{align}
This geometry has the correct boundary conditions to contribute to the semiclassical wavefunction,
\begin{align}
\langle\CL; h,\bh | e^{-\tau_0(L_0 + \bL_0)} \CL_\Sigma | 0 \rangle &\approx
e^{-I_{\rm wedge}} 
\end{align}
with the $I_{\rm wedge}$ the renormalized on-shell action of the geometry in \eqref{wedgepic}, including boundary terms appropriate to the microcanonical state on the bulk slice (which consist of a Hayward term plus an additional corner term coming from the rotation \cite{Chua:2023srl}). The action of the wedge geometry $I_{\rm wedge}$ is related in a straightforward way to that of the full black hole.
Thus using \eqref{summarybh} we can obtain $I_{\rm wedge}$ and the semiclassical wavefunction. Applying the result for the action derived in \cite{Chua:2023srl} this immediately gives the wavefunction
\begin{align}\label{wf}
\langle\CL; h,\bh | e^{-\tau_0(L_0 + \bL_0)} \CL_\Sigma | 0 \rangle
&\approx
\left|
\sqrt{C_0^{\CL}(h)\rho_0(h)} 
e^{-\tau_0 h}
\right|^2
\end{align}
We emphasize that this is a gravity calculation of a \textit{bulk} overlap. We have given a precise definition of $|\CL; h,\bh\rangle$ in the bulk by a boundary value problem. We have not given any precise meaning to this state in the dual CFT, but it is naturally interpreted as a semiclassical approximation to the state $\CL_\Sigma |0\rangle$ after projecting onto a microcanonical energy and angular momentum window. 

The equation \eqref{wf} is essentially \eqref{summarybh} cut in half. Indeed, by gluing together two copies of the wavefunction and integrating over the bulk data $(h,\bh)$ imposed on the extremal surface, we recover exactly the equation \eqref{summarybh}, including the effects of rotation. However, \eqref{wf} contains more information because $(h,\bh)$ do not need to be tuned to the saddlepoint.

\subsection{Wormhole amplitudes}\label{ss:wormholek}
The wavefunction \eqref{wf} can now be used as a building block to calculate wormhole amplitudes. When two or more wedges are glued together at an extremal surface, that gravitational action acquires an extra contribution from the Hayward term \cite{Takayanagi:2019tvn} and the angular momentum corner term \cite{Chua:2023srl}. This leads to the following gluing rule for the on-shell action of a spacetime that can be built by gluing together $2k$ wedges at an extremal surface:
\begin{align}\label{gluingrule}
I_{\rm total} = (k-1)S +  \sum_{i=1}^{2k} I_{\rm wedge}^{(i)}
\end{align}
where $S$ is one quarter the area of the extremal surface. This formula holds when the full geometry is smooth, i.e., the total angle of all the wedges adds up to $2\pi$. 
Note that when all the terms are added together on the right-hand side of \eqref{gluingrule}, the overall coefficient of the entropy term is independent of $k$ such that the amplitude is proportional to $e^S$, for any $k$.

\begin{figure}
\begin{center}
\begin{align*}
\vcenter{\hbox{
\includegraphics[width=1.6in]{figures/k6wormhole.pdf}
}}
=
 \int dh d\bh \, e^{(1-k)S}
\vcenter{\hbox{
\includegraphics[width=1.6in]{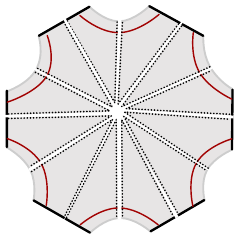}
}}
\end{align*}
\end{center}
\caption{\small The wormhole with $k$ boundaries is constructed by gluing together $2k$ copies of the wavefunction and integrating over the parameters $(h,\bh)$ specifying the length and twist of the extremal surface in the center. Shown here for $k=6$. The entropy prefactor comes from the corner terms in \eqref{gluingrule}, and is mostly canceled by a contribution $e^{kS}$ from the sum of wedge actions. \label{fig:glue6}}
\end{figure}


We are now ready to calculate the gravity amplitude corresponding to the averaged CFT observable
\begin{align}\label{definezk}
Z_k(\{ \tau_0^{(i)}, \btau_0^{(i)}\})
&:= 
\left. \overline{
\langle \CL_\Sigma^{\dagger} e^{-\tau_0^{(1)} L_0 - \btau_0^{(1)} \bL_0} \CL_\Sigma \rangle
\cdots
\langle \CL_\Sigma^{\dagger} e^{-\tau_0^{(k)} L_0 - \btau_0^{(k)} \bL_0} \CL_\Sigma \rangle
}\right|_{\rm connected}
\end{align}
where $i=1\dots k$. On the gravity side the appropriate boundary condition consists of $k$ asymptotic boundaries on which we insert line defects at Im $z = \pm \mbox{Im\ } \tau_0^{(i)}$, with the basepoints shifted in the transverse direction by a relative displacement Re $\tau_0^{(i)}$. This is the observable discussed in the introduction, and the saddlepoint is illustrated in \eqref{intro6boundary}. To calculate the gravitational action, we fix $(h,\bh)$ at the Euclidean horizon, slice this saddlepoint into $2k$ copies of the wavefunction \eqref{wf}, and then integrate over $(h,\bh)$.  The case with $k=6$ boundaries is illustrated in figure \ref{fig:glue6}.
When $\tau_0^{(i)} = \tau_0$ is real and identical on all boundaries, this is the wormhole constructed in \cite{Chandra:2022fwi}. For complex parameters, independent on each boundary, it is a new wormhole solution.
Using the wedge action \eqref{wf} and applying the gluing rule \eqref{gluingrule} we obtain the wormhole amplitude
\begin{align}
Z_k &\approx e^{-I_{\rm grav}} \approx \left|
\int_{c/24}^\infty dh \rho_0(h) C_0^{\CL}(h)^{k} e^{-2 h\sum_{i=1}^k \tau_0^{(i)}  }\right|^2
\label{zkgrav}
\end{align}
To be clear, this is a bulk calculation in semiclassical gravity, not a boundary calculation. We have used the microcanonical wavefunction in order to simplify the calculation of the on-shell action, but this is just a shortcut --- analogous results were obtained without the use of microcanonical wavefunctions in \cite{Chandra:2022bqq} for point particles and in \cite{Chandra:2022fwi} for thin shells, and the results are equivalent. Note that the result \eqref{zkgrav} applies to rotating shells, and to the case where the $\tau_0^{(i)}$ are different on every boundary. This calculation would be much more difficult using the methods of \cite{Chandra:2022bqq} (see the discussion there of the action of quasifuchsian wormholes, which are 2-boundary wormholes with different moduli). For point defects, those calculations were later simplified greatly using the technology of Virasoro TQFT \cite{Collier:2023fwi, Collier:2024mgv}.

This calculation was at the level of the classical action. But the 1-loop results in \cite{Chandra:2022bqq} and the all-orders calculations using Virasoro TQFT in \cite{Collier:2023fwi} strongly suggest that if we replace the exponentials by Virasoro blocks in \eqref{zkgrav}, then this should give the exact perturbative gravity amplitude on this topology. See also the proposal in \cite{hhpaper} for a perturbatively exact wavefunction in 3D gravity in the Hartle-Hawking state.

\subsection{The CFT ensemble interpretation}
The final answer for the wormhole \eqref{zkgrav} is written in a way that makes it almost trivial to find the CFT ensemble interpretation. The ensemble is specified in terms of the statistics of the matrix elements of the line defect between Virasoro primaries, $\langle i | \CL_\Sigma | j\rangle$, and the overlaps $\langle 0 |\CL_\Sigma| i\rangle$. As in \cite{Chandra:2022bqq} we define the ensemble by assuming heavy states have Cardy entropy, and treat the matrix elements as Gaussian random variables with zero mean and variance given by the universal formula formula for defect matrix elements:
\begin{align}\label{variance}
\overline{ | \langle 0 | \CL_\Sigma | i \rangle|^2} &\approx C_0^{\CL}(h_i) C_0^{\CL}(\bh_i) \\
\overline{ | \langle i | \CL_\Sigma | j\rangle|^2} &\approx C_0^{\CL}(h_i, h_j) C_0^{\CL}(\bh_i, \bh_j) \notag
\end{align}
In previous sections, the overline was interpreted as an average over nearby primaries with similar conformal weights, in a single CFT. Now we are using the bar to signify an ensemble average over CFT data, and \eqref{variance} defines the variance of the ensemble. Higher moments are calculated by Gaussian contractions. In this section we are assuming the defect has $m>1$ to avoid complications with light intermediate states.

In this ensemble, the observable $Z_k$ defined in \eqref{definezk} is calculated by inserting the spectral decomposition into each 2-point function, then averaging over the ensemble according to \eqref{variance}:
\begin{align}
&\overline{
\langle \CL_\Sigma^{\dagger} e^{-2\tau_0^{(1)} L_0 - 2\btau_0^{(1)} \bL_0} \CL_\Sigma \rangle
\cdots
\langle \CL_\Sigma^{\dagger} e^{-2\tau_0^{(k)} L_0 - 2\btau_0^{(k)} \bL_0} \CL_\Sigma \rangle
}
\\
&\qquad = \sum_{n_1\dots n_k} \overline{ 
\langle n_1 | \CL_\Sigma\rangle^*
\langle n_1 | \CL_\Sigma\rangle
\cdots
\langle n_k | \CL_\Sigma\rangle^*
\langle n_k | \CL_\Sigma\rangle\notag
}
\exp\left[-2\sum_{i=1}^k\left( \tau_0^{(i)} h_{n_i} + \btau_0^{(i)}\bh_{n_i} \right) \right]
\end{align}
Descendants are subleading at large $c$, so we have used simple exponentials (scaling blocks) here rather than Virasoro conformal blocks.
 The fully connected, cyclic Gaussian contraction is:
\begin{align}
Z_k &= \sum_n \left(\overline{| \langle n| \CL_\Sigma \rangle|^{2}}\right)^k
\exp\left[-2\sum_{i=1}^k\left( \tau_0^{(i)} h_n + \btau_0^{(i)}\bh_{n} \right) \right] \ .
\end{align}
Now we use the variance \eqref{variance}, and replace the sum over states by an integral over weights, $\sum_n \to \int_{c/24}^{\infty} dh d\bh \rho_0(h)\rho_0(\bh)$. The result is exactly the gravity answer, \eqref{zkgrav}. 

A nearly identical calculation of the wormhole with $k$ torus boundaries, and two line defects on each boundary, probes the matrix elements between two heavy states as in the second line of \eqref{variance}. Again the gravity answer agrees with the ensemble average.

The conclusion is that semiclassical gravity agrees with the ensemble average over CFT data, with Gaussian statistics.

\subsection{The 2-boundary Maldacena-Maoz wormhole}
    The case with two boundaries, $k=2$, is the Maldacena-Maoz wormhole \cite{Maldacena:2004rf}. The general analysis above applies to this case, but we will explain it in a bit more detail to make the connection to \cite{Chandra:2022bqq, Collier:2023fwi}.\footnote{See also \cite{Abajian:2023bqv}, in which the Maldacena-Maoz wormhole arises in a surprising and novel way when one constructs the bulk solution with heavy operators inserted at the boundary.} The metric of the 2-boundary wormhole, with equal moduli on the two ends, is \cite{Maldacena:2004rf}
\begin{align}\label{metricmm}
ds^2 = d\rho^2 + \cosh^2 \rho e^{\Phi} |dz|^2 \ .
\end{align}
The Einstein equation requires $e^{\Phi} |dz|^2 $ to be hyperbolic away from matter sources. Including thin shells, one finds as described in section \ref{secblackholes} that the Liouville field is sourced by the line defect $L_\Sigma$. 

In section \ref{secLiouville} we considered the case where $e^{\Phi}|dz|^2$ is a hyperbolic metric on the disk, and in this case the two boundaries components at $\rho \to \pm \infty$ are glued together at the boundary of the disk. That solution has only one asymptotic boundary. By contrast, the 2-boundary wormhole is obtained from a hyperbolic metric on a compact surface. 

For example, there is a 2-boundary wormhole with a single thin shell going across, and Liouville solution given by \eqref{phi1pt}. This calculates the ensemble average of $|\langle \CL_{\Sigma} \rangle_{\rm sphere}|^2$. Since $D_\Sigma$ is a charged line operator, $\langle D_\Sigma \rangle=0$. Therefore, we can interpret the wormhole amplitude contributing to $\overline{|\langle D_\Sigma \rangle|^2}$ as a global charge violating amplitude, along the lines of \cite{Chen:2020ojn,Hsin:2020mfa,Bah:2022uyz}. Also, since $\langle D_\Sigma \rangle$ is the $g$-function of the line defect in a large-$c$ holographic CFT, we see that the $g$-function is random with vanishing mean and variance set by the $g$-function of the corresponding line defect $L_\Sigma$ in Liouville CFT calculated in \eqref{defect1}.

Similarly, we can use the Liouville solution on a sphere with one line defect and two conical defects described in Appendix \ref{applineconic} to construct a two-boundary wormhole that computes the variance of the ensemble,
\begin{equation}
\begin{split}
  \overline{|\bra{i}D_\Sigma \ket{j}|^2}\approx |C_0^D(h_i,h_j)|^2 & \approx Z_{\text{grav}}\left [
\vcenter{\hbox{
\begin{overpic}[width=2in,grid=false]{figures/defectC0wormhole.pdf}
\end{overpic}
}} \right ]\\
&\approx \mathcal{N}^2\exp\left[\frac{c}{3}c_L(\gamma(h_i),\gamma(h_j))+S(h_i)+S(h_j)\right]
\end{split}
\end{equation}
Here $i,j$ denote scalar primary operators in the dual CFT of weights $h_i=\frac{c}{6}\eta_i(1-\eta_i)$ and $h_j=\frac{c}{6}\eta_j(1-\eta_j)$ below the black hole threshold. In the last line, the Cardy factors are pure phases, which cancel the phases in the analytic continuation of $c_L$ below threshold, so that the exponent is real. Similarly when written in terms of $C_0^D$, for scalar states below threshold, the magnitude is
\begin{align}
|C_0^D(h_i, h_j)| = \exp\left[ \frac{c}{6}c_D(\gamma(h_i), \gamma(h_j)) + S(h_i) + S(h_j) \right] \ . 
\end{align}
In the figure above, the wormhole is drawn in the hyperobolic metric and involves a shell and two particles going through it. The metric on this wormhole is given by \eqref{metricmm} with the Liouville field given by \eqref{phi1ptdef}. The wormhole exists as a saddle only when the generalized Gauss-Bonnet constraint \eqref{genGB} is satisfied. The gravitational action for the wormhole is renormalised in a way similar to that described in \cite{Chandra:2022bqq} for the sphere 3-point wormhole, which corresponds to dividing the unrenormalised partition function by the two-point functions of the local operators. One can perhaps use the idea of \cite{Abajian:2023bqv} to construct a two-boundary wormhole that computes the ensemble average of $|\bra{i}D_\Sigma \ket{j}|^2$ when one or both of the local operators are above the black hole threshold.

Another example is the wormhole with two defects going through. In this case, which is the wormhole of section \ref{ss:wormholek} with $k=2$, the metric is \eqref{metricmm}, with the Liouville field in \eqref{phi2cyl}. This wormhole calculates the ensembled-averaged product of two 2-point functions,
\begin{align}
Z_2(\tau_0^{(1)}, \btau_0^{(1)}, \tau_0^{(2)}, \btau_0^{(2)})
= 
\overline{
\langle \CL_\Sigma^{\dagger} e^{-2\tau_0^{(1)} L_0 - 2\btau_0^{(1)} \bL_0} \CL_\Sigma \rangle
\langle \CL_\Sigma^{\dagger} e^{-2\tau_0^{(2)} L_0 - 2\btau_0^{(2)} \bL_0} \CL_\Sigma \rangle
}
\end{align}
in the special kinematics with $\tau_0^{(2)} = -\btau_0^{(1)}$, such that the operator insertions on the two ends of the wormhole are mirror images. The generalization to independent moduli is achieved by the discussion above in section \ref{ss:wormholek}.

This amplitude is, in fact, a product of Liouville correlators, just like the Maldacena-Maoz wormholes for local operators studied in \cite{Chandra:2022bqq}.  Recall that the universal matrix element $C_0^{\CL}$ is related to the matrix element of the Liouville line defect by  
\begin{align}
 C_0^{\CL}(h)  =  \rho_0(h)^{-1/2} \left| \langle 0|\hat{L}_{\Sigma}| h\rangle \right| , 
\end{align}
as shown in section \ref{secbootstrap}. (Recall that $\hat{L}_\Sigma$ is the Liouville defect normalized to match the identity block, \eqref{defLhat}.) In the wormhole amplitude \eqref{zkgrav} derived from both gravity and the CFT ensemble, we set $k=2$ and find
\begin{align}
Z_2 \approx \left| \int_{c/24}^{\infty} dh |\langle 0|\hat{L}_\Sigma|h\rangle|^2 e^{-2h(\tau_0^{(1)} + \tau_0^{(2)})}\right|^2
\end{align}
When expressed in terms the Liouville line defect, all factors of the Cardy entropy have disappeared from the integrand. The right-hand side is a product of Liouville defect 2-point functions, so we have 
\begin{align}
Z_2  =&\overline{
\langle \CL_\Sigma^{\dagger} e^{-2\tau_0^{(1)} L_0 - 2\btau_0^{(1)} \bL_0} \CL_\Sigma \rangle
\langle \CL_\Sigma^{\dagger} e^{-2\tau_0^{(2)} L_0 - 2\btau_0^{(2)} \bL_0} \CL_\Sigma \rangle
}\\
&\qquad \approx 
 \langle \hat{L}_\Sigma e^{-2( \tau_0^{(1)} +\tau_0^{(2)})L_0} \hat{L}_\Sigma \rangle
 \langle \hat{L}_\Sigma e^{-2( \btau_0^{(1)}+\btau_0^{(2)})L_0} \hat{L}_\Sigma\rangle
\end{align}
The left-hand side is an ensemble average in the compact CFT, and the right-hand side is calculated in Liouville. Note the rearrangement of moduli on the two sides of the equation, which was also seen in \cite{Chandra:2022bqq,Collier:2023fwi}. Based on the results of those papers we anticipate that this formula actually gives the exact gravity amplitude for this topology, though here we have only matched the classical action.

\ \\ 

\ \\

\noindent \textbf{Acknowledgments}\\
We thank Scott Collier, Clay Cordova, Zohar Komargodski, Yuya Kusuki, Alex Maloney, Baur Mukhametzhanov, Keivan Namjou, David Simmons-Duffin, and Wayne Weng for discussions. The work of JC and TH is supported by NSF grant PHY-2014071. We also acknowledge support by grant NSF PHY-1748958 to the Kavli Institute for Theoretical Physics (KITP) where some of this work was done. V.M. is supported in part by the Murata Family Fellowship at McGill University.

\appendix


\section{Liouville formulas} \label{liouvilleformulas}

We use ZZ conventions \cite{Zamolodchikov:1995aa,Zamolodchikov:2001ah} with the Liouville cosmological constant chosen as $\mu = \frac{1}{4\pi b^2}$. The Liouville reflection amplitude is
\begin{align}
s_L(P) &=\left( \frac{\Gamma(b^2)}{4b^2\Gamma(1-b^2)}\right)^{-2iP/b} \frac{\Gamma(1+2iP/b) \Gamma(1+2ibP) }{ \Gamma(1-2iP/b)\Gamma(1-2ibP)} \ . 
\end{align}
For real $P$ this is a pure phase. The DOZZ structure constants \cite{Dorn:1994xn,Zamolodchikov:1995aa} satisfy for each entry
\begin{align}
C_L(P_1, P_2, -P) = -s_L(-P) C_L(P_1, P_2, P) \ . 
\end{align}
The ZZ wavefunction is
\begin{align}
\langle ZZ | P \rangle\!\rangle &= i s_L(P)^{1/2} \sqrt{\rho_0(P)}\\
\langle\!\langle P | ZZ \rangle &= \langle ZZ| -P \rangle\!\rangle =  -i s_L(P)^{-1/2}  \sqrt{\rho_0(P)}
\end{align}
where $\rho_0$ is the modular transformation of the Virasoro vacuum character, i.e. Cardy density of states,
\begin{align} \label{rho0}
\rho_0(P) &= | \langle ZZ|P\rangle\!\rangle |^2 
= 2^{3/2} \sinh(2\pi b P ) \sinh(2 \pi P/b)
\end{align}
This satisfies the fusion transformation
\begin{align}
\chi_0(-1/\tau) = \int_{-\infty}^\infty dP \rho_0(P)\chi_{h_{P}}(\tau) \ . 
\end{align}
The one-point function on the disk is
\begin{align}
\langle V_\alpha(z) \rangle_{ZZ} &= \frac{U(\alpha)}{ |z - \bz|^{2h_\alpha} }
\end{align}
with
\begin{align}
U(P) &=- \left( \frac{\Gamma(b^2)}{4b^2\Gamma(1-b^2)}\right)^{-\alpha/b} \frac{2iP(b+b^{-1})\Gamma(1+b^2)\Gamma(1 + b^{-2})}{ \Gamma(1-2iP/b)\Gamma(1-2iPb) }
\end{align}
This satisfies the reflection identity
\begin{align}
U(-P) &= -s_L(-P)U(P)
\end{align}
Define the constant
\begin{align}
a_b &= \left( \frac{\Gamma(b^2)}{4\Gamma(1-b^2)}\right)^{-1/2 - 1/(2b^2)} \frac{b \Gamma(2 + b^{-2})\Gamma(1 + b^2)}{2^{3/4}\pi}
\end{align}
This is formally related to the reflection coefficient of the vacuum state upon analytically continuing $P \to P_{vac} = \frac{i}{2}(b+b^{-1})$,
\begin{align}
a_b = -1/ (\rho_0(P_{vac})s_L(P_{vac}))^{1/2} \ , 
\end{align}
which is equivalent to $U(P_{vac}) = 1$.  In the semiclassical limit
\begin{align}
\log a_b \sim \frac{c}{6}(\log 2 - 1) \ ,
\end{align}
and the condition $U(P_{vac})=1$ corresponds to the choice of constant added in \eqref{LiouvdiskZZ}.
Some useful relations are
\begin{align}
U(P) &= a_b  s_L(P) \langle\!\langle P |ZZ\rangle = -i a_b s_L(P)^{1/2} \sqrt{\rho_0(P)} \\
\langle ZZ|P\rangle\!\rangle &= -a_b^{-1}U(P) 
\end{align}
The identity block for the 4-point function on the plane, as described around \eqref{F0ZZ} in the main text, is 
\begin{align}
{\cal F}_{\id} &= \frac{ \langle ZZ | V_{\alpha_1}(z_1) V_{\alpha_2}(z_2)|0\rangle }{U(P_1)U(P_2)}
\end{align}
Doing the OPE on the two vertex operators gives the representation \cite{Zamolodchikov:2001ah}
\begin{align}
{\cal F}_{\id} &= \frac{1}{4\pi} \int_{-\infty}^{\infty} dP \frac{C_L(P_1, P_2, P)U(-P)}{U(P_1)U(P_2)} \widetilde{{\cal F}}_P
\end{align}
$\widetilde{\cal F}_P$ is the conformal block in the dual channel cut across the ZZ boundary.
Using the reflection properties of the OPE coefficients and 1-point functions this can be recast as
\begin{align}
{\cal F}_{\id} 
&= \frac{1}{4\pi}\int_0^\infty dP \frac{C_L(P_1, P_2, P)U(-P) + C_L(P_1, P_2, -P)U(P)}{U(P_1)U(P_2)} \widetilde{{\cal F}}_P\\
&= -\frac{1}{2\pi} \int_0^{\infty} dP \frac{C_L(P_1, P_2, P)U(P)s_L(-P)}{U(P_1)U(P_2)} \widetilde{{\cal F}}_P\\
&= \frac{1}{2\pi} \int_0^\infty dP \rho_0(P) a_b^2 \frac{C_L(P_1,P_2,P)}{U(P_1)U(P_2)U(P)} \widetilde{{\cal F}}_P
\end{align}
Comparing to \eqref{fourtran} we see that the two-point function in an Ishibashi state is
\begin{align}\label{ishibashi2}
\langle\!\langle P | V_{\alpha_1}(z_1) V_{\alpha_2}(z_2)|0\rangle
&= -\frac{1}{a_b} C_L(P_1, P_2, -P) \widetilde{\cal F}_P
\end{align}

\section{Liouville solutions with line defects and conical defects} \label{seclineconical}

In this appendix, we give some examples of Liouville correlators of line defects in the presence of conical defect insertions. The Liouville actions computed here serve to provide some consistency checks for the matrix elements of the line operator $L_{\Sigma}$ derived in section \ref{secLiouville} when analytically continued to below the threshold. In addition, the geometries discussed in this section can be used to construct 2-boundary wormholes in 3D gravity with the gravitational action of the wormholes expected to be given by two copies of the corresponding Liouville actions computed in this appendix. 

\subsection{The 1-point function of line defect in a conical defect background} \label{applineconic}

We generalize the calculation of section \ref{defsphone} to compute the 1-point function of the line defect on the sphere with conical defects inserted at the poles. Working in the cylinder frame, the Liouville solution used to compute the resulting matrix element between conical defect states of conformal weights $h=\frac{c}{6}\eta(1-\eta)$ and $h'=\frac{c}{6}\eta'(1-\eta')$ is:
\begin{align}\label{phi1ptdef}
\Phi = \begin{cases}
-2\log \left( \frac{1}{1-2\eta}\sinh\left((1-2\eta)(y+ A)\right)\right) & y > 0\\
-2\log \left( \frac{1}{1-2\eta'}\sinh\left((1-2\eta')(A'-y)\right)\right)  & y< 0
\end{cases}
\end{align}
$A$ and $A'$ are determined by solving the continuity and junction conditions for the Liouville field across the line defect at $y=0$. Now, we need to modify the regulated Liouville action to incorporate the effect of operator insertions at $y\to \pm \infty$,
\begin{multline}\label{cyl1pactiondef}
S_L = \frac{1}{4\pi}\int_{\Gamma} d^2z\left( \p \Phi \bar{\p}\Phi +e^{\Phi}\right) - \frac{m}{4\pi} \int_\Sigma dz \Phi+ \frac{(1-2\eta)}{4\pi}\int_{\Gamma_+} \!\! dz \Phi + \frac{(1-2\eta')}{4\pi}\int_{\Gamma_-} \!\! dz \Phi\\ + \frac{T}{2}(1-2\eta)^2+\frac{T}{2}(1-2\eta')^2+2(1-\log2)
\end{multline}
where $\Gamma$ is the region $y \in [-T,T]$ and $\Gamma_{\pm}$ are the boundaries at $y = \pm T$ (both oriented toward increasing Re $z$), with $T \to \infty$ at the end. The boundary terms ensure that the action has a good variational principle given the asymptotics
\begin{equation}
   \partial_y \Phi \to 
\begin{cases}
   -2(1-2\eta)  & y \to \infty \\
    2(1-2\eta')  & y \to -\infty
\end{cases}
\end{equation}
The on-shell action evaluates to 
\begin{multline} \label{lineconical}
    S_L=-(1-2\eta)\sinh^{-1}(\frac{1-2\eta}{r_0})-(1-2\eta')\sinh^{-1}(\frac{1-2\eta'}{r_0})+m-m\log(r_0)\\+(1-2\eta)(\log(1-2\eta)+\log 2-1 )+(1-2\eta')(\log(1-2\eta')+\log 2-1 )+2(1-\log 2)
\end{multline}
where $r_0$ satisfies
\begin{equation}
    r_0=\frac{(1-2\eta)}{\sinh((1-2\eta)A)}=\frac{(1-2\eta')}{\sinh((1-2\eta')A')}
\end{equation}
In terms of $r_0$, the junction condition can be expressed as
\begin{equation}
   m=\sqrt{(1-2\eta)^2+r_0^2}+\sqrt{r_0^2+(1-2\eta')^2}
\end{equation}
We see that a solution to this junction condition exists only if the parameters obey the following inequality,
\begin{equation} \label{genGB}
  \eta+\eta'+\frac{m}{2}>1
\end{equation}
This is a generalization of the Gauss-Bonnet constraint for a hyperbolic metric on a sphere to include line defects.
Note that the on-shell Liouville action in \eqref{lineconical} can be expressed as
\begin{equation}
  e^{-\frac{c}{6}S_L}\approx -a_b^2\rho_0(P)^{1/2}s_L(P)^{1/2}\rho_0(P')^{1/2}s_L(P')^{1/2}\exp\left[\frac{c}{6}c_L(\gamma(h),\gamma(h'))+\frac{1}{2}S(h)+\frac{1}{2}S(h')\right]
\end{equation}
On setting $\eta'=0$, we get
\begin{equation}
  e^{-\frac{c}{6}S_L}\approx -i a_b\rho_0(P)^{1/2}s_L(P)^{1/2}\exp\left[\frac{c}{6}c_L(\gamma(h))+\frac{1}{2}S(h)\right]
\end{equation}
We define the matrix element $\bra{h}L_{\Sigma}\ket{0}$ with a relative normalisation between the vertex operator and the state $\ket{h}$ below the threshold so that
\begin{equation}
   \bra{h}L_{\Sigma}\ket{0} \approx -i \rho_0(P)^{-1/2}s_L(P)^{-1/2}\exp\left[\frac{c}{6}c_L(\gamma(h))+\frac{1}{2}S(h)\right]
\end{equation}
consistent with \eqref{finalc1}.

\subsection{The 2-point function of line defect in a conical defect background}

Now, we compute the two-point function of the line defect in a conical defect background of defect strength $\eta$ so that $h=\overline{h}=\frac{c}{6}\eta(1-\eta)$. The Liouville solution in this case is given by
\begin{align}\label{phi2cyldef}
\Phi = \begin{cases}
-2\log \left( \frac{1}{1-2\eta}\sinh\left((1-2\eta)(y-\tau_0 + A)\right)\right) & y>\tau_0  \\
-2\log \left(\frac{1}{r_H}\cos(r_H y) \right)& |y| < \tau_0 \\
-2\log \left( \frac{1}{1-2\eta}\sinh\left((1-2\eta)(-y-\tau_0 +A)\right)\right) & y < -\tau_0
\end{cases}
\end{align}
The semiclassically renormalised Liouville action with conical defects at $y=\pm \infty$ is given by
\begin{multline}
    S_L=\frac{1}{4\pi}\int_{\Gamma} d^2z(\partial\Phi\overline{\partial}\Phi+e^\Phi)-\frac{m}{4\pi}\int_{\Sigma_1} dz \Phi-\frac{m}{4\pi}\int_{\Sigma_2} dz \Phi\\+\frac{(1-2\eta)}{4\pi}\int_{ \Gamma_+} dz \Phi +T(1-2\eta)^2+2(1-\log 2)
\end{multline}
where $\Sigma_{1,2}$ are the line defects at $y=\pm \tau_0$ and $\Gamma$ is the cylinder cutoff at $y=\pm T$ denoted $\Gamma_+$. The on-shell Liouville action is given by
\begin{multline}
    S_L=2\bigg (-\frac{\tau_0 r_H^2}{2}+(1-2\eta)^2\frac{\tau_0}{2}-(1-2\eta)\sinh^{-1}(\frac{1-2\eta}{r_0})+m-m\log(r_0)\\+(1-2\eta)(\log(1-2\eta)+\log 2-1 )+1-\log 2\bigg)
\end{multline}
where $r_0$ satisfies
\begin{equation}
    r_0=\frac{(1-2\eta)}{\sinh((1-2\eta)A)}=\frac{r_H}{\cos(r_H \tau_0)}
\end{equation}
In terms of $r_0$, the junction condition can be expressed as
\begin{equation}
   m=\sqrt{(1-2\eta)^2+r_0^2}+\sqrt{r_0^2-r_H^2}
\end{equation}

\section{The Ward identity for the continuum identity block}

In this appendix, we derive the Ward identity obeyed by the continuum identity block which we integrated to evaluate the block in section \ref{secblocks}.

\subsection{Continuum identity block on the sphere} \label{secwardsph}

 The CFT stress tensor on the sphere in the cylinder frame in the presence of local primary operator insertions is
\begin{equation}
    T^{\text{cyl}}_{zz}(z)=\frac{c}{24}+\sum_i \frac{h_i}{(z-z_i)^2}+\frac{h_i}{(z-z'_i)^2}-\frac{b_i}{z-z_i}-\frac{b_i'}{z-z_i'}+\dots
\end{equation}
where the $\dots$ denote non-singular terms. The first term is the Casimir energy density and is necessary to reproduce the canonical form of the stress tensor on the plane. The operator insertions are at $z_i=-i\tau_0+\theta_i$ and $z_i'=i\tau_0+\theta_i$. By symmetry along the angular direction, the accessory parameters should be constant $b_i=b$ and $b_i'=b'$. By inversion (time reflection) symmetry, $b'=-b$. We also take the weights of all the operators to be equal $h_i=h$. Transforming to the plane ($w=e^{-iz}$) using
\begin{equation}
    T^{\text{pl}}_{ww}(w)=(\frac{\partial w}{\partial z})^{-2}(T^{\text{cyl}}_{zz}(z)-\frac{c}{12}\{w,z\})=-\frac{1}{w^2}T_{zz}^{\text{cyl}}(z)+\frac{c}{24w^2}
\end{equation}
and expanding near the poles we get
\begin{equation}
    T_{ww}(w)=\sum_i \frac{h}{(w-w_i)^2}+\frac{h}{(w-w_i')^2}-\frac{h+ib}{w_i(w-w_i)}-\frac{h-ib}{w_i'(w-w_i')}+\dots
\end{equation}
Here $w_i=e^{-\tau_0+i\theta_i}$ and $w_i'=e^{\tau_0+i\theta_i}$.
Comparing to the canonical form of the stress tensor on the plane with operator insertions, we can read off the accessory parameters,
\begin{equation}
    c_i=\frac{h+ib}{w_i} \quad \quad c_i'=\frac{h-ib}{w_i'}
\end{equation}
We can compute the energy by integrating the stress tensor around the unit circle,
\begin{equation}
    h+\frac{c}{24}=\frac{1}{2\pi i }\oint_{|w|=1}dw w T_{ww}=-iNb
\end{equation}
where $N$ is the number of operators inserted on each circle.
Now, we can derive the Ward identity for the semiclassical block $f$ defined by $\mathcal{F}=e^{-\frac{c}{6}f}$,
\begin{equation}
    \frac{c}{6}\partial_{\sigma}f=\sum_i \left(c_i\frac{\partial w_i}{\partial \sigma_0}+c_i'\frac{\partial w_i'}{\partial \sigma_0}\right )=-2iNb
\end{equation}
where $\sigma=\tau_0-i\theta$ is a complex modulus where $\theta$ is a twist. In deriving the above relation, we used the familiar form of the Ward identity relating derivatives of the block with respect to moduli to the accessory parameters,
\begin{equation}
    \frac{c}{6}\partial_{w_i} f=c_i \quad \quad \frac{c}{6}\partial_{w_i'} f=c_i'
\end{equation}
Thus, we have the Ward identity for the semiclassical continuum identity block,
\begin{equation}
    \frac{c}{12}\partial_\sigma f=h+\frac{c}{24}
\end{equation}

\subsection{Continuum identity block on the torus} \label{wardtorus}

Now, we derive the Ward identity for the continuum identity block on the torus using a procedure very similar to the one we followed for the case of the cylinder two-point function. We start with the stress tensor on the torus in the cylinder frame,
\begin{equation}
    T^{\text{cyl}}_{zz}(z)=\frac{c}{24}+\sum_i \frac{h_i}{(z-z_i)^2}+\frac{h_i}{(z-z'_i)^2}-\frac{b_i}{z-z_i}-\frac{b_i'}{z-z_i'}+\dots
\end{equation}
Here, $\text{Im}z\in (-\frac{\beta}{2},\frac{\beta}{2})$ with the boundaries identified possibly with a twist $\chi$. To this end, we define a complexified modulus, $\tau=\beta-i\chi$.
The operator insertions are at $z_i=-i\tau_0+\theta_i$ and $z_i'=i\tau_0+\theta_i$. Transforming to the plane ($w=e^{-iz}$) and expanding near the poles we get
\begin{equation}
    T_{zz}(z)=\sum_i \frac{h}{(w-w_i)^2}+\frac{h}{(w-w_i')^2}-\frac{h+ib}{w_i(w-w_i)}-\frac{h-ib}{w_i'(w-w_i')}+\dots
\end{equation}
Here $w_i=e^{-\tau_0+i\theta_i}$ and $w_i'=e^{\tau_0+i\theta_i}$. We restrict to the annulus region $|w|\in (e^{-\frac{\beta}{2}},e^{\frac{\beta}{2}})$.
Comparing to the canonical form of the stress tensor on the plane with operator insertions, we can read off the accessory parameters,
\begin{equation}
    c_i=\frac{h+ib}{w_i} \quad \quad c_i'=\frac{h-ib}{w_i'}
\end{equation}
We can relate the accessory parameters to a difference in conformal weights as follows,
\begin{equation}
    h-h'=\frac{1}{2\pi i }\oint_{|w|=e^{-\tau_0}+\delta}dw w T_{ww}-\frac{1}{2\pi i }\oint_{|w|=e^{-\tau_0}-\delta}dw w T_{ww}=-iNb
\end{equation}
where $N$ is the number of operators inserted on each circle and $\delta$ is a small positive constant.
Now, we can derive the Ward identity for the semiclassical identity block $f$, $\mathcal{F}=e^{-\frac{c}{6}f}$ which is function of two complex moduli $\tau=\beta-i\chi$ and $\sigma=\tau_0-i\theta$,
\begin{equation}
    \frac{c}{6}\partial_{\sigma}f=\sum_i \left(c_i\frac{\partial w_i}{\partial \sigma}+c_i'\frac{\partial w_i'}{\partial \sigma}\right )=-2iNb
\end{equation}
We are taking the derivative holding $\tau$ fixed.
Thus, we have the Ward identity for the semiclassical continuum identity block on the torus,
\begin{equation}
    \frac{c}{12}\partial_{\sigma}f|_{\tau}=h-h'
\end{equation}
Note that we are taking the derivative holding $\beta$ fixed.
The above Ward identity and the symmetry condition that the block is invariant under $\sigma \xleftrightarrow{} \frac{\tau}{2}-\sigma$ determines the block up to terms fixed by operator normalisation. 

\bibliographystyle{ourbst}
\bibliography{refs.bib}

\end{document}